\documentclass[
 reprint,
superscriptaddress,
preprintnumbers,
nofootinbib,
 amsmath,amssymb,
 aps,
 twocolumn,
]{revtex4-2}
\usepackage{graphicx}
\usepackage{dcolumn}
\usepackage{bm}
\usepackage{xcolor}
\usepackage{amsmath}	
\usepackage{gensymb} 
\usepackage{hyperref}
\hypersetup{colorlinks=true, linkcolor=blue, citecolor=blue, filecolor=blue, urlcolor=blue}
\usepackage[mathlines]{lineno}
\usepackage{xspace}
\usepackage{makecell}

\newcommand{\be}{\begin{equation}}
\newcommand{\ee}{\end{equation}}
\newcommand\bea{\begin{eqnarray}}
\newcommand\eea{\end{eqnarray}}
\newcommand\isd{\textsc{ISD}}
\newcommand\nside{\texttt{NSIDE}}
\newcommand\Ntpl{N_{\mathrm{tpl}}}
\newcommand\enet{\textsc{ENet}}
\newcommand\Cl{C_\ell}
\newcommand\badd[1]{b(\theta)_{\mathrm{add}}^{#1}}

\newcommand{\DNF}{\ifmmode \mathtt{DNF} \else \texttt{DNF} \fi}

\newcommand{\dchisqsub}[1]{\Delta\chi^2_{#1}}

\newcommand{\deltasub}[1]{\delta_{\rm{#1}}}

\newcommand{\decals}{\texttt{DECaLS{}}}

\newcommand{\maglim}{\textsc{MagLim}}
\newcommand{\redmagic}{\textsc{RedMaGiC}}
\newcommand{\joint}{\textsc{Joint}}
\newcommand{\gold}{\textsc{Gold}}
\newcommand{\dnf}{\textsc{DNF}}

\newcommand{\maglimpp}{\textsc{MagLim}\texttt{++}}

\newcommand{\isdthresh}{T_{\rm thresh}}

\newcommand{\cosmosis}{\textsc{CosmoSiS}}
\newcommand{\sompz}{\textsc{SOMPz}}
\newcommand{\healpix}{\textsc{HealPix}}

\def\LCDM{$\Lambda$CDM}

\newcommand\threextwopt{3$\times$2pt}
\newcommand\twoxtwopt{2$\times$2pt}
\newcommand\zsigma{\hat{\sigma}_z}
\newcommand\zmc{z_{NN}}
\newcommand\zpred{\hat{z}}
\newcommand\wtheta{w(\theta)}
\newcommand\dnfzn{\texttt{DNF\_ZN}}
\newcommand\dnfz{\texttt{DNF\_Z}}
\newcommand\dnfzsigma{\texttt{DNF\_ZSIGMA}}

\newcommand\aap{A\&A}                
\newcommand\aaps{A\&AS}              

\newcommand\apjs{ApJS}               
\newcommand\jcap{J.~Cosmology Astropart. Phys.} 
\newcommand\mnras{MNRAS}             
\begin{document}

\preprint{DES-2024-0876}
\preprint{FERMILAB-PUB-26-0030-PPD}

\title[DES Y6 \maglimpp{} Galaxy Clustering]{Dark Energy Survey Year 6 Results: \maglimpp{} Lens Sample Selection and Measurements of Galaxy Clustering}


\author{N.~Weaverdyck}
\email{NWeaverdyck@lbl.gov}
\affiliation{Berkeley Center for Cosmological Physics, Department of Physics, University of California, Berkeley, CA 94720, US}
\affiliation{Lawrence Berkeley National Laboratory, 1 Cyclotron Road, Berkeley, CA 94720, USA}

\author{M.~Rodríguez-Monroy}
\email{Martin.Rodriguez@inv.uam.es}
\affiliation{Instituto de F\'{i}sica Te\'{o}rica UAM/CSIC, Universidad Aut\'{o}noma de Madrid, 28049 Madrid, Spain}
\affiliation{Laboratoire de physique des 2 infinis Ir\`ene Joliot-Curie, CNRS Universit\'e Paris-Saclay, Bât. 100, F-91405 Orsay Cedex, France}
\affiliation{Ruhr University Bochum, Faculty of Physics and Astronomy, Astronomical Institute, German Centre for Cosmological Lensing, 44780 Bochum, Germany}

\author{J.~Elvin-Poole}
\affiliation{Department of Physics and Astronomy, University of Waterloo, 200 University Ave W, Waterloo, ON N2L 3G1, Canada}

\author{I.~Sevilla-Noarbe}
\affiliation{Centro de Investigaciones Energ\'eticas, Medioambientales y Tecnol\'ogicas (CIEMAT), Madrid, Spain}

\author{A.~Porredon}
\affiliation{Centro de Investigaciones Energ\'eticas, Medioambientales y Tecnol\'ogicas (CIEMAT), Madrid, Spain}
\affiliation{Ruhr University Bochum, Faculty of Physics and Astronomy, Astronomical Institute, German Centre for Cosmological Lensing, 44780 Bochum, Germany}

\author{S.~Avila}
\affiliation{Centro de Investigaciones Energ\'eticas, Medioambientales y Tecnol\'ogicas (CIEMAT), Madrid, Spain}

\author{S.~Lee}
\affiliation{Jet Propulsion Laboratory, California Institute of Technology, 4800 Oak Grove Dr., Pasadena, CA 91109, USA}
\affiliation{Department of Physics and Astronomy, Ohio University, Clippinger Labs, Athens, OH 45701}

\author{W.~Riquelme}
\affiliation{ICTP South American Institute for Fundamental Research, IFT-UNESP, S\~{a}o Paulo, SP 01440-070, Brazil}

\author{M.~Tabbutt}
\affiliation{Physics Department, 2320 Chamberlin Hall, University of Wisconsin-Madison, 1150 University Avenue Madison, WI  53706-1390}

\author{D.~Huterer}
\affiliation{Department of Physics, University of Michigan, Ann Arbor, MI 48109, USA}

\author{J.~Prat}
\affiliation{Nordita, KTH Royal Institute of Technology and Stockholm University, Hannes Alfv\'ens v\"ag 12, SE-10691 Stockholm, Sweden}

\author{J.~De~Vicente}
\affiliation{Centro de Investigaciones Energ\'eticas, Medioambientales y Tecnol\'ogicas (CIEMAT), Madrid, Spain}

\author{J. Mena-Fern{\'a}ndez}
\affiliation{Universit\'e Grenoble Alpes, CNRS, LPSC-IN2P3, 38000 Grenoble, France}

\author{M.~Crocce}
\affiliation{Institut d'Estudis Espacials de Catalunya (IEEC), 08034 Barcelona, Spain}
\affiliation{Institute of Space Sciences (ICE, CSIC),  Campus UAB, Carrer de Can Magrans, s/n,  08193 Barcelona, Spain}

\author{C.~S{\'a}nchez}
\affiliation{Departament de F\'{\i}sica, Universitat Aut\`{o}noma de Barcelona (UAB), 08193 Bellaterra, Barcelona, Spain}
\affiliation{Institut de F\'{\i}sica d'Altes Energies (IFAE), The Barcelona Institute of Science and Technology, Campus UAB, 08193 Bellaterra (Barcelona) Spain}

\author{G.~M.~Bernstein}
\affiliation{Department of Physics and Astronomy, University of Pennsylvania, Philadelphia, PA 19104, USA}

\author{E.~Henning}
\affiliation{Department of Physics, William Jewell College, Liberty MO 64068, USA}

\author{R.~Cawthon}
\affiliation{Oxford College of Emory University, Oxford, GA 30054, USA}

\author{A.~J.~Ross}
\affiliation{Center for Cosmology and Astro-Particle Physics, The Ohio State University, Columbus, OH 43210, USA}

\author{T.~M.~C.~Abbott}
\affiliation{Cerro Tololo Inter-American Observatory, NSF's National Optical-Infrared Astronomy Research Laboratory, Casilla 603, La Serena, Chile}

\author{M.~Aguena}
\affiliation{INAF-Osservatorio Astronomico di Trieste, via G. B. Tiepolo 11, I-34143 Trieste, Italy}
\affiliation{Laborat\'orio Interinstitucional de e-Astronomia - LIneA, Av. Pastor Martin Luther King Jr, 126 Del Castilho, Nova Am\'erica Offices, Torre 3000/sala 817 CEP: 20765-000, Brazil}

\author{S.~S.~Allam}
\affiliation{Fermi National Accelerator Laboratory, P. O. Box 500, Batavia, IL 60510, USA}

\author{O.~Alves}
\affiliation{Department of Physics, University of Michigan, Ann Arbor, MI 48109, USA}

\author{F.~Andrade-Oliveira}
\affiliation{Physik-Institut, University of Zürich, Winterthurerstrasse 190, CH-8057 Zürich, Switzerland}

\author{D.~Bacon}
\affiliation{Institute of Cosmology and Gravitation, University of Portsmouth, Portsmouth, PO1 3FX, UK}

\author{K.~Bechtol}
\affiliation{Physics Department, 2320 Chamberlin Hall, University of Wisconsin-Madison, 1150 University Avenue Madison, WI  53706-1390}

\author{E.~Bertin}
\affiliation{CNRS, UMR 7095, Institut d'Astrophysique de Paris, F-75014, Paris, France",0000-0002-3602-3664}
\affiliation{Sorbonne Universit\'es, UPMC Univ Paris 06, UMR 7095, Institut d'Astrophysique de Paris, F-75014, Paris, France",0000-0002-3602-3664}

\author{J.~Blazek}
\affiliation{Department of Physics, Northeastern University, Boston, MA 02115, USA}

\author{S.~Bocquet}
\affiliation{University Observatory, LMU Faculty of Physics, Scheinerstr. 1, 81679 Munich, Germany}

\author{D.~Brooks}
\affiliation{Department of Physics \& Astronomy, University College London, Gower Street, London, WC1E 6BT, UK}

\author{R.~Camilleri}
\affiliation{School of Mathematics and Physics, University of Queensland,  Brisbane, QLD 4072, Australia}

\author{A.~Carnero~Rosell}
\affiliation{Instituto de Astrofisica de Canarias, E-38205 La Laguna, Tenerife, Spain}
\affiliation{Laborat\'orio Interinstitucional de e-Astronomia - LIneA, Av. Pastor Martin Luther King Jr, 126 Del Castilho, Nova Am\'erica Offices, Torre 3000/sala 817 CEP: 20765-000, Brazil}
\affiliation{Universidad de La Laguna, Dpto. Astrofísica, E-38206 La Laguna, Tenerife, Spain}

\author{J.~Carretero}
\affiliation{Institut de F\'{\i}sica d'Altes Energies (IFAE), The Barcelona Institute of Science and Technology, Campus UAB, 08193 Bellaterra (Barcelona) Spain}

\author{F.~J.~Castander}
\affiliation{Institut d'Estudis Espacials de Catalunya (IEEC), 08034 Barcelona, Spain}
\affiliation{Institute of Space Sciences (ICE, CSIC),  Campus UAB, Carrer de Can Magrans, s/n,  08193 Barcelona, Spain}

\author{A.~Choi}
\affiliation{NASA Goddard Space Flight Center, 8800 Greenbelt Rd, Greenbelt, MD 20771, USA}

\author{L.~N.~da Costa}
\affiliation{Laborat\'orio Interinstitucional de e-Astronomia - LIneA, Av. Pastor Martin Luther King Jr, 126 Del Castilho, Nova Am\'erica Offices, Torre 3000/sala 817 CEP: 20765-000, Brazil}

\author{M.~E.~da Silva Pereira}
\affiliation{Hamburger Sternwarte, Universit\"{a}t Hamburg, Gojenbergsweg 112, 21029 Hamburg, Germany}

\author{T.~M.~Davis}
\affiliation{School of Mathematics and Physics, University of Queensland,  Brisbane, QLD 4072, Australia}

\author{H.~T.~Diehl}
\affiliation{Fermi National Accelerator Laboratory, P. O. Box 500, Batavia, IL 60510, USA}

\author{C.~Doux}
\affiliation{Department of Physics and Astronomy, University of Pennsylvania, Philadelphia, PA 19104, USA}
\affiliation{Universit\'e Grenoble Alpes, CNRS, LPSC-IN2P3, 38000 Grenoble, France}

\author{A.~Drlica-Wagner}
\affiliation{Department of Astronomy and Astrophysics, University of Chicago, Chicago, IL 60637, USA}
\affiliation{Fermi National Accelerator Laboratory, P. O. Box 500, Batavia, IL 60510, USA}
\affiliation{Kavli Institute for Cosmological Physics, University of Chicago, Chicago, IL 60637, USA}

\author{T.~Eifler}
\affiliation{Department of Astronomy/Steward Observatory, University of Arizona, 933 North Cherry Avenue, Tucson, AZ 85721-0065, USA",0000-0002-1894-3301}
\affiliation{Department of Astronomy/Steward Observatory, University of Arizona, 933 North Cherry Avenue, Tucson, AZ 85721-0065, USA",0000-0002-1894-3301}

\author{S.~Everett}
\affiliation{California Institute of Technology, 1200 East California Blvd, MC 249-17, Pasadena, CA 91125, USA}

\author{A.~Evrard}
\affiliation{Department of Astronomy, University of Michigan, Ann Arbor, MI 48109, USA",0000-0002-4876-956X}
\affiliation{Department of Physics, University of Michigan, Ann Arbor, MI 48109, USA",0000-0002-4876-956X}

\author{B.~Flaugher}
\affiliation{Fermi National Accelerator Laboratory, P. O. Box 500, Batavia, IL 60510, USA}

\author{J.~Garc\'ia-Bellido}
\affiliation{Instituto de Fisica Teorica UAM/CSIC, Universidad Autonoma de Madrid, 28049 Madrid, Spain}

\author{M.~Gatti}
\affiliation{Kavli Institute for Cosmological Physics, University of Chicago, Chicago, IL 60637, USA}

\author{E.~Gaztañaga}
\affiliation{Institut d'Estudis Espacials de Catalunya (IEEC), 08034 Barcelona, Spain",0000-0001-9632-0815}
\affiliation{Institute of Cosmology and Gravitation, University of Portsmouth, Portsmouth, PO1 3FX, UK",0000-0001-9632-0815}
\affiliation{Institute of Space Sciences (ICE, CSIC),  Campus UAB, Carrer de Can Magrans, s/n,  08193 Barcelona, Spain",0000-0001-9632-0815}

\author{G.~Giannini}
\affiliation{Institute of Space Sciences (ICE, CSIC),  Campus UAB, Carrer de Can Magrans, s/n,  08193 Barcelona, Spain",0000-0002-3730-1750}
\affiliation{Kavli Institute for Cosmological Physics, University of Chicago, Chicago, IL 60637, USA",0000-0002-3730-1750}

\author{D.~Gruen}
\affiliation{University Observatory, LMU Faculty of Physics, Scheinerstr. 1, 81679 Munich, Germany}

\author{G.~Gutierrez}
\affiliation{Fermi National Accelerator Laboratory, P. O. Box 500, Batavia, IL 60510, USA}

\author{S.~R.~Hinton}
\affiliation{School of Mathematics and Physics, University of Queensland,  Brisbane, QLD 4072, Australia}

\author{D.~L.~Hollowood}
\affiliation{Santa Cruz Institute for Particle Physics, Santa Cruz, CA 95064, USA}

\author{K.~Honscheid}
\affiliation{Center for Cosmology and Astro-Particle Physics, The Ohio State University, Columbus, OH 43210, USA}
\affiliation{Department of Physics, The Ohio State University, Columbus, OH 43210, USA}

\author{B.~Jain}
\affiliation{Department of Physics and Astronomy, University of Pennsylvania, Philadelphia, PA 19104, USA",0000-0002-8220-3973}

\author{T.~Kacprzak}
\affiliation{Department of Physics, ETH Zurich, Wolfgang-Pauli-Strasse 16, CH-8093 Zurich, Switzerland}

\author{K.~Kuehn}
\affiliation{Australian Astronomical Optics, Macquarie University, North Ryde, NSW 2113, Australia}
\affiliation{Lowell Observatory, 1400 Mars Hill Rd, Flagstaff, AZ 86001, USA}

\author{O.~Lahav}
\affiliation{Department of Physics \& Astronomy, University College London, Gower Street, London, WC1E 6BT, UK}

\author{J.~L.~Marshall}
\affiliation{George P. and Cynthia Woods Mitchell Institute for Fundamental Physics and Astronomy, and Department of Physics and Astronomy, Texas A\&M University, College Station, TX 77843,  USA}

\author{F.~Menanteau}
\affiliation{Center for Astrophysical Surveys, National Center for Supercomputing Applications, 1205 West Clark St., Urbana, IL 61801, USA}
\affiliation{Department of Astronomy, University of Illinois at Urbana-Champaign, 1002 W. Green Street, Urbana, IL 61801, USA}

\author{R.~Miquel}
\affiliation{Instituci\'o Catalana de Recerca i Estudis Avan\c{c}ats, E-08010 Barcelona, Spain}
\affiliation{Institut de F\'{\i}sica d'Altes Energies (IFAE), The Barcelona Institute of Science and Technology, Campus UAB, 08193 Bellaterra (Barcelona) Spain}

\author{J.~J.~Mohr}
\affiliation{University Observatory, LMU Faculty of Physics, Scheinerstr. 1, 81679 Munich, Germany}

\author{J.~Muir}
\affiliation{Department of Physics, University of Cincinnati, Cincinnati, Ohio 45221, USA}
\affiliation{Perimeter Institute for Theoretical Physics, 31 Caroline St. North, Waterloo, ON N2L 2Y5, Canada}

\author{J.~Myles}
\affiliation{Department of Astrophysical Sciences, Princeton University, Peyton Hall, Princeton, NJ 08544, USA}

\author{R.~Nichol}
\affiliation{Institute of Cosmology and Gravitation, University of Portsmouth, Portsmouth, PO1 3FX, UK}

\author{R.~L.~C.~Ogando}
\affiliation{Centro de Tecnologia da Informa\c{c}\~ao Renato Archer, Campinas, SP, Brazil - 13069-901\\
Observat\'orio Nacional, Rio de Janeiro, RJ, Brazil - 20921-400\\}

\author{A.~Palmese}
\affiliation{Department of Physics, Carnegie Mellon University, Pittsburgh, Pennsylvania 15312, USA}

\author{M.~Paterno}
\affiliation{Fermi National Accelerator Laboratory, P. O. Box 500, Batavia, IL 60510, USA}

\author{W.~J.~Percival}
\affiliation{Department of Physics and Astronomy, University of Waterloo, 200 University Ave W, Waterloo, ON N2L 3G1, Canada}
\affiliation{Perimeter Institute for Theoretical Physics, 31 Caroline St. North, Waterloo, ON N2L 2Y5, Canada}

\author{A.~A.~Plazas~Malag\'on}
\affiliation{Kavli Institute for Particle Astrophysics \& Cosmology, P. O. Box 2450, Stanford University, Stanford, CA 94305, USA}
\affiliation{SLAC National Accelerator Laboratory, Menlo Park, CA 94025, USA}

\author{R.~Rosenfeld}
\affiliation{ICTP South American Institute for Fundamental Research\\ Instituto de F\'{\i}sica Te\'orica, Universidade Estadual Paulista, S\~ao Paulo, Brazil}
\affiliation{Laborat\'orio Interinstitucional de e-Astronomia - LIneA, Av. Pastor Martin Luther King Jr, 126 Del Castilho, Nova Am\'erica Offices, Torre 3000/sala 817 CEP: 20765-000, Brazil}

\author{E.~Rykoff}
\affiliation{Kavli Institute for Particle Astrophysics \& Cosmology, P. O. Box 2450, Stanford University, Stanford, CA 94305, USA",0000-0001-9376-3135}
\affiliation{SLAC National Accelerator Laboratory, Menlo Park, CA 94025, USA",0000-0001-9376-3135}

\author{S.~Samuroff}
\affiliation{Department of Physics, Northeastern University, Boston, MA 02115, USA}
\affiliation{Institut de F\'{\i}sica d'Altes Energies (IFAE), The Barcelona Institute of Science and Technology, Campus UAB, 08193 Bellaterra (Barcelona) Spain}

\author{E.~Sanchez}
\affiliation{Centro de Investigaciones Energ\'eticas, Medioambientales y Tecnol\'ogicas (CIEMAT), Madrid, Spain}

\author{D.~Sanchez Cid}
\affiliation{Centro de Investigaciones Energ\'eticas, Medioambientales y Tecnol\'ogicas (CIEMAT), Madrid, Spain}
\affiliation{Physik-Institut, University of Zürich, Winterthurerstrasse 190, CH-8057 Zürich, Switzerland}

\author{E.~Sheldon}
\affiliation{Brookhaven National Laboratory, Bldg 510, Upton, NY 11973, USA",0000-0001-9194-0441}

\author{N.~Sherman}
\affiliation{Institute for Astrophysical Research, Boston University, 725 Commonwealth Avenue, Boston, MA 02215, USA",0000-0001-5399-0114}

\author{M.~Smith}
\affiliation{Physics Department, Lancaster University, Lancaster, LA1 4YB, UK}

\author{M.~Soares-Santos}
\affiliation{Physik-Institut, University of Zürich, Winterthurerstrasse 190, CH-8057 Zürich, Switzerland}

\author{E.~Suchyta}
\affiliation{Computer Science and Mathematics Division, Oak Ridge National Laboratory, Oak Ridge, TN 37831}

\author{M.~E.~C.~Swanson}
\affiliation{Center for Astrophysical Surveys, National Center for Supercomputing Applications, 1205 West Clark St., Urbana, IL 61801, USA}

\author{T.~Gregory}
\affiliation{Kavli Institute for Cosmological Physics, University of Chicago, Chicago, IL 60637, USA",0000-0002-3730-1750}

\author{D.~Thomas}
\affiliation{Institute of Cosmology and Gravitation, University of Portsmouth, Portsmouth, PO1 3FX, UK}

\author{C.~To}
\affiliation{Department of Astronomy and Astrophysics, University of Chicago, Chicago, IL 60637, USA}

\author{D.~L.~Tucker}
\affiliation{Fermi National Accelerator Laboratory, P. O. Box 500, Batavia, IL 60510, USA}

\author{V.~Vikram}
\affiliation{Central University of Kerala, Kasaragod, Kerala, India}

\author{M.~Yamamoto}
\affiliation{Department of Astrophysical Sciences, Princeton University, Peyton Hall, Princeton, NJ 08544, USA}
\affiliation{Department of Physics, Duke University Durham, NC 27708, USA}

\author{B.~Yanny}
\affiliation{Fermi National Accelerator Laboratory, P. O. Box 500, Batavia, IL 60510, USA}

\collaboration{DES Collaboration}

\date{\today}

\begin{abstract}
Galaxy clustering is a sensitive probe of the expansion history and growth of structure of the universe, and key degeneracies can be broken by combining these data with measurements of cosmic shear and galaxy-galaxy lensing (a so-called \threextwopt{} analysis). {The largest and least biased statistical samples of galaxies for use in clustering analyses can be collected photometrically through large imaging surveys. However, selecting} clean photometric {sub}samples {for cosmology} are crucial for avoiding contamination that can bias cosmological constraints. Here we present the \maglimpp{} galaxy sample, selected to optimize for cosmological constraining power and incorporating an array of novel quality cuts to identify and remove residual contamination. This sample comes from the full six years of observations from the Dark Energy Survey. We present measurements of the two-point angular clustering ($w(\theta)$) of 9,186,205 galaxies distributed over 4031 sq. degrees and in six tomographic redshift bins {centered at $\bar{z}\approx$ [0.31, 0.44, 0.62, 0.78, 0.90, 1.01]}. These measurements are used as part of the \threextwopt{} and other DES Y6 legacy cosmological analyses in companion works. We describe the battery of null tests and mitigation schemes implemented to address observational, astrophysical, and methodological systematics in the analysis. The resulting $w(\theta)$ measurements have a S/N = 149 (90.2 for linear scales only), which we use to place galaxy-clustering-only constraints on the matter density of the Universe, $\Omega_m=0.311^{+0.023}_{-0.035}$, and amplitude of galaxy clustering in each redshift bin, $b_i\sigma_8=[1.16^{+0.04}_{-0.06},\ 1.40^{+0.04}_{-0.06},\ 1.57^{+0.04}_{-0.06},\ 1.59^{+0.04}_{-0.05},\ 1.50^{+0.04}_{-0.05},\ 1.74^{+0.06}_{-0.08}]$.
\end{abstract}

\maketitle


\section{Introduction}
Measurements of the large scale structure (LSS) of the Universe are a key probe of cosmology, providing constraints on both the expansion history of the Universe and how cosmic structure grows over time \citep[e.g.,][]{Dodelson:2020bqr,Huterer:2022dds}. The last few decades have seen a rapid increase in the number, volume and depth of LSS surveys, heralding an era in which LSS surveys alone can provide cosmological parameter constraints that are competitive and complementary with those  from the cosmic microwave background (CMB) and Type Ia supernovae (SNIa). The modern era of LSS surveys was effectively ushered in by the Two-degree Field Galaxy Redshift Survey \cite{2DFGRS:2001zay} and Sloan Digital Sky Survey \cite{SDSS:2000hjo}, including its extensions the Baryon Oscillation Sky Survey (BOSS; \cite{BOSS:2012dmf}) and the extended BOSS (eBOSS; \cite{eBOSS:2015jyv}). These were followed by three major contemporary photometric surveys: the Dark Energy Survey\footnote{\url{https://www.darkenergysurvey.org/}} (DES; \cite{DES:2005,DES-3x2:2018,DES:2021wwk}), the Kilo-Degree Survey\footnote{\url{http://kids.strw.leidenuniv.nl/}} (KiDS; \cite{Kuijken2015,Heymans2020}), and the Hyper Suprime-Cam (HSC) Subaru Strategic Program\footnote{\url{https://www.naoj.org/Projects/HSC/}} \cite{Aihara2017,Hikage2018,Hamana2019}, alongside large spectroscopic surveys like the Dark Energy Spectroscopic Instrument  (DESI\footnote{\url{https://www.desi.lbl.gov/}};  \cite{DESI:2016fyo,DESI:2016igz}), Prime Focus Spectrograph (PFS) survey on the Subaru telescope \cite{Tamura:2016wsg}, and Hobby Eberly Telescope Dark Energy Experiment (HETDEX\footnote{\url{https://hetdex.org/}}; \cite{Gebhardt:2021vfo}). Space-based missions Euclid\footnote{\url{https://www.esa.int/Science_Exploration/Space_Science/Euclid}} \cite{Euclid:2024yrr}, SPHEREx \cite{dore2015cosmologyspherexallskyspectral} and the Nancy Grace Roman Space Telescope\footnote{\url{https://roman.gfc.nasa.gov/}} \cite{Eifler:2020vvg} will complement the upcoming ground-based Legacy Survey of Space and Time\footnote{\url{https://www.lsst.org/}} (LSST) on the Vera Rubin Observatory \cite{LSST:2008ijt}, bringing unprecedented statistical power for understanding our Universe.

Photometric LSS surveys such as DES, HSC, KiDS, Euclid and LSST capture images through broad photometric filters across a wide area of the sky, enabling the detection and shape measurement of millions of galaxies at high density and significant depth. While these surveys excel at mapping large volumes of the Universe {and avoid the complex target selection required for spectroscopic surveys}, they provide only coarse information on object type and redshift, whose precise determinations often {do} require spectroscopic data. Cosmological analyses with photometric data therefore typically rely on two-dimensional angular summary statistics from subsamples of galaxies that are grouped into tomographic redshift bins, where the overall radial density of objects can be modeled in aggregate. 
Such ``tomographic" analyses contain a rich trove of information useful for probing cosmological models because of how the distributions of galaxies and their shapes are sensitive to the expansion history and growth of structure. 
These are commonly probed through two-point statistics like the two-point correlation function or power spectrum. 

The autocorrelation of galaxy shapes, or ``cosmic shear", probes the integrated matter density distribution through weak gravitational lensing, and is particularly sensitive to $S_8 = \sigma_8\sqrt{\Omega_m/0.3}$, where $\sigma_8$ is the RMS amplitude of matter density fluctuations at 8~Mpc/$h$ and $\Omega_m$ the total matter energy density today. The autocorrelation of galaxy positions (``galaxy clustering") probes density fluctuations themselves, but mediated through the so-called galaxy bias, which relates the clustering of galaxies to the clustering of the underlying matter field, which is dominated by dark matter. The cross-correlation of positions and shapes (``galaxy-galaxy lensing") helps break degeneracies between $\sigma_8$ and galaxy bias ($b$), as well as other nuisance parameters, and so all three are commonly combined into a so-called \threextwopt{} analysis.

Cosmic shear requires very high number densities and redshifts over large areas, but is relatively robust to contamination from stars, {survey inhomogeneity} or variations in the effective selection function for galaxies \cite{decade_iii, decade_iv}. The shear signal comes from all intermediate redshifts (weighted by the lensing kernel) out to the source galaxies, so the signal is only weakly dependent on the width of the source galaxy redshift distributions. Details on the definition and estimation of the ``source" galaxy sample and redshift distribution for cosmic shear can be found in \cite{y6-metadetect, y6-sourcepz, y6-piff}. 

In contrast, the galaxy clustering signal is significantly enhanced with narrow redshift distributions $n(z)$, and is quite sensitive to variations in the selection function, contamination from interlopers, and errors in the estimated $n(z)$. The galaxy sample used for {this clustering measurement} (sometimes termed ``lens" galaxies for their role as lenses for the source galaxies in the galaxy-galaxy lensing measurement) {thus} has more stringent requirements for uniformity in the selection, and hence preferentially consists of brighter, easier-to-model galaxies.

In the Year 3 (Y3) DES \threextwopt{} analysis, two lens samples were used: the \maglim{} sample \cite{y3maglimdef}, which had a broad selection designed to have minimal color cuts but a wider range of galaxy types and less precise photometric redshift determination, and the \redmagic{} sample \cite{redmagicSV}, with $\sim1/4$ the number of galaxies but {designed to have higher redshift precision via} a photometric template to select primarily Luminous Red Galaxies (LRGs).\footnote{A third galaxy clustering sample was designed specifically for measuring Baryon Acoustic Oscillations (BAO) separately from the \threextwopt{} analyses, see \cite{y3-baosample, y3-baomocks, y3-BAOkp}.} In practice, the stricter \redmagic{} selection resulted in residual systematic errors from non-uniform selection that were challenging to mitigate, and the $\maglim{}$ sample was used for most cosmological inference. 

{Systematic errors in the galaxy density are introduced via astrophysical contamination from different objects or foregrounds, or by spatially varying observational conditions that modulate the true selection function.}
Controlling systematics like these is a key challenge in modern cosmological analyses, as they limit our ability to fully exploit the statistical precision of our observations. {Careful planning and execution of observing strategies to maximize uniformity and subsequent selection of bright galaxies significantly above the detection limit can help minimize the impact of observational systematics. Nonetheless,} non-uniformity in the selection function of the galaxy sample and contamination of the sample from interlopers (e.g., stars or galaxies assigned to incorrect redshift bins) are two of the primary systematics in galaxy clustering analyses
both of which have strong angular dependence. The process of mitigating these systematics when present in the catalog has a long history \citep{SDSS:2001wbd,Ho:2012vy,Ross_2011,Ross_2012,Huterer:2012zs,Ross:2016gvb,Awan:2016zuk}, with an increasingly sophisticated set of statistical \citep{Leistedt:2013gfa,Elsner:2015aga} and machine-learning tools \citep{Rezaie:2019vlz,Weaverdyck:2020mff, sanchez_dark_2022}.
Removing such contaminants altogether is considerably preferable to weighting the data post-selection, but in practice the processes of masking, sample selection, and systematic weights correction are often undertaken serially and independently, rendering the treatment of such systematics suboptimal. We take a more integrated approach for the DES Y6 analyses, identifying and removing regions of the footprint and color-space that are most likely to be contaminated and ill-suited for systematics weights correction, resulting in a considerably cleaner sample (see \citet*{y6-mask}).

The goal of this paper is to introduce and describe \maglimpp{}, a new and improved galaxy sample that uses a series of novel quality cuts to identify and remove contamination from the baseline \maglim{} selection used in DES Y3. This results in a new sample with high statistical power that is significantly more robust to contamination than \maglim{}. When combined with the novel masking approach described in \cite{y6-mask} and data-driven characterization of the galaxy selection function, we demonstrate that the \maglimpp{} sample and its clustering measurements are significantly hardened against systematic contamination. The new sample provides a core pillar of the Legacy DES Y6 \threextwopt{} analysis;  {cosmological constraints and companion works contributing to the \threextwopt{} analysis can be found in \cite{y6-1x2pt, y6-2x2pt, y6-3x2pt, y6-balrog, y6-cardinal, y6-gglens, y6-imagesims, y6-gold, y6-clusters, y6-extensions, y6-lenspz, y6-magnification, y6-mask, y6-metadetect, y6-methods, y6-nzmodes, y6-piff, y6-ppd, y6-sourcepz, y6-wz}.}

The organization of this paper is as follows. In Sec.~\ref{sec:data_processing_and_qual_cuts}, we describe the \maglimpp{} selection and quality cuts in detail, including the use of a self-organizing map (SOM) to characterize the color-space and identify compact regions that are disproportionately likely to be mismodeled and contaminate the two-point measurements. We also describe the implementation of a color-based star-galaxy separation approach (detailed in \cite{stargalsep}) that is tuned specifically to each redshift bin and significantly reduces residual stellar contamination, a systematic difficult to mitigate via galaxy weights. In Sec.~\ref{sec:sysmitigation} we describe our approach for characterizing variation of the selection function of the sample through two different and complementary regression-based approaches to produce galaxy weights, including survey property (SP) maps as spatial templates and the mocks we use to characterize expected biases the process might impart. 
Sec.~\ref{sec:clustering} presents the angular clustering measurements and in Sec.~\ref{sec:null_tests} we describe various null tests we perform {and their unblinding criteria}, as well as estimator and covariance corrections we apply to further harden the inference against the different choices for producing galaxy weights.
Finally, in Sec.~\ref{sec:cosmo_constraints} we constrain $\Omega_m$ and  $b_i\sigma_8$ using galaxy clustering alone and in combination with other probes. The appendices include supplemental materials and additional studies that will be useful for sample selection and cleaning for Stage IV surveys.

\section{\maglimpp{} catalog construction} \label{sec:data_processing_and_qual_cuts}
\subsection{Gold catalog base and \textsc{Joint} footprint}

{We use the DES Y6 Gold data collection \cite{y6-gold}, which is based on the final DES data release \cite{DESDR2} and includes survey property (SP) maps, multi-epoch dereddened photometry for objects using Bulge-Disk fitting (BDF), general purpose star-galaxy classification, and catalog quality flags. Each object is also assigned several photometric redshift properties from the Directional Neighbourhood Fitting (\dnf) algorithm \cite{DeVicente2016, Toribio_San_Cipriano_2024}, which fits a local linear model to estimate the redshift for each object using its 80 nearest spectroscopic neighbors in (a transformation of) the \textit{griz} color space. Stored photo-$z$ properties include a mean predicted redshift (\dnfz, or $\hat{z}$), the nearest neighbor training redshift (\dnfzn, or $\zmc$), and an estimated photo-$z$ uncertainty (\dnfzsigma, or $\zsigma$).
The final Gold object catalog has 669 million objects distributed over 4,923\,deg$^2$ and a median coadd depth of $i = 23.8$\,mag \cite{y6-gold}.}

{We use the \textsc{Joint} footprint defined in \citet*{y6-mask}, which is based on the Y6 \gold{} footprint \cite{y6-gold} (indicating good photometry, absence of foregrounds and at least two exposures in each band) and incorporates a number of additional cuts based on survey properties that could affect the fidelity of our shear and clustering measurements. The footprint is restricted to regions with $10\sigma$ depth $i_{ \textrm{depth}} > 22.2$ to ensure high significance detection of the faintest galaxies in the \maglimpp{} sample. Additional cuts (novel to Y6) remove regions that are outliers in any of a large number of survey properties, ensuring greater homogeneity of the footprint. If the variation of any of these properties causes systematic contamination of the galaxy sample, this clipping of the tails ensures that (1) the most extreme contamination is removed and (2) the perturbative contamination model in Sec.~\ref{sec:sysmitigation} better describes the data. Regions of the sky where the density is least likely to be accurately reconstructed via systematic weights are removed via the Leverage cut \cite{Weaverdyck:2020mff}. The final \joint{} footprint is 4,031 deg$^2$, and we refer the reader to \citet*{y6-mask} for further details on the survey property maps and footprint construction.}

\subsection{Base selection}
We perform a similar selection to the Y3 selection of the \maglim{} catalog \cite{y3maglimdef}, but significantly improve the quality of the sample through the application of two key quality cuts, resulting in what we call \maglimpp{}. The base \textsc{MagLim} selection uses extended objects\footnote{The \texttt{EXTMASH} = 4 selection is a stricter tier for extendedness than was available in Y3, which only went up to \texttt{EXTMASH} = 3, see \cite{y6-gold}.} from the DES Y6 Gold catalog \cite{y6-gold} in a given redshift range with a photo-$z$ dependent $i$-band limiting magnitude:
\bea
    {\texttt{EXTMASH} = 4} \\
    {0.2 < \zpred < 1.2} \\
    17.5 < i < 18 + 4\zpred \label{eq:ilim_maglim}
\eea

{We remove objects with extreme colors by enforcing
\be
\{g-r,\ r-z,\ i-z\} \in [-1, 4]
\ee}

\begin{figure}
    \centering
    \includegraphics[width=1\linewidth]{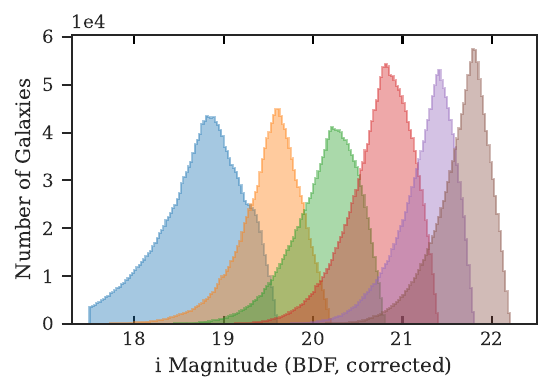}
    \caption{Distribution of \maglimpp{} galaxies, selected via a photo-$z$-dependent $i$-band limiting magnitude given by Eq.~\ref{eq:ilim_maglim}}
    \label{fig:maglim_hist}
\end{figure}

\begin{figure}
    \centering
\includegraphics[width=1\linewidth]{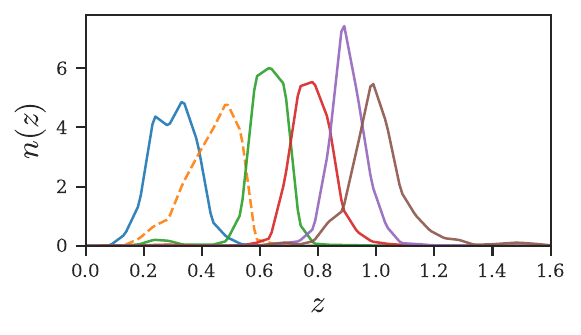}
    \caption{Redshift distribution of \maglimpp{} galaxies, as estimated in \cite{y6-lenspz}. Bin 2 is dashed since $\gamma_t$ and $w(\theta)$ measurements using this bin are discarded in the Y6 \threextwopt{} cosmology analyses in order to meet unblinding criteria (c.f. \cite{y6-3x2pt}).}
    \label{fig:nz}
\end{figure}

We apply additional selections based on
\begin{enumerate}
    \item the NIR star-galaxy separation approach conditioned on each redshift bin as described in \cite{stargalsep}, using data from unWISE \cite{Meisner_2019} (see Sec.\ref{sec:stargal}) and
    \item a novel, non-parametric SOM-based approach to identify regions in high-dimensional color-space with a high density of systematic contamination from interlopers (see Sec.\ref{sec:somcuts}).
\end{enumerate}

These additional steps render the resultant galaxy catalog significantly more pure, removing the dominant source of additive systematic contamination. While multiplicative systematics modulate the observed density field (e.g., due to varying depth, point-spread function, etc.), additive systematics correspond to interloping objects that are misclassified as the galaxies we are attempting to select for our sample. These objects dilute the observed density field as well as imprint spurious fluctuations, which are difficult to fully correct using galaxy weights \cite{stargalsep, Hern_ndez_Monteagudo_2025, Berlfein:2024uwi}. Identifying and removing these objects at the catalog-level makes our galaxy weights more sensitive to the remaining angular systematics and allows us to treat them as fully multiplicative.

Fig.~\ref{fig:maglim_hist} shows the distribution of resultant \maglimpp{} galaxies as a function of the dereddened \textit{i}-band magnitude, and Fig.~\ref{fig:nz} the inferred $n(z)$ (as measured in \citet*{y6-lenspz}).

\subsection{Star-galaxy separation}\label{sec:stargal}
We follow the approach of \citet{stargalsep}, which demonstrated that using NIR data from unWISE can provide a means of identifying stars that is complementary to the morphological \texttt{EXTMASH} star-galaxy classifier used in the DESY3 and DESY6 Gold catalogs.\footnote{The XGB star-galaxy classifier in \citet{bechtol2025darkenergysurveyyear} is complementary morphological classifier that uses machine learning to {more optimally separate the classes}. This classifier was still in development when the Y6 \maglimpp{} catalog was constructed.} 

{All but the youngest stellar populations ($\sim1$My)} tend to have a bump at $1.6\mu$m, which shifts into the infrared for galaxies at higher redshift {\cite{Sawicki_2002}}, making NIR photometry a powerful differentiator between stars and galaxies with otherwise similar optical photometry \cite{Prakash_2015}. In their comparison of star-galaxy separation methods on DES Y1 data, \citet{y1-sg} found that using $J-K_s$ band NIR data from VHS had the best performance, but had limited area where it could be applied. 
\decals{} and DESI both use NIR data from WISE, defining LRG samples with a
cut in $r-z$ and $z-W1$, where $W1$ is the 3.368$\mu$m band of unWISE photometry. \decals{} contains forced $W1$ photometry for all objects and uses DES imaging (though a different photometric pipeline) in the DES footprint. We cross-match Y6 \maglim{} galaxies with their \decals{} DR9 counterparts by position and import the W1 photometry. As noted in \cite{y6-mask}, for Y6 we have masked areas of the footprint removed by the DESI LRG mask\footnote{\url{https://data.desi.lbl.gov/public/ets/vac/lrg veto mask/v1/}} \cite{Zhou_2023},
which applies a larger mask radius around stars, and is important for getting correct W1 photometry given the typical size of the WISE PSF ($\sim6$"). This was done as part of the masking process; any quantities quoted for residual stellar contamination removed via catalog-level quality cuts are computed using the final footprint, after all footprint-level masks were applied.

\begin{figure*}
    \centering
    \includegraphics[width=1\linewidth]{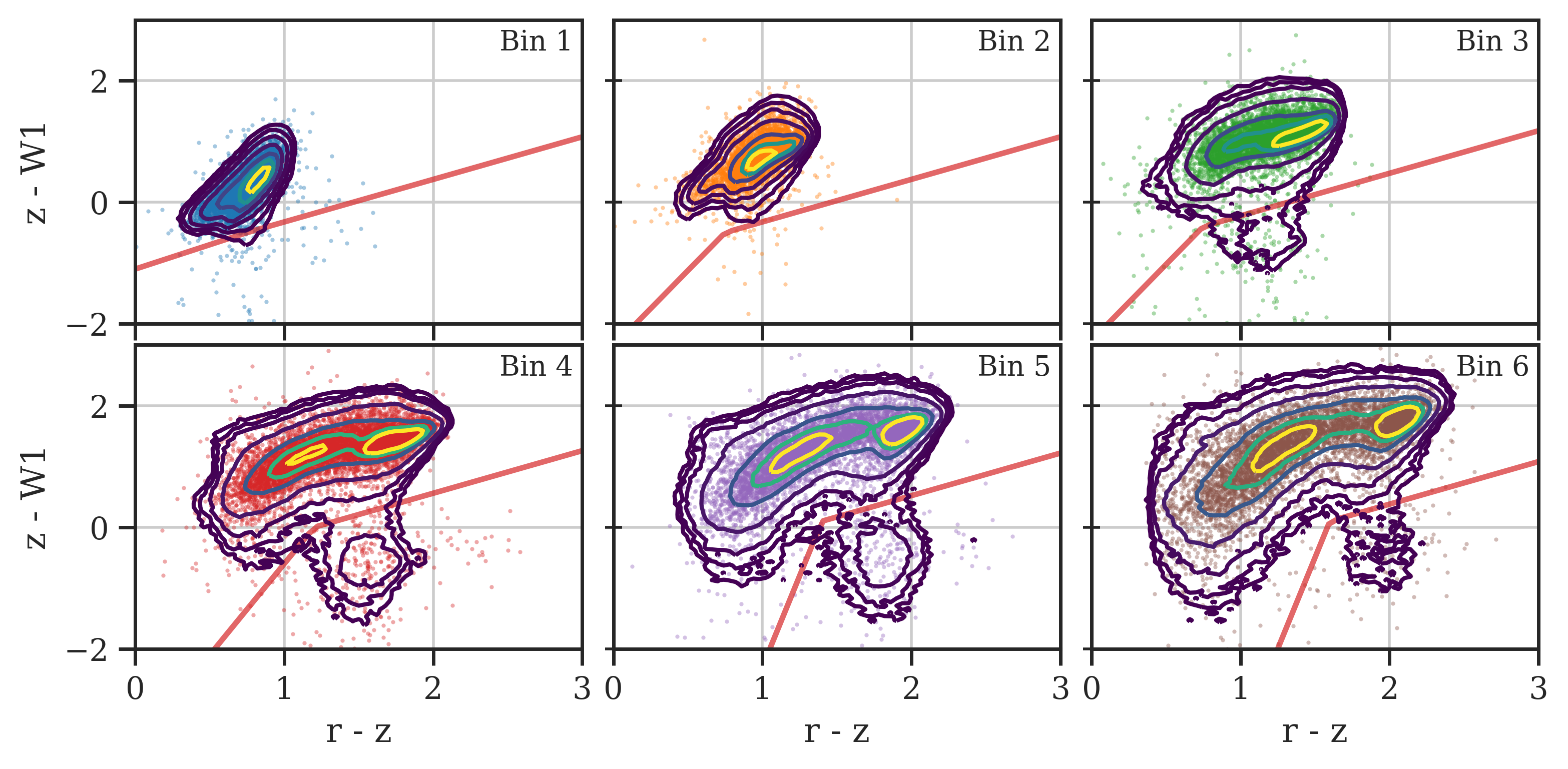}
    \caption{Color distributions of \maglim{} galaxies in each redshift bin, along with the star-galaxy cut (red) optimized to isolate and remove residual stellar contamination. Iso-density contours are $[25, 50, 75, 90, 95, 97, 98] \%$-iles, and points subsampled by a factor of 200.}
    \label{fig:color_stargal}
\end{figure*}

The bin-optimized approach of \cite{stargalsep}, exploits the fact that the optimal cut separating galaxies and stars will differ across redshift bins, as the whole color distribution of objects in $r-z$ and $z-W1$ itself varies strongly with photometric redshift. We therefore define a piece-wise linear star-galaxy cut for each redshift bin which traces the ``valley'' between the star and galaxy modes in the color-space. This is illustrated in Fig.~\ref{fig:color_stargal}, which shows the density of objects in the color-space, along with the bin-optimal cut for separating galaxies from residual contamination of stars in each bin. Note that the iso-density contours are $[25, 50, 75, 90, 95, 97, 98] \%$-iles,  extending far into the tails and with the ``stellar mode" containing $\sim [1.1, 0.6, 2.0, 3.3, 3.2, 1.3]\%$ of the total objects in each bin. We use the same process described in \citet{stargalsep}, starting with the LRG star-galaxy cut in \citet{DESI:2022gle} and adjusting the slopes and knee-point to trace the valley of the observed density with each increasing redshift bin.

\begin{figure}
    \centering
    \includegraphics[width=1\linewidth]{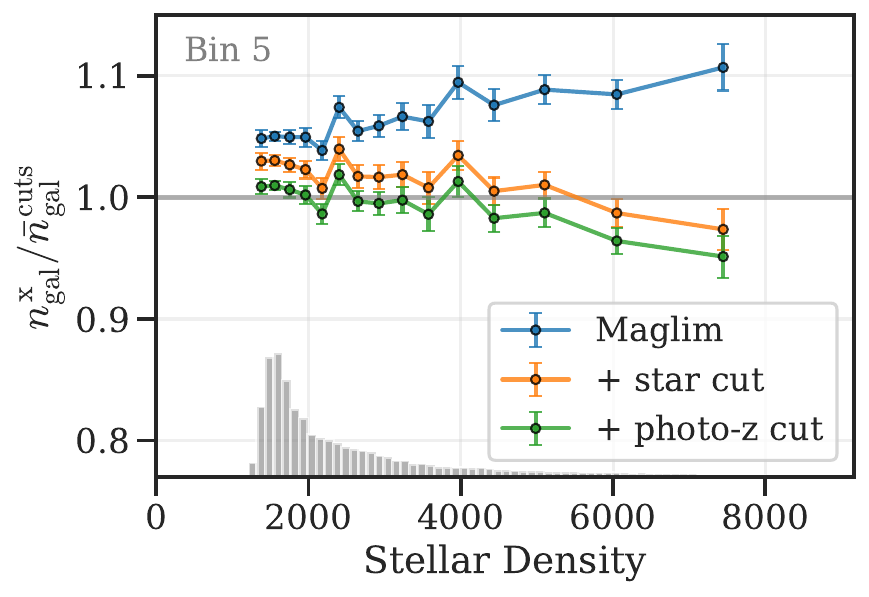}
    \caption{Example of how the density of objects in as a function of stellar density changes as one moves from the base Y6 \maglim{} sample (blue), to applying bin-optimized star-galaxy separation (orange), to additionally applying photo-$z$ quality cuts to get the final \maglimpp{} sample (green). Once stellar contamination is removed, the impacts of stellar obscuration becomes apparent. 
    Density is normalized to the average density for \maglimpp{}, and no galaxy weights have been applied. Gray bars indicate the area distribution of stellar density in the footprint.
    }
    \label{fig:stardens_1d}
\end{figure}

Applying this redshift-bin optimized star galaxy selection cuts $1.94\%$ of the objects in the sample.\footnote{Note that contamination estimates quoted by star-galaxy classifiers will in general depend strongly on test set they use, and in particular on the actual fraction of \textit{true} stars in the sample. Thus such estimates should not be assumed to generically represent the residual stellar contamination rate when applied to other samples that have different characteristics, such as a different fraction of stars, especially since the stellar density varies strongly across the sky.} While a small fraction of true galaxies are removed, \cite{stargalsep} showed that the removed objects are dominated by true stars, and we expect this to be even more true with the improved photometry of DES Y6 relative to Y3. 

Furthermore, by removing the stellar contamination, \textit{the remaining stellar-density dependent systematic becomes dominated by obscuration rather than a mixture of contamination and obscuration}. Stellar contamination is an \textit{additive} systematic (largely independent of true galaxy density), while obscuration is \textit{multiplicative}, and while both effects are traced by maps of stellar density, they have opposite effects on the observed density field to first order. When both are significant, the observed trend of galaxy density with stellar density reflects a non-trivial combination of these competing effects, complicating efforts to estimate the selection function's dependence on stars.
Removing contamination via color cuts renders stellar obscuration more readily detectable, which can then be corrected self-consistently using our standard multiplicative weights (Sec.~\ref{sec:sysmitigation}). 

Fig.~\ref{fig:stardens_1d} demonstrates this for Bin 5, where the blue curve shows a sample dominated by stellar contamination via a positive trend of the observed galaxy density with star density, whereas the negative trend of obscuration is revealed after applying the bin-optimized star-galaxy separation (orange).

\subsection{SOM-based quality selection} \label{sec:somcuts}
We apply further quality cuts through a Self-Organizing Map (SOM) of the \textit{griz} color-space. The SOM functions as a sort of combination dimensionality-reduction and clustering algorithm --- it learns a 2-D manifold of the 4-D color-space traced by the \maglim{} sample, and galaxies are assigned to cells that partition this 2-D space. 
One of the primary benefits of the SOM approach is that it enables the visualization of higher-dimensional spaces.
For instance, we can use a 2-D heatmap to plot summary statistics of the galaxies in each cell to visualize how galaxy properties vary across the space, without \textit{a priori} confining ourselves to specific color axes that may or may not be relevant.

We adopt the SOM algorithm first described in \citet{S_nchez_2020}, with several improvements over traditional SOM algorithms:
\begin{enumerate}
    \item The SOM does \textit{not} use periodic boundaries, which are not well motivated for mapping the $griz$ color-space of galaxies. 
    \item The uncertainty on flux measurements is factored into the computed ``distance" between galaxies and SOM cells so that low SNR fluxes don't dominate the cell assignment. 
    \item The distance measure is Euclidean in log-flux at high SNR but transitions to Euclidean in linear flux at low SNR. This makes the metric essentially Euclidean for color differences (i.e., ratios of fluxes) when fluxes are well measured, but prevents it from diverging when they are not. 
    \item We marginalize over a scalar of the overall flux when determining the distance, with a user-defined Gaussian prior on the amplitude. This enables the user to define how much to weight total flux differences vs.\ only color differences when learning the manifold and assigning objects to cells. This is valuable especially for photo-$z$'s, which are likely to depend rather strongly on color, but much less so on overall magnitude. However there is still useful information in the overall magnitude, both in terms of the underlying galaxy populations and the fidelity of the photo-$z$ algorithm, such that it is desirable to maintain \textit{some} sensitivity to it when training the SOM. The associated parameter $\sigma_{\ln s}$ essentially determines the width in magnitude of each SOM cell. This provides a more nuanced solution to a common problem faced by those constructing SOMs for photo-$z$ purposes, which is whether to use just colors (i.e., magnitude differences) as features, or to also include one or more overall magnitudes, which can significantly change the resulting SOM.
\end{enumerate}

We find the SOMs produced via this approach to be significantly more well behaved and less stochastic than those without these improvements.

We train a $48\times48$ {cell} SOM on the dereddened \textit{griz} BDF fluxes and uncertainties of the preliminary \maglim{} selection, with a Gaussian prior on an overall flux scale factor $s$ of $\sigma_{\ln s}=0.4$. The resulting SOM varies smoothly in color, magnitude and photo-$z$, despite being blind to the latter. 

Our goal is to identify regions in the SOM (and thus color-space) that are disproportionately likely to have systematic contamination. We do this by identifying cells where the photo-$z$ estimation is very poor --- these objects are likely to be assigned to the incorrect bin of photo-$z$, and to have malcharacterized $p(z)$. 

We must define a summary statistic of the photo-$z$ distributions of galaxies in each cell in order to define cuts; we use three different and complementary measures of photo-$z$ dispersion {using outputs from \dnf{}}, which account for uncertainty within each galaxy as well as within each cell: 
\begin{enumerate}
    \item $\sigma_{\rm cell}(\zmc)$: the standard deviation of $\zmc\ $ for galaxies in the cell,
    \item $\langle\zsigma\rangle_{\rm cell}$: the  average photo-$z$ standard deviation of galaxies in the cell, and
    \item $\langle |\zpred - \zmc| \rangle_{\rm cell}$: the average absolute difference between each galaxy's mean and nearest neighbor photo-$z$ estimates (see also \cite{Toribio_San_Cipriano_2024}).
\end{enumerate}

\begin{figure}
    \centering
    \includegraphics[width=1\linewidth]{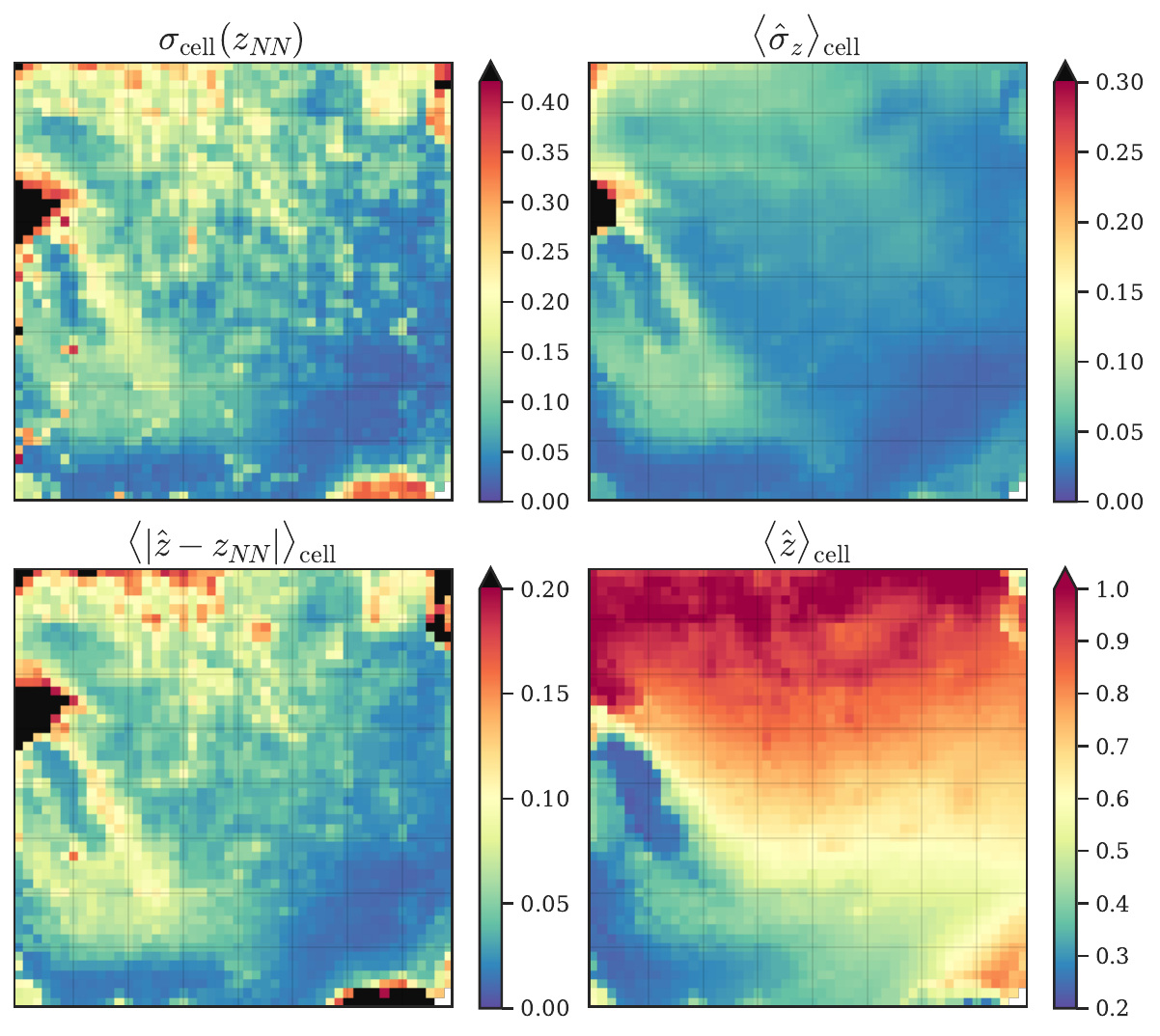}
    \caption{Three different measures of photo-$z$ uncertainty for galaxies in self-organizing map cells (see Sec.~\ref{sec:somcuts}), except for the lower right subplot, which shows the average predicted photo-$z$ of galaxies in each cell. The different measures largely agree, but have different sensitivity to outliers. Objects in cells with photo-$z$ error above a certain threshold are cut from the sample (black).}
    \label{fig:sigz_som}
\end{figure}

These statistics are shown for the SOM in Fig.~\ref{fig:sigz_som}. In the lower right subplot, one can see how the the average photo-$z$ in each cell smoothly varies through the SOM. 

\begin{figure}
    \centering
    \includegraphics[width=1\linewidth]{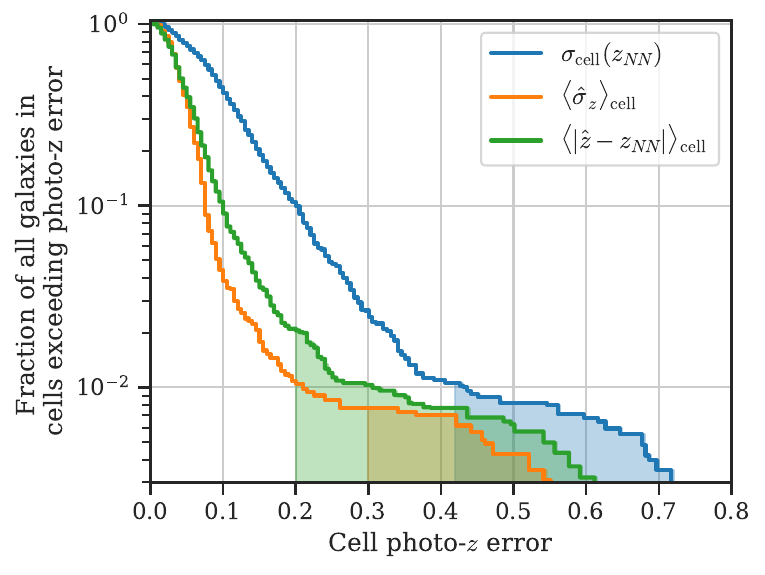}
    \caption{Cumulative fraction of galaxies in SOM cells below a given photo-$z$ uncertainty for each of the uncertainty metrics. Shaded regions indicate the final selection threshold for each measure, with galaxies removed if they fall within a cell that exceeds any one of the thresholds. Threshold values were selected by to correspond to a flattened part of the curve (i.e., a large jump in the error) that corresponds to compact regions in SOM space (black regions in Fig.~\ref{fig:sigz_som}).}
    \label{fig:sigz_hist}
\end{figure}

We identify cells that are outliers in each of these three statistics and remove any galaxies that populate these cells. We tune the outlier threshold such that the flagged cells correspond to compact regions in the SOM and remain largely at the edges. This minimizes the surface area of the cuts in the high dimensional color-space, which otherwise present opportunities for objects to scatter in and out of the selection through variations in the photometry or survey properties. 
Outliers are determined by inspecting the cumulative fraction of galaxies that populate cells below a given photo-$z$ error, and identifying the tails where the photo-$z$ error increases significantly relative to the number of galaxies (i.e., a flattening of the curves in Fig.~\ref{fig:sigz_hist}) while fulfilling the compactness criteria above. 

The resultant cuts remove $1.66\%$ of objects and are indicated by black regions in Fig.~\ref{fig:sigz_som} and the shaded regions in Fig.~\ref{fig:sigz_hist}. While the number of objects removed is relatively small, subsequent investigation of these objects through cross-matching with the VISTA Hemisphere Survey\footnote{https://doi.org/10.18727/archive/57} indicate that these objects are dominated by QSOs, which correspond to the region on the left side of the SOM removed by all three statistics, and which is clearly a nexus in color space where the photo-$z$'s are degenerate (note the convergence of wildly different $\langle z\rangle$ in nearby cells in the lower right plot). A smaller number of the removed objects are stars, corresponding to the lower and upper right regions.\footnote{These cuts were applied before the bin optimized star-galaxy separation of Sec.~\ref{sec:stargal}.} The aggregate cuts are the union of the SOM cuts for each of the three different dispersion measures.

\section{Mitigating Residual Systematics}\label{sec:sysmitigation}
We assess and mitigate spurious fluctuations in the observed \maglimpp{} galaxy density via corrective galaxy weights. These capture and remove systematic variations in the selection function due to survey properties. The improved Y6 photometry \cite{bechtol2025darkenergysurveyyear}, masking \cite{y6-mask}, and quality cuts (Sec.~\ref{sec:data_processing_and_qual_cuts}) result in a significantly more uniform selection function compared to Y3, requiring correspondingly smaller galaxy weights.

Unaddressed, these spurious density fluctuations can bias clustering measurements and cosmological parameters \cite{DES:2021bat}, typically boosting the observed power spectrum \citep{Huterer_2013, Shafer_2015, Weaverdyck:2020mff}. 

We model the observed density field as a function of the true field and a set of systematic template maps ($t_i$) which we trace potential sources of contamination (e.g., dust, PSF variations, stars, etc.):
\be
\delta_{\text{obs}} \sim \mathcal{F}(t_0, t_1,...t_{N_{\text{tpl}}}, \delta_{\text{true}}) + \delta_{\text{true}},
\ee
where $\mathcal{F}$ is estimated directly from the data via regression.

Successful galaxy weighting requires balancing three key requirements in the systematic model $\mathcal{F}$:
\begin{enumerate}
    \item It must have sufficient freedom to capture the dominant contamination modes;
    \item It shouldn't have so much freedom that statistical over-correction becomes significant compared to the original systematics;
    \item It should not be able to fit true LSS beyond chance correlation from cosmic variance.
\end{enumerate}

Given the highly uniform Y6 photometry \cite{Abbott_2021}, we expect systematic fluctuations to be small and well-described perturbatively via Taylor expansion, provided we can model the spatial dependence of the contamination. We therefore opt to allow more model freedom primarily in this spatial dependence, i.e., through a large template library, rather than allowing complex functional forms for $\mathcal{F}(t_i)$ such as via machine learning \cite{rezaie_local_2023, Chaussidon_2021, sanchez_dark_2022}.  

Since the aggressive quality cuts and star-galaxy separation in Sec.~\ref{sec:data_processing_and_qual_cuts} should remove the dominant source of additive contamination, we treat remaining systematics as multiplicative and construct galaxy weights as $w_k = (1+\hat{\mathcal{F}}_k)^{-1}$, with all galaxies in healpixel \cite{Gorski_2005} $k$ assigned the same weight when computing correlation functions and other summary statistics. Our methods for estimating $\hat{\mathcal{F}}$ are described in Sec.~\ref{sec:weights}, and are designed to take data-driven approaches to meeting requirements (1) and (2) (see \cite{Weaverdyck:2020mff} for a review of other approaches).

Meeting requirement (3) is crucial, since any ability to fit $\deltasub{g}$ directly will result in a strong (but false) detection of contamination and will significantly bias clustering and galaxy-galaxy lensing estimates in a manner that won't be detected or calibrated via tests on mocks. The risk of this increases with both the number of templates and with the inclusion of highly correlated templates when they are derived from the same data used to measure the density field. Different sensitivities to the local density would render the \textit{difference} between these two templates as effectively a tracer of the LSS signal itself. 

\textit{We therefore caution against simultaneously using multiple estimators for the same systematic, e.g., multiple dust maps, or estimates of the local sky background level.} See App.~\ref{sec:lss_in_weights_app} for details, including a case study with Y3 DES galaxy samples demonstrating these effects.

\subsection{Template rejection via external tracers}\label{sec:template_rejection}
To help ensure that our set of templates does not include true LSS modes, we first vet a larger set of potential templates against LSS tracers with which they should generally not correlate.\footnote{This was done early in the analysis, with a slightly larger mask than the final one described in \cite{y6-mask}.}

\begin{figure*}
    \centering
    \includegraphics[width=1\textwidth]{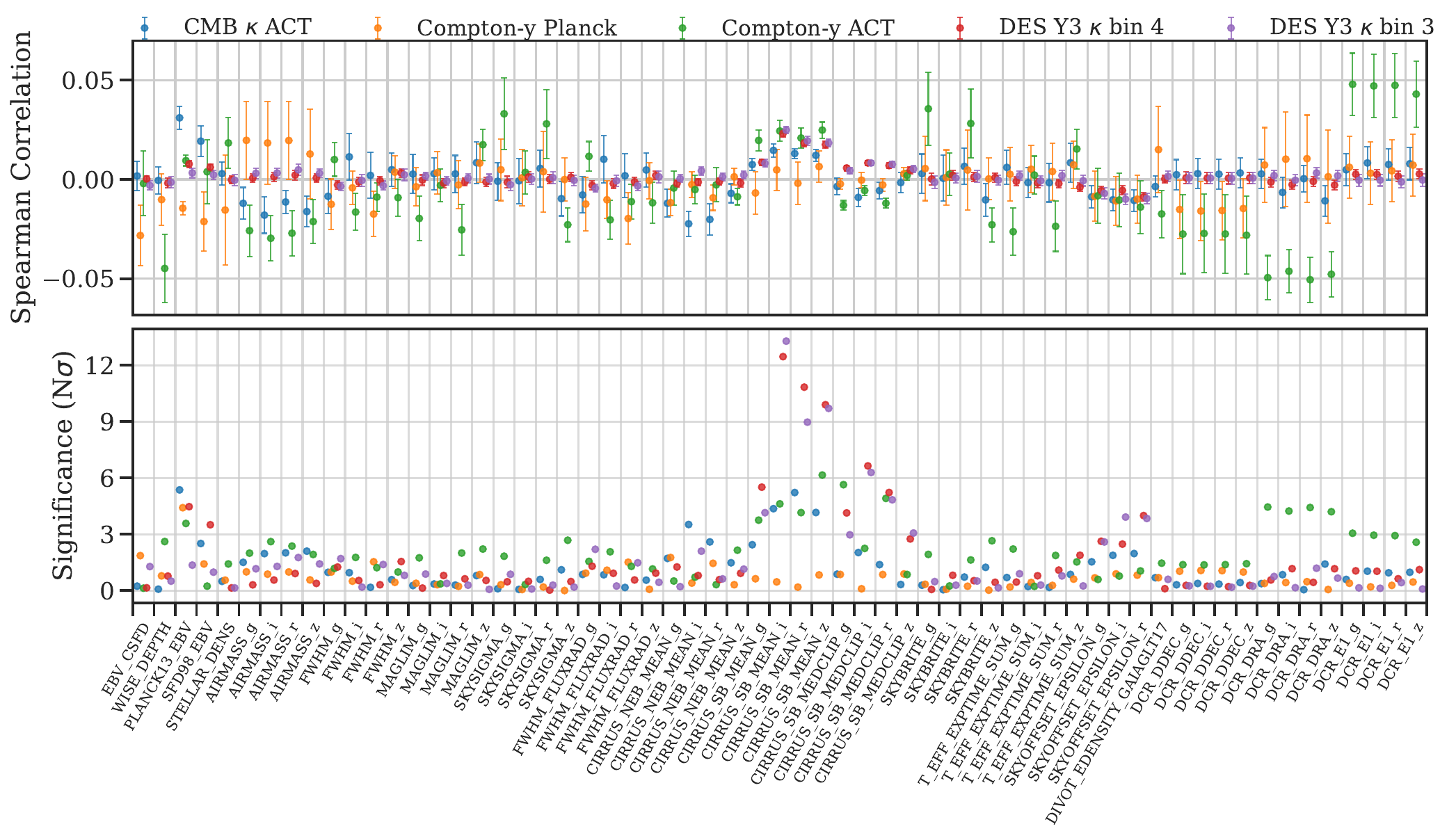}
    \caption{Correlation of SP maps with external LSS tracers. We identify and remove maps that consistently correlate with external LSS tracers, as this is an indication that the maps themselves may contain LSS.}
    \label{fig:sp_cull}
\end{figure*}

We compute the (weighted) Spearman correlation coefficient of the LSS tracer against each SP map at \healpix\. \nside{} 1024\footnote{Corresponding to a pixel side length of $\sqrt{A_{\rm pix}}=3.4'$.}, using the fractional area of each pixel as its weight, and compute the jackknife uncertainty with 40 patches. We use the Spearman rank correlation, as this is less affected by extreme values and  also captures non-linear dependencies; Fig.~\ref{fig:sp_cull} shows the significance of the correlations. 

A significant detection could indicate either that the template contains LSS or that the LSS tracer contains residual systematics that the template traces (or it is simply a noise fluctuation); therefore we look for indications of correlation in multiple LSS tracers and exercise care to not over-interpret any single signal or tracer.

The LSS tracers we use are:
\begin{enumerate}
    \item The standard Compton-$y$ map\footnote{\url{https://lambda.gsfc.nasa.gov/product/act/actadv_dr6_compton_maps_get.html}} combined from Planck and ACT DR6 tracing the line-of-sight integrated electron pressure via  the thermal Sunyaev-Zel'dovich effect as described in \citep{coulton2023atacamacosmologytelescopehighresolution}.
    \item The ACT DR6 CMB lensing mass map\footnote{\url{https://lambda.gsfc.nasa.gov/product/act/actadv_dr6_lensing_maps_get.html}} as described in \citep{Qu_2024, Madhavacheril_2024}. We create an SNR-weighted pixel mass map similar to Fig.~3 of \cite{Madhavacheril_2024}. The $\kappa$ map is provided as harmonic coefficients, which we filter to harmonic modes $40 \lesssim \ell \lesssim 1250$ by applying a top-hat filter convolved with a Gaussian of $\sigma_\ell=5$. We then apply a Wiener filter using a theory lensing power spectrum\footnote{\url{https://phy-act1.princeton.edu/public/data/dr6_lensing_v1/misc/clkk.txt}} and the provided noise array, ($f^{\rm wiener}_{\ell} = C_\ell^{\kappa \kappa, th}/(C_\ell^{\kappa \kappa, th} + N_\ell^{\kappa \kappa})$) and convert to a \healpix \ \nside{} 1024 map via \texttt{healpy.alm2map}. 
    \item The DES Y3 weak lensing $\kappa$ mass map \cite{Jeffrey_2021} for the two highest source redshift bins. 
\end{enumerate}

We find that the various \texttt{CIRRUS\_SB} maps have correlations with almost all the LSS tracers at high significance, and as such we remove them from our list of templates used to generate systematic weights across the footprint. However, we do use these maps for masking, identifying thresholds via visual inspection above which pixels are removed from the footprint for both \texttt{CIRRUS\_SB\_MEAN} and \texttt{CIRRUS\_NEB} maps (see \cite{y6-mask}). The \texttt{CIRRUS\_NEB} maps do not show as significant correlation, but these are very highly skewed maps corresponding to the output of a convolutional neural net used to predict the probability of a pixel containing cirrus. These maps were generated in an exploratory manner and applied rather far outside the conditions in which they were trained; they are therefore useful for flagging potentially problematic areas, but their continuous output should not be used as a spatial template without further testing. {Curiously, the Planck reddening map (\texttt{PLANCK13\_EBV}) \cite{dust_planck14} shows significant ($>3\sigma$) correlation with four of the five LSS tracers, vs.\ only one of the five for the SFD reddening map (\texttt{SFD98\_EBV}) \cite{Schlegel_1998}.}

Interestingly, we find that the ACT Compton-y map  shows a consistently significant correlation with maps characterizing the effects of differential chromatic aberration, DCR\_DRA and DCR\_E1. {None of the other LSS tracers indicate a signal, and this} may indicate some residual systematics in the ACT Compton-y maps; see e.g., \citet{liu_tsz_25} who find significant foreground contamination in the the fiducial ACT Compton-y maps, especially from Cosmic Infrared Background (CIB). While not shown, we additionally check for signs of LSS in combinations of our fiducial template library by regressing the external tracers against all maps simultaneously and looking for large, anti-correlated amplitudes of correlated templates, verifying the absence of such.

{In the end, we down-select to a base set of 19 fiducial templates as described in \citet*{y6-mask}, which are well motivated to potentially impact the selection function while passing the aforementioned tests and providing non-redundant information. These include three DES-band independent maps (extinction, stellar density, WISE depth) and four DES-band dependent maps (mean airmass, mean seeing, mean depth, and sky level dispersion) for each band in \textit{griz}.}

\subsection{Weights} \label{sec:weights}
\subsubsection*{\enet{}} \label{sec:enet}
\enet{} is a systematics mitigation approach introduced in \cite{Weaverdyck:2020mff} that was found to produce similarly high quality results to the \textit{Iterative Systematics Decontamination} (\isd{}) method used in Y1 \cite{y1-wthetapaper} and Y3 \cite{y3-galaxyclustering} (and considerably improved compared to standard linear regression), but in a fraction of the time ($\mathcal{O}$(10 seconds) vs.\ $\mathcal{O}$(1 day)). We refer the reader to Sec.~V of \cite{Weaverdyck:2020mff} for details, and only briefly outline the approach here, highlighting improvements made in this work. 
\enet{} performs elastic net regression in the full $\Ntpl$-dimensional space to estimate the systematic contamination contribution from all SP maps simultaneously:
\be
1+\delta_{\text{obs}} =
\left(1+\sum_{i=1}^{\Ntpl} \alpha_it_{i}\right)\left(1+\deltasub{gal}\right),
\ee
where we use pixelized maps and $\delta_{\text{obs},\ k}=n_k/\bar{n} - 1$ is the overdensity in pixel $k$, computed from the galaxy number density ($n$), count ($N$) and area ($A$) of pixel $k$: $n_k=N_k/A_k$, and a bar ($\bar{\ }$) indicates a map-wide average. $\alpha_i$ is the contamination amplitude for template $t_i$, with the systematic weight for pixel $k$ computed as $w_k = \left(1+\sum_{i=1}^{\Ntpl} \hat{\alpha}_i t_{ik}\right)^{-1}$.

{The estimated contamination amplitudes $\hat a_i$ come from maximizing the log-likelihood:}
\begin{multline}
\mathcal{L(\alpha)} \propto -\sum_{k=1}^{N_{\rm pix}} \left(\frac{A_k^2}{N_k + 2}\right) 
\left(\delta_{\text{obs},\ k}
- \sum_{i=1}^{\Ntpl} \alpha_it_{ik}\right)^2\\+ \sum_{i=1}^{\Ntpl} \left(\lambda_1 |\alpha_i| + \frac{\lambda_2}{2}\alpha_i^2\right). \label{eq:loglike}
\end{multline}

\begin{figure}
    \centering
    \includegraphics[width=1\linewidth]{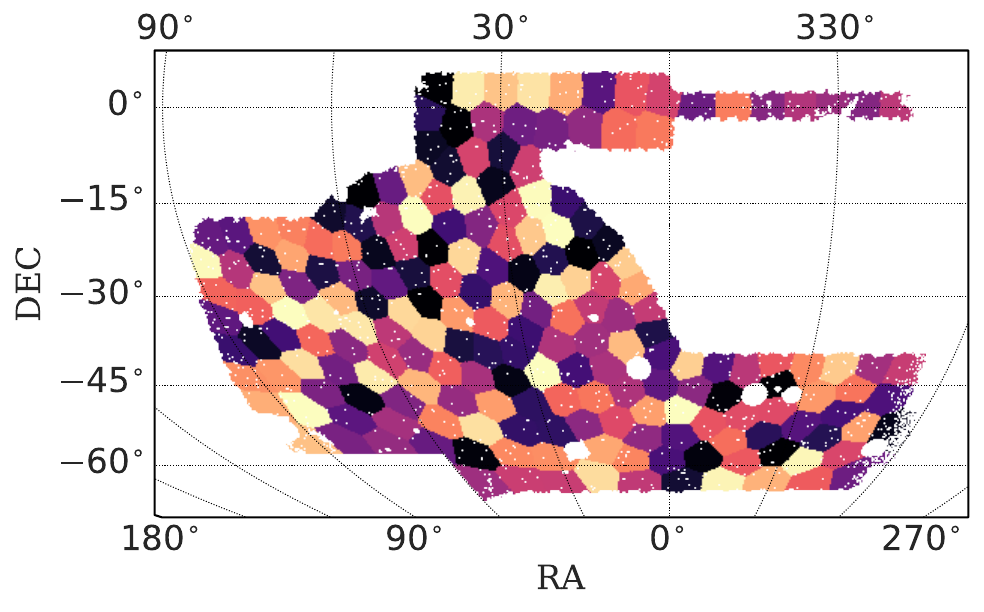}
    \caption{200 jackknife patches of the footprint used for cross-validation in the \enet{} weights and for estimating the covariance of various statistics throughout this work. Each patch has diameter $\sim5^\circ$.}
    \label{fig:jackknife_patches}
\end{figure}

This is equivalent to a Gaussian likelihood in the pixel overdensity, with zero-centered priors on the contamination amplitude of each template. The shape of the prior is a weighted-average of a Gaussian and Laplace distribution, with the strength and form (how Gaussian vs.\ Laplacian) controlled by the $\lambda_1$ and $\lambda_2$ parameters. The best prior is inferred from the data itself via cross validation to maximize predictive power and minimize overfitting. 
We perform 200-fold cross-validation, wherein we regress the observed overdensity against our templates with a given zero-centered prior using 199 patches, then compute the root-mean-square error of the prediction of the overdensity on the (held-out) 200th patch. We repeat for all 200 patches (shown in Fig.~\ref{fig:jackknife_patches}) and use the average RMSE as the aggregate prediction error score for a given prior. We repeat the entire process for 100 configurations of prior width and shape, selecting the one with the best predictive score. 

The prior widths range from completely uninformative (equivalent to standard linear regression using all templates, \textit{a la} Mode Deprojection, with significant nulling of true LSS) to effectively a delta function on zero where no template cleaning is performed at all (and thus leaving any systematics). This effectively lets the data choose how many templates to use for cleaning, based on maximizing predictive power on hold-out samples. This has been shown to significantly lower error on the estimated power spectrum when compared to ordinary least squares regression \cite{Weaverdyck:2020mff}. 

Because we are working with samples where the systematic contamination is small ($\lesssim 10\%$), we restrict how much model freedom we allow for the contamination for a given template value, modeling it perturbatively, and instead spend our modeling degrees of freedom by using a relatively large template library since these trace spatial patterns measured by various survey properties and thus provide a clear motivation for vectors of contamination. 
This is complemented by aggressive masking of pixels at the extremes of an even larger superset of survey property maps. 
{This is done for the individual maps, as well as in the combined $\Ntpl$-dimensional  template space, where the latter is achieved by cutting areas with large ``Leverage," a high-dimensional outlier statistic (see \cite{Weaverdyck:2020mff} for theoretical background and \cite{y6-mask} for details of implementation of the Leverage mask).}
These cuts remove regions that are both (a) more likely to be more contaminated, and not well described by a lower-order approximation, and (b) \textit{a priori} have larger influence in determining the contamination model, potentially sacrificing accuracy for the bulk of the footprint in the interest of better approximating the contamination in a few small regions.
This helps ensure the contamination is well-modeled perturbatively.

We implement several additional improvements over the \enet{} implementations described in \cite{Weaverdyck:2020mff} and \cite{DES:2021bat}. 

\begin{itemize}
    \item We construct spatially compact cross-validation patches through \textit{k}-means clustering of observed \nside{} 256 pixel centroids on the celestial sphere, resulting in patches with diameter $\sim$5\degree, encompassing the largest scales used for the Y6 \threextwopt analysis.
    \item We train the template amplitudes using \nside{} 256 pixels, resulting in average pixel galaxy counts of $\sim14.6 - 20.3$ across redshift bins, but evaluate the weights at \nside{} 1024 to obtain higher spatial resolution.\footnote{In principle if the templates have significant small scale power between the training and prediction pixel scales ($\theta \in [3.4', 13.7']$), this could result in extrapolation outside of the $\nside{}=256$ template data-space in which the fits were trained. We verified that the weights predicted at \nside{} 256 and 1024 have \textit{very} similar distributions, including out to the tails, indicating negligible extrapolation. The \nside{} 512 Leverage mask {(Sec.~\ref{sec:data_processing_and_qual_cuts})} likely helps to mitigate concerns of extrapolation, by identifying and removing pixels that are near the extremes in the high dimensional template space at an intermediate scale. Note that with a more non-linear contamination model, even minor extrapolation in the template space can have a major impact on the predicted weights.}
    \item We account for both the observed area of each pixel and the Poisson scatter of measured galaxy counts in the pixel covariance when fitting for the weights. In practice, this is applied as a per-pixel inverse-variance weight:
    \be
    W_{kk} \propto \frac{A_k^2}{N_k + 2},
    \ee
    where $A_k$ is the area and $N_k$ the galaxy count in pixel $k$. The second term in the denominator effectively regularizes the covariance with a galaxy count of 2 and ensures that empty pixels maintain a non-zero variance. 
\end{itemize}

In principle the likelihood in Eq.~\ref{eq:loglike} could be made more exact by accounting for the covariance between pixels or diagonalizing it and performing the fits in harmonic space \cite{Weaverdyck:2020mff}, but we expect the gains from this to be marginal. Furthermore, our implementation of the likelihood is not strictly Poissonian, but rather applies inverse (Poisson) variance weighting to an otherwise Gaussian likelihood. This accounts for the heteroskedasticity of the observations while still permitting significant flexibility to assess the robustness of the weights to alternative methods for computing the weights via the same code base (see App.~\ref{sec:more_weights_tests}).

\begin{figure*}
    \centering
    \includegraphics[width=1\linewidth]{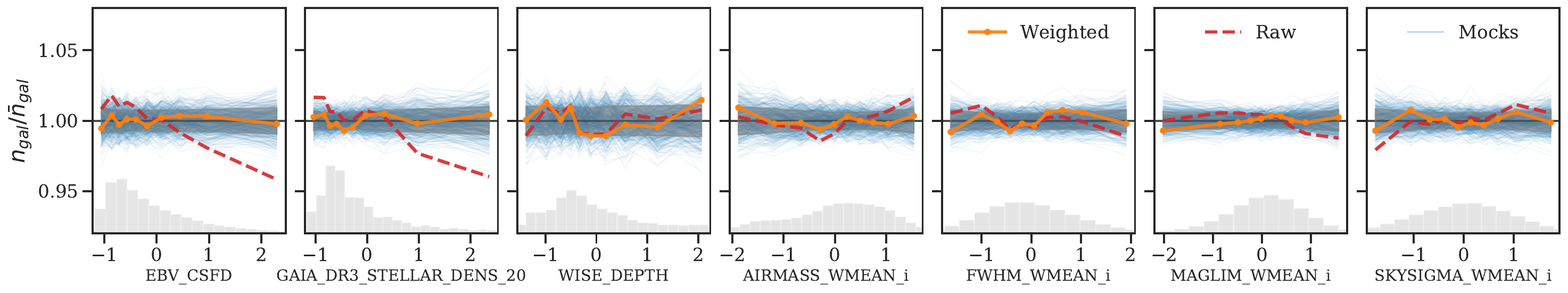}
    \caption{Example density trends vs.\ survey property maps for redshift bin 4. We show three band-independent SP maps as well as the band-dependent SP maps for $i$ (other bands look similar). We show the average observed (red dashed) and weighted (orange solid) density in equal-area bins. We show bin 4 as it shows a stronger observed density trend than all other bins except bin 6, highlighting the impact of the weights. Note the scale of the y-axis, which highlights the uniformity of the lens sample, with weights applying typical corrections of only $\sim2\%$. See Fig.~\ref{fig:ngal_sp_1d_allbins} for the equivalent plot for all redshift bins.}
    \label{fig:ngal_sp_1d}
\end{figure*}

\subsubsection*{ISD Weights} \label{sec:isd}
DES Y1 and Y3 analyses used the \isd{} method for weights generation, which performs a series of binned 1D fits of the observed number density against each template, with stepwise weighting to remove the most significant trend with each iteration.
We briefly review the approach and highlight improvements made for this analysis.

Each template map is first degraded to \nside{} 512
and pixels assigned to 10 equal-width bins in survey property map value. We compute the average observed density in each bin and perform a least-squares fit with a model $\mathcal{{F}}_i^{\rm 1D}$:
\be
\deltasub{obs}^{\rm binned} = \mathcal{\hat{F}}_i^{\rm 1D}(t_i^{\rm binned}) + \epsilon, \label{eq:isd_1d}
\ee
with a covariance ${\rm Cov}[\deltasub{obs}^{\rm binned}, \deltasub{obs}^{\rm binned}]$ generated from 1000 uncontaminated lognormal mocks (see App.~\ref{sec:mocks} for details on mocks). The $\Delta\chi^2$ improvement over the null hypothesis ($\mathcal{F}=0$) is used as a summary statistic, and the significance of this improvement for template $i$ at step $j$ is characterized as $S_{ij} = \Delta\chi_{ij}^2/\Delta\chi^2_{68}$, where $\Delta\chi^2_{68}$ is the 68$^{\rm th}$ percentile of $\Delta\chi^2$ computed on an independent set of 1000 uncontaminated mocks.

For every step $j$, Eq.~\ref{eq:isd_1d} is fit for every template and the most significant fit (i.e., largest $S_{ij}$) is used to compute intermediate weights at $\nside$ 4096 resolution as $w_j = 1/(1+\mathcal{\hat{F}}_{ij}^{\rm 1D})$, effectively removing the marginal dependence on that template. The process is then repeated, using weighted galaxies for the new fit in Eq.~\ref{eq:isd_1d}. The weight for each galaxy at step $j$ is thus the product of the intermediate weights from each previous step: $w_{\rm tot}^j = \prod_{k=0}^{j-1}w_k$, and the process stops after all maps are below a given significance threshold, $\isdthresh{}$.

For the Y6 analysis, we use a third-order polynomial for $\mathcal{F}_{\rm 1D}$, providing more model freedom than in Y3 or Y1. This is possible thanks to the improved masking described in \cite{y6-mask}, which removes extreme template values that otherwise cause problems when evaluating $\hat{\mathcal{F}}_{\rm 1D}$ (which has been fitted on binned values) on the raw 4096 pixels.
This additional freedom is useful because we are identifying marginal trends of the density against each template, and even if the number density only responds linearly to a conditional change of a given template, nonlinear spatial dependencies with other contaminating templates will result in nonlinear dependence in the \textit{marginal} 1D fits. This is in contrast to \enet{}, which performs simultaneous \textit{conditional} fits to all templates. 
Some models that are flagged strongly when using the cubic marginal fits but not when using the linear fits are $\texttt{FWHM\_WMEAN\_z}$, $\texttt{SKYSIGMA\_WMEAN\_z}$, and $\texttt{MAGLIM\_WMEAN\_r}$.
We use a stopping threshold of $\isdthresh{}=2$, though test the impact of varying this threshold (or of adopting linear-only weights) in Sec.~\ref{sec:null_tests}.

In comparison to \enet{}, \isd{} is sensitive to smaller scales, accounts for (binned) pixel covariances, and fits higher-order contamination models, but is less sensitive to linear combinations of templates, more dependent on mock assumptions and method parameter choices (e.g., stopping threshold) rather than having these calibrated on the data, and has higher temporal and computational costs.

\subsubsection*{}
Both \isd{} and \enet{} produce sets of weights that remove significant dependence of the galaxy density on our fiducial templates, but determine these weights and significance in complementary ways. By default, we show \enet{} weights, as it is significantly faster to generate them and test variations of the analysis. We analytically marginalize over the choice of weights and its impact on the galaxy clustering measurement by adding a term to the $\wtheta$ covariance (see Sec.~\ref{sec:bias_and_cov_corrections}).

Fig.~\ref{fig:ngal_sp_1d} shows the dependence of the observed galaxy density in the fourth redshift bin on a set of typical survey properties (we show $i$-band for the filter-dependent templates, as this is the main band for selection). Density is computed in ten equal-area deciles of each template, with the red dashed curve showing the raw dependence without weights and orange the dependence after applying weights. The same trends for uncontaminated mocks are shown as blue lines, with the 68\% range shaded. Gray histograms show the distribution of sky area for each survey property. 
While we detect trends with high significance, the size of the dependence is considerably smaller in Y6 than in Y3, with root-mean-square weights corrections of $\sigma^{\rm Y6}_{\rm weight} = [2.5, 1.9, 1.7, 3.0, 1.4, 3.9]\ \%$ across redshift bins, as compared to $\sigma^{\rm Y3}_{\rm weight} = [1.7, 3.9, 2.8, 5.6, 6.1, 6.0]\ \%$ in Y3. 

Note that while the corrections are linear in each template for the \enet{} weights conditional on all other templates being held fixed, the \textit{marginal} dependence can change quite non-linearly after application of the corresponding weights. Furthermore, it is possible to make the weighted curve arbitrarily tight around $n/\bar{n}=1$ by adding more freedom to $\mathcal F(t_i)$, but this is in fact undesirable as it comes from removing actual LSS. What we want is a weighted curve that is consistent with cosmic variance, without having abnormally large or small dependence on the templates. We characterize this consistency with cosmic variance in Sec.~\ref{sec:null_tests}, finding good agreement and no evidence of residual systematics. 
Plots analogous to Fig.~\ref{fig:ngal_sp_1d} for all bins can be found in Fig.~\ref{fig:ngal_sp_1d_allbins}

Characteristic statistics for the final \maglimpp{} sample are given in Table~\ref{tab:maglimsamp}.

\begin{table*}
    \centering
    \begin{tabular}{||c|c|c|c|c|c|c||} \hline
 Redshift Bin& \thead{Nominal\\ Redshift Selection}& $\bar{z}$ & $\sigma_{\mathrm{MAD}}(z)$ &\thead{Number of\\ Galaxies}&\thead{Angular Density\\ $[$arcmin$^{-2}]$}&\thead{Weights dispersion\\ $\sigma[w_i]$}\\ \hline \hline
         1&  [0.20, 0.40]& 0.306 & 0.097 & 1852538 &0.1277& 0.025\\ \hline 
         2&  [0.40, 0.55]& 0.435 & 0.11 & 1335294 &0.0920& 0.019\\ \hline 
         3&  [0.55, 0.70]& 0.624 & 0.085 & 1413738 &0.0974& 0.017\\ \hline 
         4&  [0.70, 0.85]& 0.778 & 0.093 & 1783834 &0.1229& 0.030\\ \hline 
         5&  [0.85, 0.95]& 0.903 & 0.081 & 1391521 &0.0959& 0.014\\ \hline 
         6&  [0.95, 1.05]& 1.011 & 0.131 & 1409280 &0.0971& 0.039\\ \hline
    \end{tabular}
    \caption{Statistics of final \maglimpp{} sample. Redshift statistics are computed using the datavector $n(z)$ \citep{y6-lenspz, y6-wz}, and do not include any posterior-inferred corrections via nuisance parameters. $\sigma_{\rm MAD}(z)$ is a robust standard deviation {characterizing the population $n(z)$}, computed as 1.483$\times$ the median absolute deviation {(note this is not the same as the typical per-object redshift uncertainty)}. $\sigma[w_i]$ is the standard deviation of object-level systematic weights.}
    \label{tab:maglimsamp}
\end{table*}

\section{Galaxy Clustering: Theory and Measurement} \label{sec:clustering}
\subsection{Theory}\label{sec:clustering_modeling}

For a given cosmology, one can compute the expected angular power spectrum of galaxy density fluctuations as:

 \begin{align}
  \nonumber   C_{DD}^{ij} (\ell)=&\frac{2}{\pi}\int d \chi_1\,W^i_{g}(\chi_1)\int d\chi_2\,W^j_{g}(\chi_2)\\
     &\int\frac{dk}{k}k^3 P_{gg}(k,\chi_1,\chi_2)j_\ell(k\chi_1)j_\ell(k\chi_2)\,,
 \label{eq:Cl-DD}
\end{align}
where $\chi$ is the comoving distance, $W_{g}^i = n_g^i(z)\, d z/d\chi$ the normalized window function for galaxies in tomographic bin $i$, and $P_{gg}(k,\chi_1,\chi_2)$ is the 3D galaxy power spectrum (see \citet*{y6-methods} for more detail). This density term is the dominant contributor to the observed galaxy power spectrum $C_{{g}{g}}^{ij} (\ell)$, which also includes contributions from redshift-space distortions (RSD) and magnification (c.f. \cite{y6-magnification}). Notably, we do not use the common Limber approximation for this, instead using \texttt{FFTLOG} \cite{Fang_2020}.

The predicted angular correlation function is then given by 
\begin{equation} 
    w^{ij}(\theta) = \sum_\ell \frac{2\ell+1}{4\pi}P_\ell(\cos\theta) C^{ij}_{{gg}}(\ell),
\label{eq:2pt_w}
\end{equation}
where $P_\ell$ are the Legendre polynomials.

There is a small-scale limit beyond which uncertainties on our theoretical modeling dominate our statistical errors due to the breakdown of perturbation theory, uncertainties in baryonic feedback models, etc. We therefore remove scales from our analysis below a given $\theta_{\rm min}^i$ for each redshift bin $i$. The process for determining scale cuts is described in \citet*{y6-methods}, including how these scale cuts vary depending on the theoretical modeling prescription.
Unless otherwise noted, we compute $\chi^2$ values using the non-linear galaxy bias model scale cuts, as these encompass a larger range of scales than the analyses using linear bias. Scale cuts are indicated in figures by gray shading, and only scales up to $\theta < 250$ arcmin are used in the cosmology analysis\footnote{{The maximum scale of 250 arcmin was chosen so as to have the same angular binning as the $\xi_\pm$ and $\gamma_t$ measurements. Tests going to $\theta<1000$ arcmin showed little gain in constraining power and while we find no indication of issues at these larger scales, we have not included them in our validation tests for cosmology.}}.

We use the \cosmosis{}\footnote{\url{https://cosmosis.readthedocs.io}} framework for producing theory predictions and sample the posterior of cosmological and nuisance parameters via the \texttt{Nautilus} sampler \cite{lange_nautilus_2023}. Full details of the modeling can be found in \cite{y6-methods}.

\subsection{Measurement}\label{sec:clustering_measurement}

We use the Landy-Szalay estimator \cite{landy_szalay} of $\wtheta$:
\begin{equation}
\tilde{w}(\theta) = \frac{DD(\theta) - 2DR(\theta) + RR(\theta)}{RR(\theta)},
\end{equation} 
where the normalized pair counts of galaxies ($D$) and randoms ($R$) for a given separation $\theta$ are given by:
\begin{align}
 \quad DD(\theta) &= \frac{N_{DD}(\theta)}{N_D(N_D-1)/2} \\
DR(\theta) &= \frac{N_{DR}(\theta)}{N_D N_R} \\
RR(\theta) &= \frac{N_{RR}(\theta)}{N_R(N_R-1)/2},
\end{align}
{and we use a tilde ($\tilde\ $) to indicate an estimator.} The randoms catalog for each redshift bin is Poisson sampled from a uniform density across the footprint, with $N_R=50\times N_D$.

\begin{figure*}
    \centering
    \includegraphics[width=\textwidth]{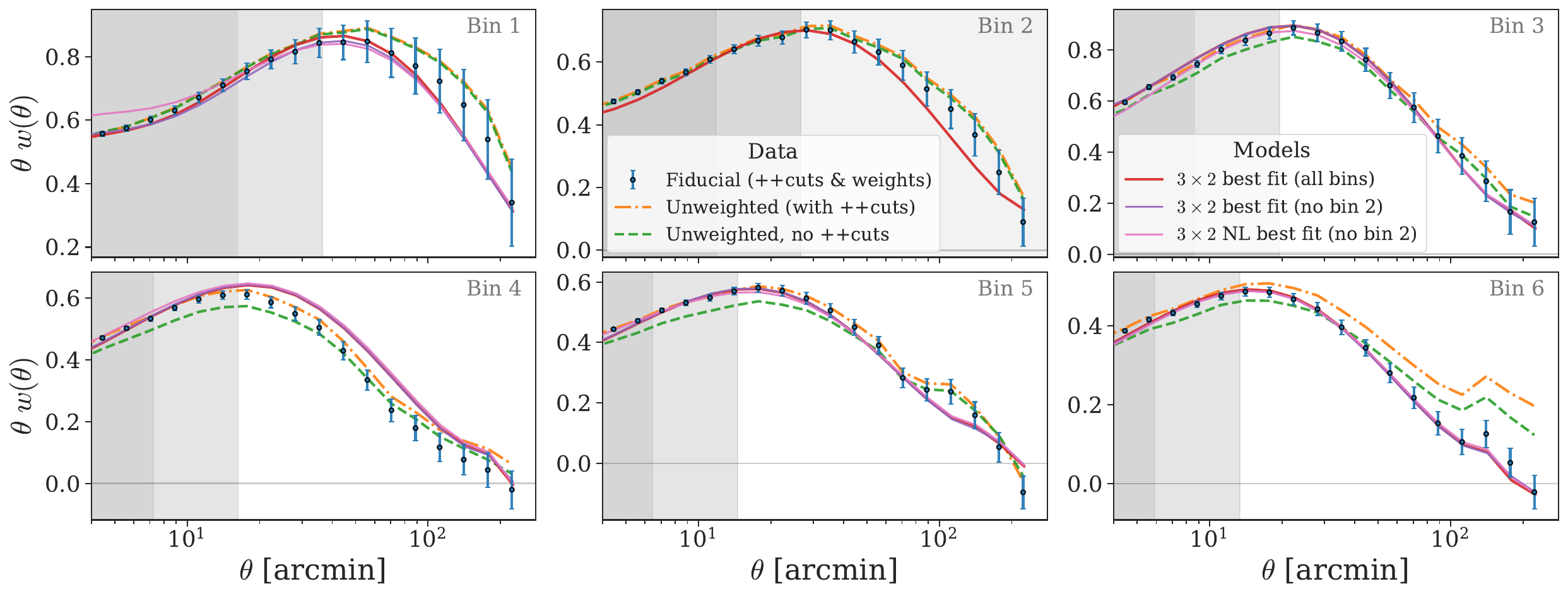}
    \caption{$w(\theta)$ measurements of \maglimpp{} (blue), including weights and debias correction (Sec.~\ref{sec:bias_and_cov_corrections}) as used for cosmological analysis in \cite{y6-3x2pt, y6-2x2pt}. In orange we show without systematic weights applied, and in green dashed the raw measurements without the quality cuts and improved star-galaxy separation described in Sec.~\ref{sec:data_processing_and_qual_cuts}. The red line is the best fit \threextwopt{} \LCDM{} cosmology prediction using all bins with linear galaxy bias. {Note that errors are strongly correlated, so coherent offsets between model and data like in bin 4 do not necessarily imply large inconsistency (see App.~\ref{sec:wtheta_correlation}).} We also show the best-fits when dropping bin 2 and adopting non-linear bias, for comparison, {finding consistent results}. Dark (light) shaded regions indicate scales removed for the nonlinear (linear) cosmology analyses, and bin 2 is fully shaded, as it was removed for the final analyses to meet unblinding criteria (see \cite{y6-3x2pt}).}
    \label{fig:wtheta}
\end{figure*}

In practice, we take the common approach of estimating $w(\theta)$ in logarithmic bins\footnote{This is equivalent to using top-hat basis functions for the estimator. See \citet{storey-fisher_two-point_2021} for a slightly more optimal approach.} of $\theta$ using \textsc{TreeCorr}\footnote{We use \textsc{TreeCorr} v5.1.1 with 30 log-spaced bins of $\theta\in [2.5,\ 2500]$ arcmin and $\texttt{bin\_slop}=0.005$. In practice, we use a maximum scale of $\theta<250'$ and minimum scales described in the text. For the $\wtheta$ consistency tests, we sometimes use $\texttt{bin\_slop}=0.02$ but have confirmed that this doesn't affect our results.}.
Fig.~\ref{fig:wtheta} shows $\tilde w(\theta)$ as measured from the data after applying weights (blue points) and used for cosmological analysis. Orange dash-dot line shows the same measurements without weights, and green dashed without the interloper and star-galaxy quality cuts described in Sec.~\ref{sec:data_processing_and_qual_cuts}. The red solid line shows the \threextwopt{} best-fit \LCDM{} theory prediction \cite{y6-3x2pt}. Error bars correspond to the analytical covariance, but note that $\tilde w(\theta)$ points are highly correlated across $\theta$, such that coherent shifts are more expected than random scatter {in the residuals} (see App.~\ref{sec:wtheta_correlation}).

Table~\ref{tab:wtheta_pval} gives $p$-values for each redshift bin for a range of models and data combinations. These are computed via $\chi^2$ statistics
\be
\chi^2 = \left[\tilde w(\theta) - w(\theta)_{\rm model}\right]^T \mathcal{C}^{-1} \left[\tilde w(\theta) - w(\theta)_{\rm model}\right],
\ee
where $\mathcal{C}$ is the $\wtheta$ portion of the fiducial analytical covariance matrix described in \cite{y6-methods}, including the LSS terms described in Sec.~\ref{sec:bias_and_cov_corrections} (we verify that using the covariance at the best-fit determined without lens bin 2 effects no meaningful change to these results), and assuming these follow a $\chi_\nu^2$ distribution with an approximated number of degrees of freedom
\begin{equation*}
\nu = N_{\rm data}^{\wtheta} - N_{\rm eff}^{3\times2} \left(\frac{N_{\rm data}^{\wtheta}}{N_{\rm data}^{3\times2}}\right)
\end{equation*}
and $N_{\rm eff}^{3\times2}\approx15$.
We see generally good agreement across bins and for all bins combined, with some tension with the model predictions for data in bin 5. 

We perform a data-only consistency test by splitting the sample in half at DEC$\lessgtr-35$ and assess consistency of the respective $\wtheta$ measurements for each redshift bin as 
\be
\chi^2 = \left[\tilde{w}(\theta)_{>-35} - \tilde{w}(\theta)_{<-35}\right]^T \left[\frac{\mathcal{C}^{-1}}{4}\right] \left[\tilde{w}(\theta)_{>-35} - \tilde{w}(\theta)_{< -35}\right],
\ee
find good consistency for each bin, with 
\begin{align*}
    p_{\rm nonlin}&=[0.86,\ 0.16,\ 0.54,\ 0.30,\ 0.64,\ 0.60]\\
    p_{\rm lin\ only}&=[0.93,\ 0.16,\ 0.80,\ 0.16,\ 0.37,\ 0.52]
\end{align*} 
with the latter restricted to only scales used with the linear bias model.

We compute a model independent\footnote{\citet{y6-3x2pt} and some previous DES papers have used the SNR definition of ${\rm(S/N)}_{\rm mod} = \left[D_{\rm model}^T\textbf{C}^{-1}\tilde D_{\rm data}\right]/\sqrt{D_{\rm model}^T\textbf{C}^{-1} D_{\rm model}}$, which can be roughly interpreted as the significance of detection after projecting the observed data onto the best-fitting model prediction. This removes most of the noise contribution and also renders the calculation technically model-dependent. We quote the model-independent version because we find it more parsimonious, but we find it gives very similar results ($\rm{(S/N)}_{\rm data} - \rm{(S/N)}_{\rm mod}\sim0.4$) to the alternative definition. A more directly comparable model-independent version would also remove the expected noise contribution, which is just $\langle\chi^2\rangle =N_{\rm dat}$, giving $\rm{(S/N)}_{\rm data,\ alt}=\sqrt{\tilde D_{\rm data}^T\textbf{C}^{-1}\tilde D_{\rm data} - N_{\rm data}}$, which shows very good  agreement: $\rm{(S/N)}_{\rm data,\ alt} - \rm{(S/N)}_{\rm mod}\sim0.05$ (this latter version is also used in \citet{sanchez_dark_2022}).} signal-to-noise-ratio for all bins via
\begin{align}
\left(\frac{S}{N}\right)_{\rm data} &= \sqrt{\tilde w^T\mathcal{C}^{-1}\tilde w},
\end{align}
finding 148.8 (90.2) for the non-linear (linear) analysis scenarios. These reduce to 142.8 (86.0) when removing bin 2.

\begin{table}
\setlength{\tabcolsep}{2pt}
\renewcommand{\arraystretch}{1.1} 
    \begin{tabular}{ccccccc}
    \hline
     Model & Bin 1 & Bin 2 & Bin 3 & Bin 4 & Bin 5 & Bin 6 \\
    \hline
    \threextwopt{} (Fid) & 0.64 & - & 0.78 & 0.14 & 0.007 & 0.97 \\ 
    \threextwopt{} (+ bin 2) & 0.54 & 0.42 & 0.72 & 0.14 & 0.007 & 0.97 \\ 
    \threextwopt{} NL & 0.79 & - & 0.66 & 0.21 & 0.007 & 0.99 \\ 
    \threextwopt{} $w$CDM & 0.61 & - & 0.74 & 0.18 & 0.005 & 0.94 \\ 
    \twoxtwopt{} & 0.69 & - & 0.79 & 0.12 & 0.008 & 0.98 \\ 
    \twoxtwopt{} NL & 0.78 & - & 0.69 & 0.20 & 0.007 & 0.99 \\ 
    \twoxtwopt{} (+ bin 2) & 0.59 & 0.50 & 0.80 & 0.17 & 0.006 & 0.96 \\ 
    \twoxtwopt{} NL (+ bin 2) & 0.80 & 0.30 & 0.72 & 0.06 & 0.012 & 0.99 \\ 
    \hline
    \end{tabular}
    \caption{Goodness of fit $p$-values for $\wtheta$ in each redshift bin, given the best-fit  prediction for each model. Bin 2 is excluded from the model fit unless otherwise noted, though we find no evidence for inconsistency of the bin 2 $\wtheta$ measurement in our null-tests. Calibrating the $p$-values for each model for multiple testing to control the false discovery rate \cite{false_discovery_rate} results in bin 5 being simply a statistical fluke at $p\in[0.031, 0.074]$. For the full six bin data vector, we find $p\in[0.11,\ 0.27]$ across models.}\label{tab:wtheta_pval}
\end{table}

When computing $\wtheta$ for the \nside{} 1024 \healpix{} lognormal mocks, we make use of a pixelized version of the estimator:

\begin{equation}
\tilde{w}(\theta) = 
\sum_{i=1}^{N_{pix}}\sum_{j=1}^{N_{pix}} \frac{(n_i-\bar{n})(n_j-\bar{n})}{\bar{n}^2} \Theta_{ij},
\end{equation} 
where $n_i$ is the galaxy number density in pixel $i$, $\bar{n}$ is the mean galaxy number density over all pixels within the footprint and $\Theta_{i, \, j}$ is a top hat function equal to $1$ when pixels $i$ and $j$ are separated by an angle $\theta$ within the bin size $\Delta \theta$ and otherwise 0. The area of each pixel (accounting for the mask) is incorporated into $n_i$ and $\bar{n}$, and used as a weight in \textsc{TreeCorr}'s \texttt{KKCorrelation} function. 
We have verified on pixelated versions of the catalog that there is good agreement between this pixel-based estimator and the full estimator above $\sim3\times$ the pixel scale.

\section{Null Tests and Estimator Corrections}\label{sec:null_tests}
{As part of the blinding procedures to avoid biasing the analysis, a number of null tests were predetermined to identify the possible presence of systematics in the datavector. We set the criteria  for satisfying these tests and proceeding with analyses before conducting the tests. Furthermore, we avoided any comparison of data vs.\ theory or plotting $\wtheta$ directly until the datavector null tests and cosmology unblinding criteria in \citet{y6-3x2pt} were satisfied.}

\subsection{Residual systematic tests ($\dchisqsub{0}$ against SP maps)}
As noted in Sec.~\ref{sec:sysmitigation}, we use a relatively large
{number of spatial templates and model the contamination perturbatively, to first or third order with \enet{} and \isd{}, respectively, while masking areas where such low-order models are more likely to break down.} 
Here we test whether our models are too simplistic by looking for evidence of residual contamination.

\textbf{Null test requirement:}
\textit{Plot distribution of $\chi^2$ values of weighted density field vs.\ SP map value for each SP map. Compare to a $\chi^2$ distribution with appropriate dof and visually inspect SP map-density trends of any strong outliers and consider adding freedom to the weights to model these.}

For each redshift bin, we check for evidence of any residual systematics (e.g., from higher-order dependence) corresponding to each SP map $i$ by checking consistency with the mean ($\bar{n}$) of the observed number density ($n$) in deciles of the SP map via a $\chi^2_{i}$ (implicit $z$ index suppressed for clarity):

\begin{equation}\label{eq:chi2_sp}
\chi^2_i = (\hat{n}_{ij} - \bar{n})^T [\mathbf{C}_i]^{-1} (\hat{n}_{ij} - \bar{n}).
\end{equation}

Here, $\hat{n}_{ij}$ is the observed number density of galaxies (with systematic weights applied) in decile bin $j$ of SP map $i$ for a given redshift bin, and $\mathbf{C}_i$ is the covariance matrix of the observed density in each decile as computed with 1000 lognormal mocks free of systematics.

The naive empirical covariance matrix for each SP map 
would be rather noisy as we are estimating $10\times10 = 100$ elements with 1000 mocks, and doing this for each of 19 SP maps for each of 6 redshift bins. Even if we debias the precision matrix for each of these covariances so they are unbiased on average, there will still be significant variance. We opt instead to use the optimal linear shrinkage covariance estimator of  \citet{LEDOIT2004365}, which minimizes the mean squared error of the elements of the covariance estimate via the Froebenius norm, and invert this to estimate the precision matrix.

Eq.~\ref{eq:chi2_sp} thus measures the significance of the weighted observed density fluctuations as a function of each SP map with respect to what is expected from cosmic variance, and we find very good consistency across all SP maps.

\begin{figure}
    \centering
    \includegraphics[width=\linewidth]{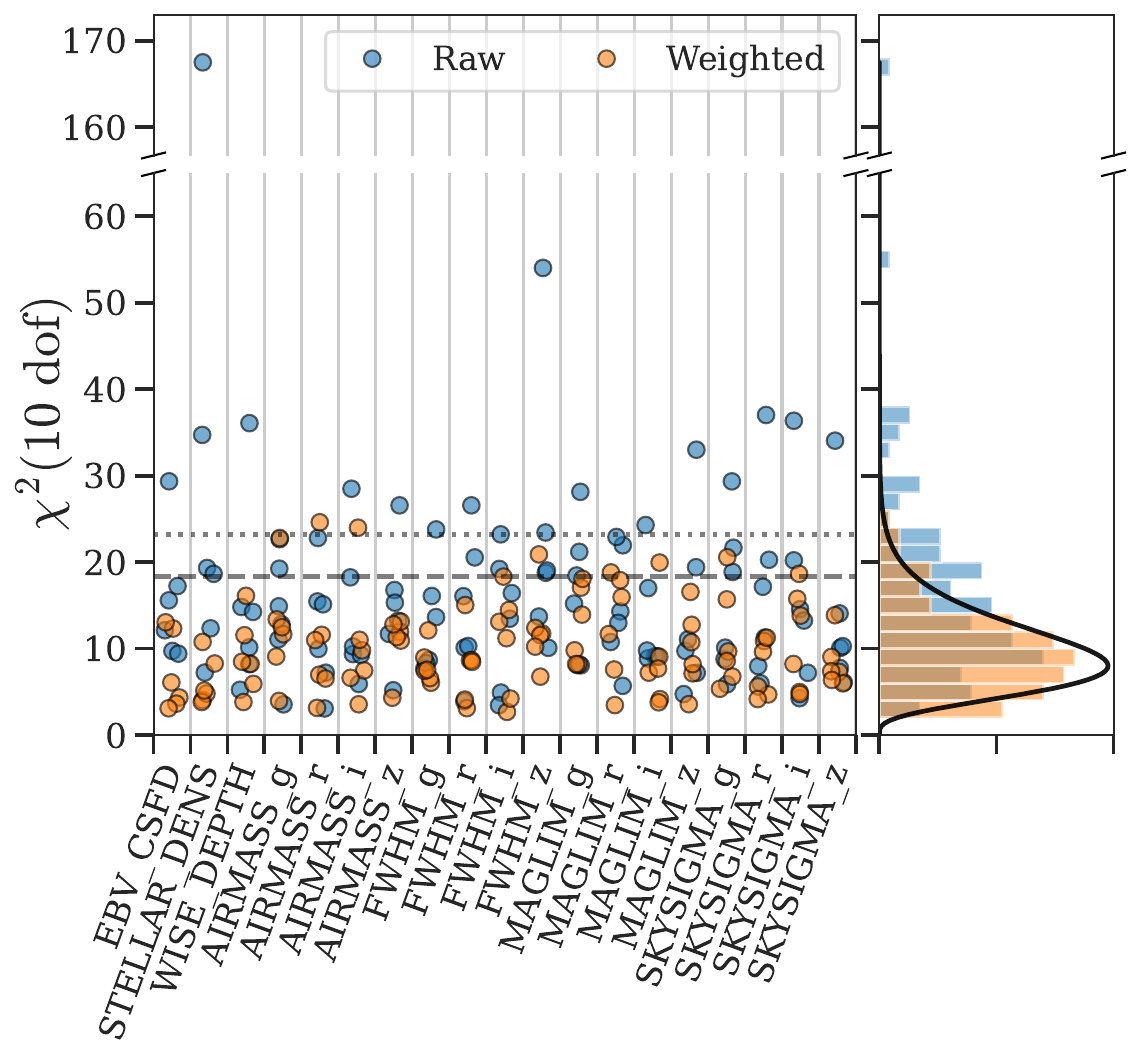}
    \caption{$\chi^2$ of raw and weighted density fields (blue, orange respectively) binned by deciles of each template. \textit{Left:} $\chi^2$ for each redshift bin, for each template (offset for visualization). Dashed (dotted) lines show thresholds of $p=0.05$ $(p=0.01)$. \textit{Right:} Histogram of results for all redshift bins and templates, as compared to a $\chi^2$ distribution with 10 degrees of freedom, which we expect in the absence of contamination. The raw data show clear evidence of contamination, which is removed after weights are applied. Note the broken axis. A KS-test finds $p_{\rm raw}=8\times10^{-11}$ and $p_{\rm weighted}=0.65$, indicating very good consistency after weighting.}
    \label{fig:chi2resid}
\end{figure}

Fig.~\ref{fig:chi2resid} shows $\chi^2$ before (blue) and after (orange) weighting for each SP map and each redshift bin. Note the broken y-axis; the blue point with $\chi^2 \sim 165$ corresponds to stellar density in the highest redshift bin before corrective weights are applied.
We perform 1-sided KS-tests of the $\chi^2$ values before and after weighting to test (in)consistency with being drawn from a $\chi^2_{\nu=10}$ distribution. For all bins combined, we find applying weights takes $p=(8\times10^{-11}) \rightarrow  0.65$. For individual redshift bins, we have $p_{\rm raw}=[0.79, 0.090, 0.021, 0.0003, 0.0054, 0.000]$ and $p_{\rm weighted} = [0.99, 0.76, 0.47, 0.73, 0.36, 0.14]$. 

As an additional test, we added several second-order templates in the \enet{} fits, but found that the additional mode removal outweighed any gain in predictive power (i.e., identification of residual systematics) and the optimal coefficients assigned to these additional templates was zero. 

\subsection{$\wtheta$ bias and covariance corrections}\label{sec:bias_and_cov_corrections}

\begin{figure*}
    \centering
    \includegraphics[width=\linewidth]{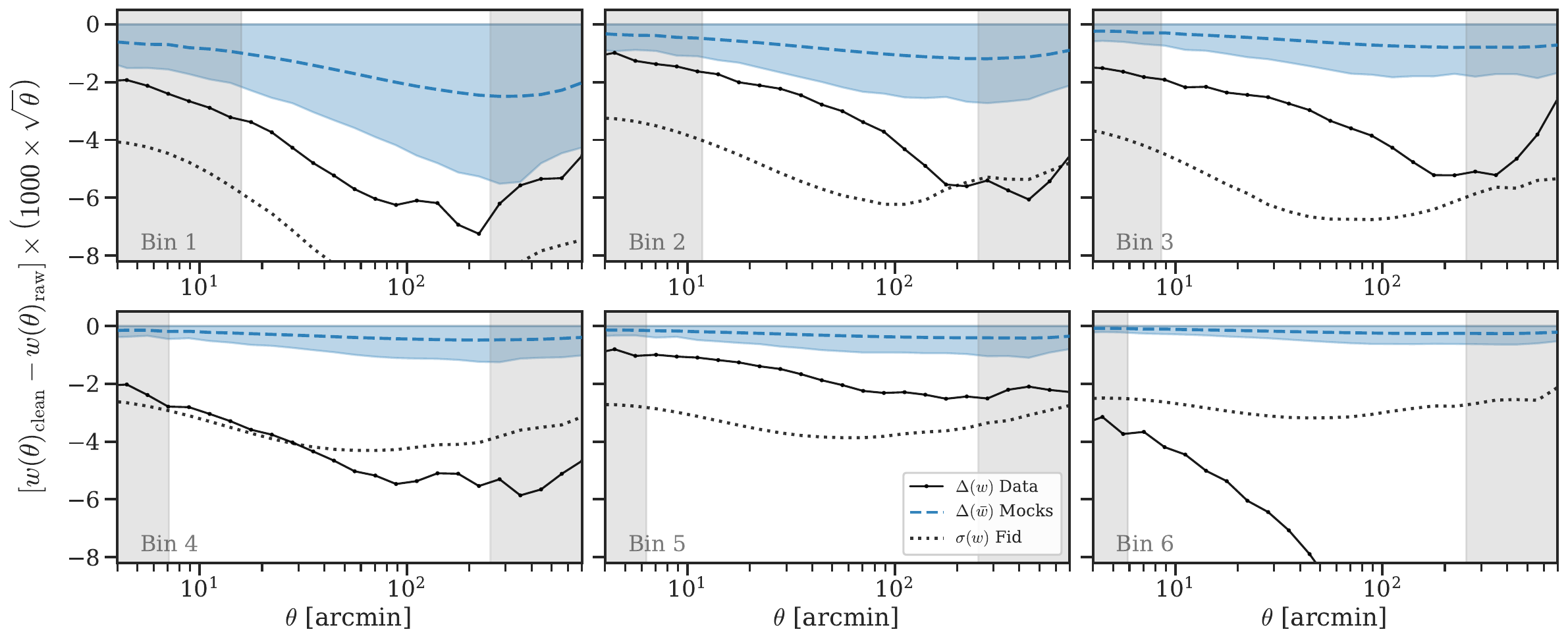}
    \caption{Suppression of $w(\theta)$ due to nulling of true LSS modes from \enet{} overcorrection, as estimated from 300 uncontaminated lognormal mocks. The dashed line gives the expectation used to debias $\hat{w}(\theta)$ via Eqs.~\ref{eq:wtheta_debias} and ~\ref{eq:weights_bias}, and the shaded region indicates the central $68\%$ spread across mock realizations. The black line shows $\Delta w(\theta)$ from \enet{} on the real data (before debiasing), indicating that impact on $\wtheta$ is likely dominated by the removal of real contamination.  Dotted lines indicate $1\sigma$ error bars from the diagonal of the covariance matrix, and gray shading indicates scales removed for cosmological analyses. Note there is very strong correlation between data points such that coherent shifts are more typical than not, see App.~\ref{sec:wtheta_correlation}.}
    \label{fig:overcorrection_mocks}
\end{figure*}

{Both of our approaches to calculating weights result in only a small removal of true LSS modes because of their respective approaches to minimize overfitting. With \enet{} this is through zero-centered contamination priors, which are calibrated on the data to maximize predictive power on hold-out samples. With \isd{}, this is through only fitting for measured contamination beyond that expected based  on realistic mocks.} This is in contrast to ordinary least squares (OLS) regression/pseudo-$\Cl{}$ mode projection schemes and especially machine learning approaches, where chance removal of LSS is much higher as well as the resultant suppression of $w(\theta)$ (see App.~\ref{sec:more_weights_tests} for a comparison). 

We can numerically estimate and remove the bias that we expect from using \enet{} and \isd{}, which we do by running the weights estimation on uncontaminated lognormal mocks and computing the mean additive bias for each redshift bin over all mocks:
\begin{align}
    \label{eq:weights_bias}
    \badd{\rm X} &\equiv \overline{\Delta w}(\theta)^{\rm bias,\ X}\\
    &= \frac{1}{N_{\rm mock}}\sum_{i=1}^{N_{\rm mock}}\left(\tilde w(\theta)^{\rm cleaned,\ X}_i - \tilde w(\theta)^{\rm true}_i\right),
\end{align}
for each weighting method X $\in \{\enet{}, \isd{}\}$, which is used to debias\footnote{The lognormal mocks are generated with a resolution of $\nside{}=1024$, so we use the pixelized $\wtheta{}$ estimator in \textsc{TreeCorr} with overdensity and mask maps at that same resolution for each bin when computing the terms in Eq.~\ref{eq:weights_bias}. Minor pixelation effects begin to appear at the smallest scales, which are wholly negligible compared to the covariance and all but removed via the scale cuts for the \threextwopt{} analysis. Never-the-less we linearly extrapolate the bias calculation to scales below $\theta\sim 14'$ ($\sim4\times$ the pixel side-length) to smooth these and avoid imprinting these minor oscillations on the data vector.} the two-point functions of the weighted data in each redshift bin:
\be
\hat{w}(\theta)^{\rm X}_{\rm data} = \tilde w(\theta)_{\rm data}^{\rm X} - \badd{\rm X}.\label{eq:wtheta_debias}
\ee

Fig.~\ref{fig:overcorrection_mocks} shows the expected bias imparted by the $\enet$ weights ($\left(\tilde w(\theta)^{\rm cleaned,\ X}_i - \tilde w(\theta)^{\rm true}_i\right)$) on the inferred $\wtheta$ using 300 uncontaminated lognormal mocks. The expected bias $\badd{\texttt{ENET}}$ used to debias $\hat{w}(\theta)$ is indicated by the dashed blue curve, whereas the shaded region shows the central $68\%$ spread across mock realizations. {More than 16\% of mocks in each bin recover $\wtheta$ almost perfectly, as indicated by the shaded boundary reaching $\wtheta_{\rm clean} - \wtheta_{\rm raw}\approx0$.  In the $\sim10\%$ of cases where this estimate is perfect,} the \enet{} algorithm shrinks all the contamination amplitudes to 0, such that weights are all exactly 1, indicating no contamination.
The black curve indicates the impact of the \enet{} weights on the real data before debiasing, such that the difference between the black and blue curves is an estimate of the real systematic contamination removed via weights. While not shown, the mean overcorrection with \isd{} weights (with $\isdthresh{}=2$) is even smaller than for \enet{}.

\begin{figure*}
    \centering
    \includegraphics[width=\textwidth]{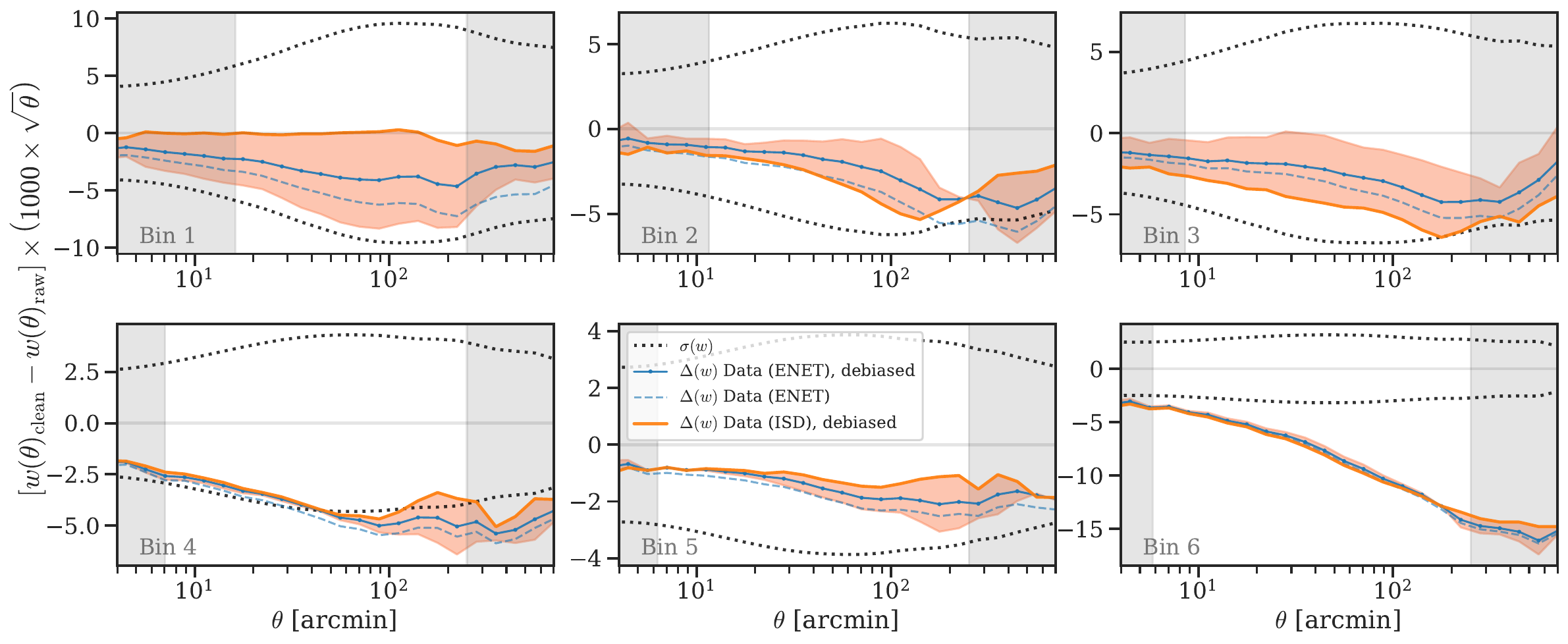}
    \caption{The estimated impact of spatial systematics corrected via systematic weights. Solid blue line shows the difference between the debiased $\wtheta$ estimate using \enet{} weights vs.\ $\wtheta^{\rm raw}$ (no weights). The orange curve shows the same thing but for $\isd$ weights. The difference between these (orange shaded) is taken to be a systematic uncertainty that is added to the covariance via Eq.~\ref{eq:weights_cov}. The dotted curves show the uncertainty from the diagonal of the covariance. The faint blue dashed line shows the estimate before the estimator is debiased via Eq.~\ref{eq:weights_bias}; the difference from the solid blue curve is the bias term $b(\theta)_{\rm add}^{\texttt{ENET}}$ and indicates the expected suppression of $\wtheta$ from removing true LSS modes in the absence of any true systematics. Note that after the quality cuts described in Secs.~\ref{sec:data_processing_and_qual_cuts} and extensive masking in \cite{y6-mask}, the systematic weights have only a relatively small impact on the sample in most redshift bins.}
    \label{fig:overcorrection}
\end{figure*}

We analytically marginalize over a mode for each redshift bin corresponding to the difference between the \enet{} and \isd{} cubic weights schemes by adding a term to the covariance matrix:
\be
\texttt{Cov}\left[\wtheta_i^k, \wtheta_j^l\right] \rightarrow \texttt{Cov}\left[\wtheta_i^k, \wtheta_j^l\right] + \Delta_{{\rm sys},i}^k \Delta_{{\rm sys},j}^l \delta_{kl} \label{eq:weights_cov}
\ee
where $\Delta_{{\rm sys},i}^k = \hat{w}(\theta_i)^{\enet}_{\rm data} - \hat{w}(\theta_i)^{\isd}_{\rm data}$, $i$ and $j$ run over $\theta$ bins and $k$ and $l$ run over redshift bins. $\delta_{kl}$ is a Kronecker delta that ensures we only modify the covariance for terms within the same redshift bin. 

This converts our systematic theoretical uncertainty on the weights methodology into statistical error, ensuring that the substitution of one weighting scheme with the other results in at most a $1\sigma$ statistical fluctuation for each redshift bin. We don't include cross-bin covariance terms as this would provide less flexibility for the data in individual bins to prefer one weights scheme vs.\ the other (it would be equivalent to marginalizing over a single parameter across all bins, rather than a parameter for each bin).

In Fig.~\ref{fig:overcorrection} we show the same impact of weights on $\wtheta$ as in Fig.~\ref{fig:overcorrection_mocks}, but comparing \enet{} (blue) and \isd{} (orange), both debiased via Eq.~\ref{eq:wtheta_debias}. The orange shading shows the difference of the two estimates, which is added to the covariance to analytically marginalize over weights choice via Eq.~\ref{eq:weights_cov}. 
After the quality cuts described in Secs.~\ref{sec:data_processing_and_qual_cuts} and extensive masking in described in \citet*{y6-mask}, the systematic weights have a relatively minor impact on the data vector relative to the uncertainty (dotted curves) in most redshift bins, with the highest redshift bin showing a much larger impact. 

We test the robustness of $\hat{w}(\theta)$ to several different analysis choices and report the significance of each shift as:
\be \label{eq:chi2alt}
\chi_{\rm alt}^2 = \left(\hat{w}(\theta)_{\rm alt} - \hat{w}(\theta)_{\rm fid}\right)^{T} \mathcal{C}^{-1} \left(\hat{w}(\theta)_{\rm alt} - \hat{w}(\theta)_{\rm fid}\right)
\ee
where $\mathcal{C} = \texttt{Cov}\left[\wtheta_{\rm fid}\right]$.
We test variations of:
\begin{enumerate}
    \item The \isd{} stopping threshold. A lower threshold removes more LSS modes but  is also more sensitive to systematics. We test setting $\isdthresh = 2 \rightarrow 1$.
    \item The \isd{} linear contamination model, fitting a linear instead of cubic polynomial of the overdensity to each template.
    \item The clustering amplitude of mocks used to characterize the overcorrection bias in Eq.~\ref{eq:weights_bias}. More LSS power typically results in more chance mode removal from templates, though the data-driven overfitting protection in \enet{} should help to mitigate this. We test the impact of our fiducial cosmology by recomputing Eq.~\ref{eq:weights_bias} with galaxy bias values changed by $\pm10\%$, which result in $\sim20\%$ variations in the amplitude of the mock galaxy clustering power spectra.
    \item Whether the mocks used to assess the debiasing term are contaminated with systematics or not. We test the case where the debias term is calculated using mocks contaminated with the fiducial \enet{} weights.
\end{enumerate}

The results of these tests are shown in Table~\ref{tab:chi2_variations}. We show $\chi^2_{\rm alt}$ for both the full $\wtheta$ as well as for each redshift bin individually. For comparison, the last row shows $\chi^2$ for $\hat{w}(\theta)_{\rm alt} = {w}(\theta)_{\rm unweighted}$, i.e., the significance of the shift from the application of weights. We find that all of these variants result in negligible impact on our results (with the exception of not applying any weights at all).

\begin{table}
    \centering
    \setlength{\tabcolsep}{5pt}
    \renewcommand{\arraystretch}{1.1} 
    \begin{tabular}{l|cccccc|c}
    \hline
    Redshift Bin: & 1 & 2 & 3 & 4 & 5 & 6 & Full \\
    \hline
    ISD T1 Linear & 0.1 & 0.2 & 0.1 & 0.0 & 0.0 & 0.0 & 0.4 \\
    ISD T2 Linear & 0.1 & 0.1 & 0.0 & 0.1 & 0.0 & 0.1 & 0.5 \\
    ISD T1 Cubic & 0.2 & 0.4 & 0.3 & 0.1 & 0.1 & 0.1 & 1.3 \\
    ISD T2 Cubic & 0.2 & 0.2 & 0.2 & 0.1 & 0.1 & 0.1 & 1.0 \\
    High amp. mocks & 0.0 & 0.0 & 0.0 & 0.0 & 0.0 & 0.0 & 0.0 \\
    Low amp. mocks & 0.0 & 0.0 & 0.0 & 0.0 & 0.0 & 0.0 & 0.1 \\
    Cont. mocks & 0.1 & 0.0 & 0.0 & 0.1 & 0.0 & 0.0 & 0.2 \\
    No debias term & 0.1 & 0.1 & 0.0 & 0.0 & 0.0 & 0.0 & 0.2 \\
    Unweighted & 0.3 & 0.6 & 0.6 & 2.8 & 0.6 & 41.3 & 45.8 \\
    \hline
    \end{tabular}
    \caption{Significance of shifts in $\wtheta{}$ when adopting different analysis choices (see Sec.~\ref{sec:clustering}), as characterized by $\chi_{\rm alt}^2$ (Eq.~\ref{eq:chi2alt}). We also show the impact of not applying the fiducial \enet{} debiasing correction or not applying weights at all, for comparison. We report the shift by redshift bin as well as for all bins combined, finding that the the analysis is robust to these choices. 
    }\label{tab:chi2_variations}
\end{table}

\subsection{Null tests against external LSS tracers}\label{sec:weights_corr_external_tracers}
Mitigation of systematics using spatial templates crucially relies on \textit{not} being able to fit any true LSS modes beyond that expected from cosmic variance. If the systematics model is able to fit true LSS modes in the map then the predicted density field and two point functions will be strongly biased, nulling existing LSS and suppressing the galaxy clustering signal. This can happen if, e.g., there are two correlated templates that trace a similar systematic, but one of these also contains a small amount of LSS.

{This phenomenon was discovered during post-unblinding tests for the DES Y3 \threextwopt{} analysis.} The template library was increased to $\sim\mathcal{O}(100)$ {templates} to see if additional systematics could be identified in the \redmagic{} sample that explained the poor \threextwopt{} fit {with that sample. A strong signal initially suggested that this larger library identified residual systematic contamination, but instead it was because weights with the expanded library contained actual LSS, as identified through strong correlations with multiple external LSS tracers. This issue with the weights was addressed through a limited principle component basis as described in \cite{y3-galaxyclustering}, but the mechanism through which LSS entered the weights was only determined later, and is detailed in App.~\ref{sec:lss_in_weights_app}.}

\textbf{Null test requirement:}\\
\textit{Compute significance of correlations of weight maps with external LSS tracers (ACT + Planck Compton-$y$, ACT DR6 CMB lensing, DES $\kappa$ bin 4), require $\le1$ tracer with $\ge3\sigma$ correlation.}

We test for LSS contamination in our weights by cross-correlating them with the external LSS maps described in Sec.~\ref{sec:template_rejection}. Correlation can arise either because the \maglim{} weights contain LSS that is also traced by the external LSS map, or because the weights successfully correct for a systematic that has \textit{not} been removed from the external map. To help differentiate these, we allow for significant ($>3\sigma$) correlation of any given weights map with at most one of the three LSS tracer maps described in Sec.~\ref{sec:template_rejection}; where correlation between the weights and multiple tracers is taken to indicate the presence of actual LSS. {Note that unlike for cosmic shear, contamination of the number density is local \cite{Alonso_2019}; as such 1-point correlation statistics are the obvious choice for assessing its presence.}

\begin{figure}
    \centering
    \includegraphics[width=1\linewidth]{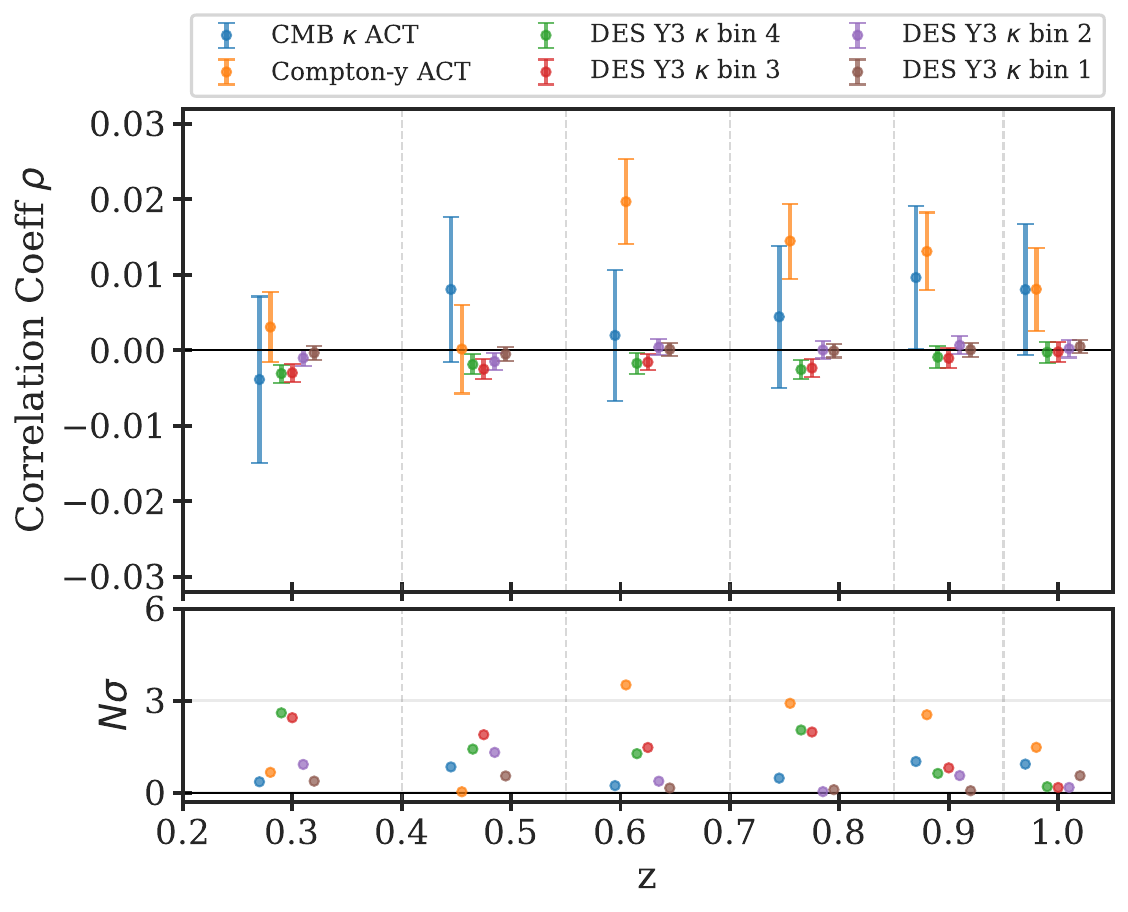}
    \caption{Correlation coefficient of fiducial weight maps for each \maglimpp{} Y6 redshift bin with external LSS tracers. Dashed vertical lines indicate bin edges. 
    We find no evidence that the weights contain LSS. We do find a mild correlation with the ACT+Planck Compton-y map, but the significance is only $>3\sigma$ for one bin, and not detected in any of the other tracers. 
    Note that correlation may appear if the weights remove a contaminant that has \textit{not} been removed from the LSS tracer. We therefore require a detection in more than one external LSS map in order to consider this null test failed. Results for ISD look very similar; see Fig.~\ref{fig:rho_extlss_rmy3} for a similar plot for a version of the DES Y3 Redmagic sample weights that contain LSS and shows large and significant correlations across external tracers.}
    \label{fig:rho_extlss}
\end{figure}

Fig.~\ref{fig:rho_extlss} shows the Pearson correlation coefficient between the weights in each redshift bin and each LSS tracer, with uncertainties computed via 200-patch jackknife (Fig.~\ref{fig:jackknife_patches}). We do not find significant evidence of LSS in the weights; only one map of one bin is detected at $>3\sigma$ (we note that the null test requirements were decided \textit{before} any such tests were performed). {While not part of the official unblinding null tests,} we also show the correlation with the lower redshift DES Y3 mass maps, and though we don't show them, we checked correlations with a version of the CMB $\kappa$ map without any filtering applied and found results consistent with no correlation. 
Results for the ISD weights are very similar, with only slight shifts in the points. 

\begin{figure}
    \centering
    \includegraphics[width=1\linewidth]{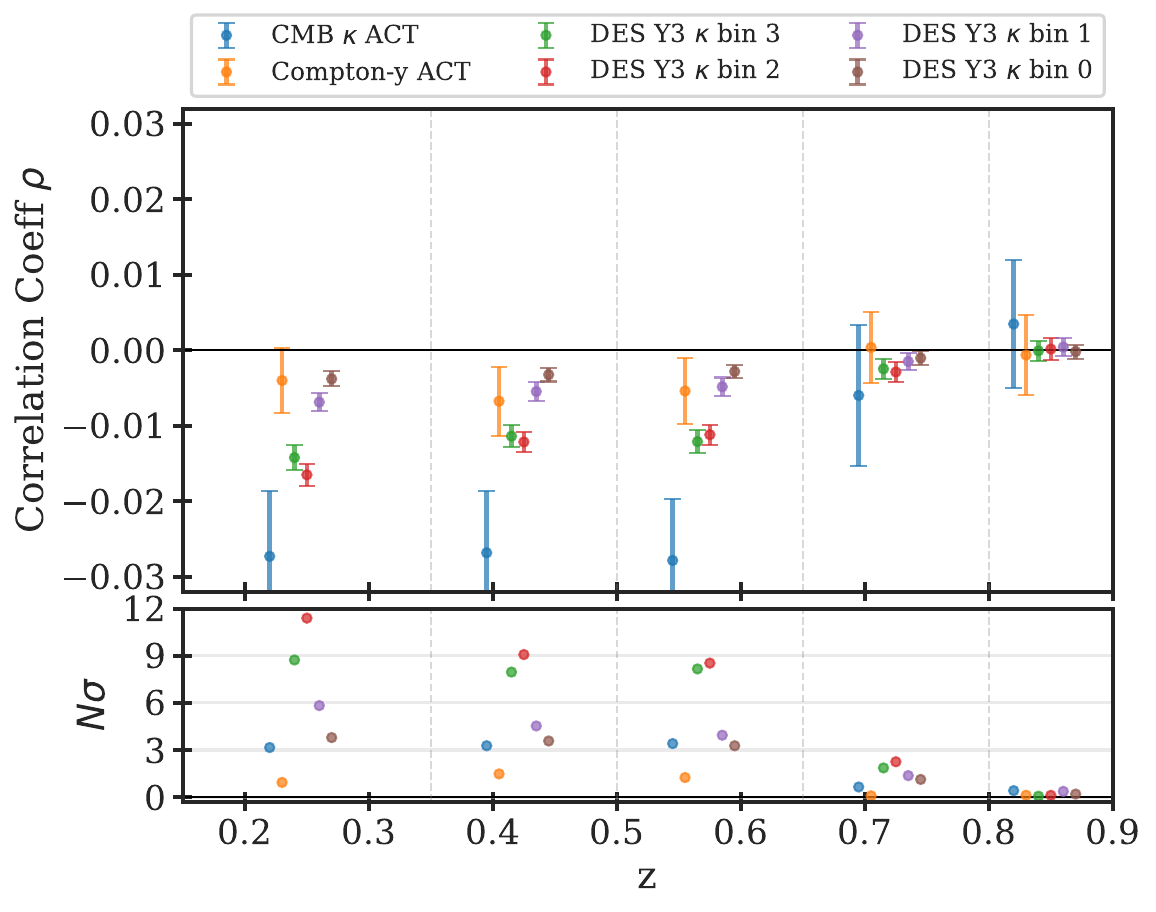}
    \caption{Same as  Fig.~\ref{fig:rho_extlss} but for  a version of the DES Y3 \redmagic{} sample weights that contain LSS. We see significant correlations across external LSS tracers (note the expanded y-axis for the bottom subplot).}
    \label{fig:rho_extlss_rmy3}
\end{figure}

For comparison, Fig.~\ref{fig:rho_extlss_rmy3} shows the same plot for the Y3 \redmagic{} weights that traced LSS.

\subsection{Spatial $n(z)$ tests}

As is typical, we assume a redshift distribution that does not systematically vary as a function of position beyond what is expected from cosmic variance. This is expected to be subdominant to the impact of systematic variations in the number count itself, but can still have non-negligible impacts on the inference of cosmological parameters \cite{lizancos_impact_2023}. 

To estimate the size and impact of such systematic variations,
we compare the model $\wtheta$ for a fiducial cosmology but with $n(z)$ coming from stacked $\dnf$ photo-$z$'s in different parts of the footprint. 
For each fiducial SP map, we stack the $\zmc$ of all galaxies in regions that lie in the top (bottom) $40\%$ of the SP map characteristic for a given redshift bin, and compute the significance of the shift in the galaxy clustering:
\be
\left[\chi^2\right]_i^t = \left[\Delta w(\theta)_i^t\right]^T\left[\frac{\mathcal {C}}{0.4}\right]^{-1}\left[\Delta w(\theta)_i^t\right]
\ee
where $\Delta w(\theta)_i^t = w\left(\theta, n(z)^t_i\right) - w\left(\theta, n(z)_{\rm full}\right)$
is the shift in the data vector when using $n(z)$ from subset $i$ defined by template $t$ rather than from the full footprint (redshift bin index is suppressed). Note that $n(z)$ is computed using the stacked $\dnf$ photo-$z$'s, not the fiducial \sompz{} estimate of the \threextwopt{} point analysis \cite{y6-3x2pt}.

Before looking at the results, we decided that if any SP map showed $\chi^2>2$, we would further investigate with a more complete \sompz{} analysis, but did not hold this to be a strict criterion for unblinding. We found $\chi^2<1$ for all SP maps, except for \texttt{y6\_FWHM\_decasu\_WMEAN\_z}, which showed $\chi^2\approx 1.5$ for both low and high regions (concentrated in the first redshift bin), which is still below the threshold of significance for concern.\footnote{Note these tests were performed early in the analysis before the covariance and scale cuts were finalized, so they used an early version of the covariance and larger set of scales as compared to other null tests.}

\section{Cosmological Constraints from Galaxy Clustering}\label{sec:cosmo_constraints}
Using only $\wtheta$, we cannot differentiate between the clustering of dark vs.\ baryonic matter and thus cannot break the degeneracy between galaxy bias and the amplitude of matter fluctuations, $\sigma_8$. However, we can measure the combined amplitude $b_i\sigma_8$, as well as probe the scale of matter-radiation equality and so constrain $\Omega_m$. We use \cosmosis{} and run the same linear bias \LCDM{} inference pipeline as described in \cite{y6-methods, y6-3x2pt} using only $\hat{w}(\theta)$ as our data vector. 

\begin{figure*}
    \centering
    \includegraphics[width=1\linewidth]{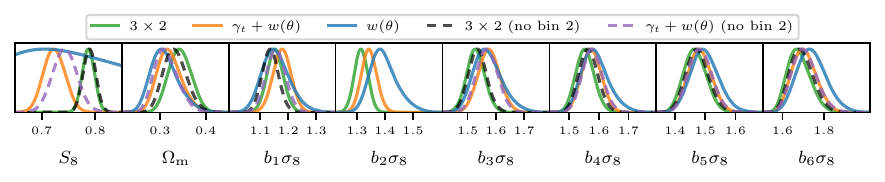}
    \includegraphics[width=1\linewidth]{
    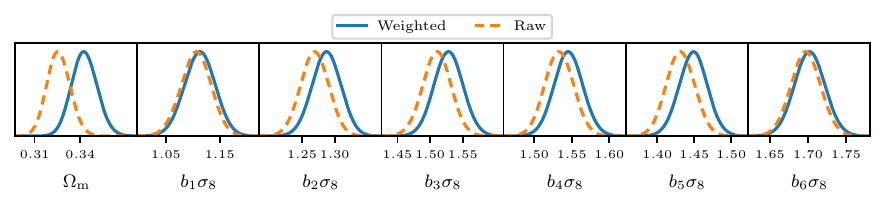}
    \caption{\textit{Top:}
    Cosmological constraints using $\wtheta$ only, $\wtheta + \gamma_t$, and $\wtheta + \gamma_t + \xi_{\pm}$ (\threextwopt{}) while varying all parameters. We show constraints on the combination $b_i\sigma_8$ for each redshift bin $i$, since this quantity is closer to what is probed by $\wtheta$ than galaxy bias alone and is largely uncorrelated across bins. We see good agreement across probes, with the exception of $S_8$ for $\wtheta+\gamma_t$, which is shifted significantly low. Dashed curves show constraints after removing bin 2.\\
    \textit{Bottom:} Constraints on $\Omega_m$ and $b_i\sigma_8$, using $\wtheta$ only and holding all other cosmological parameters fixed. Weights primarily impact the constraints on $\Omega_m$, as they suppress $w(\theta)$ on large scales.}
    \label{fig:om_bsig8_constraints}
\end{figure*}

We show marginal constraints on these parameters using all six redshift bins in the top row of Fig.~\ref{fig:om_bsig8_constraints}, and show how these constraints change with the addition of galaxy-galaxy lensing data ($\gamma_t$, orange) or both galaxy-galaxy lensing and cosmic shear data for the full \threextwopt{} datavector (green). The black dashed curve gives the constraint for the fiducial \threextwopt{} analysis with bin 2 removed, as is done for the final cosmology analysis (see App.~\ref{sec:bin2} and discussion in \cite{y6-3x2pt}).

We find consistency of the 1D posteriors across data combinations, with the {modest} exception of the $S_8$ constraint when combining $\wtheta$ and $\gamma_t$, for which a lower value is preferred. This is accompanied by strong shifts in the redshift nuisance parameters for bin 2 up to the prior boundaries, and moderate shifts in the mean redshifts of bins 1 (up) and 3 (down), {and is somewhat alleviated when bin 2 is removed.} 
For the $\wtheta$-only chain, we obtain 
\begin{equation*}
    \Omega_{\rm m} = 0.311^{+0.023}_{-0.035}
\end{equation*}
\begin{equation*}
    b_i\sigma_8 = \left[
    \begin{array}{c}
    1.16^{+0.04}_{-0.06}\\[0.5em]
    1.40^{+0.04}_{-0.06}\\[0.5em]
    1.57^{+0.04}_{-0.06}\\[0.5em]
    1.59^{+0.04}_{-0.05}\\[0.5em]
    1.50^{+0.04}_{-0.05}\\[0.5em]
    1.74^{+0.06}_{-0.08}
    \end{array}
    \right].
\end{equation*}

The bottom row of Fig.~\ref{fig:om_bsig8_constraints} shows constraints when fixing all parameters except $\Omega_m$ and $b_i\sigma_8$, and turning on and off the weights to demonstrate the impact. All other parameters are fixed to the \threextwopt{} best fit in \LCDM{} when fitting for all lens bins with the linear galaxy bias model (see the ``all bins" case in Fig.~\ref{fig:wtheta} for the corresponding theory prediction).
In this case, application of galaxy weights primarily impacts $\Omega_m$, whose conditional impact on the data vector is to suppress $\wtheta$ on large scales relative to small, which matches the impact of the weights seen in Figs.~\ref{fig:wtheta} and \ref{fig:overcorrection}.

\section{Summary and Conclusion}\label{sec:conclusion}
We have presented the \maglimpp{} galaxy sample and its angular clustering measurements in support of the DES Y6 \threextwopt{} cosmological analysis. {We placed constraints on the matter density of the Universe and amplitude of galaxy clustering across its six redshift bins.} This work represents a significant advancement in the {sample} selection and characterization and mitigation of systematic uncertainties in photometric galaxy clustering analyses while maintaining high statistical power.

The \maglimpp{} sample selection improves upon the Y3 \maglim{} selection through the two novel quality cuts designed to remove contamination from the catalog. 
First, we use a novel color-based stellar rejection scheme (described more fully in \cite{stargalsep}) that is tailored to the color distribution of objects in each redshift bin. This complements the global \texttt{EXTMASH} morphological classifier used for the Y6 Gold sample \cite{y6-gold},
{identifying and rejecting $\sim2\%$ of the total sample as residual stellar contamination}
resolving a degeneracy between the opposing effects of stellar contamination and obscuration and rendering contamination better suited for treatment with multiplicative galaxy weights. 
Second, we develop a SOM-based approach to identify and remove regions in color-space with very poor photo-$z$ characterization and a high likelihood of being contaminants. The cuts are tuned to limit their number and surface-area in color-space, while removing hard-to-characterize objects in the tails of the redshift distributions; {subsequent tests reveal that these objects are mostly QSOs}.

Combined, the SOM photo-$z$ and star-galaxy quality cuts remove approximately $3.5\%$ of the initial sample while eliminating regions with the highest contamination risk, resulting in a cleaner and more robust dataset for cosmological inference.

We also upgrade and implement two 
complementary methods for characterizing and correcting systematics in the galaxy selection function, which both show strong robustness to the nulling of real large-scale structure modes and are differently sensitive to the form and scales of potential contamination.

{We have upgraded the \isd{} method \cite{y1-wthetapaper, y3-galaxyclustering} to use non-linear contamination, and updated \enet{} \cite{Weaverdyck:2020mff} to use a more realistic likelihood and output higher resolution weights, both of which are enabled by improved masking that incorporates similar information as is used for estimating weights \cite{y6-mask}.}
We incorporate the difference between these methods into the covariance of $\wtheta$ when performing cosmological inference, thus marginalizing over the ``methodological uncertainty" of these corrections. The mean statistical suppression of the observed power spectrum from weighting is calibrated against realistic mocks and $\wtheta$ debiased for the final measurement.  We release the weights inferred from both methods to allow similar approaches for future analyses of the sample using other summary statistics. 

The sample passes a wide battery of null tests, several of which are novel. These include:
\begin{itemize}
    \item the use of external LSS tracers to cull templates and vet the resulting weights to ensure they don't inadvertently trace LSS
    \item non-parametric tests of consistency with null residual contamination of the weighted density field
    \item consistency of $\tilde w(\theta)$ between North and South regions of the footprint, split by declination
    \item spatial $n(z)$ tests to confirm negligible dependence on survey properties.
\end{itemize}

While $\wtheta$ and $\gamma_t$ measurements using lens bin 2 were removed in the \threextwopt{} analysis before unblinding, we find no evidence of abnormal behavior in this bin in our null tests (see App.~\ref{sec:bin2}). $\wtheta$ measurements show good consistency with theory predictions fit to the full data vector across a range of models and data combinations, with only bin 5 showing mild tension in goodness-of-fit. 

Using $\wtheta$ only, we place constraints on $\Omega_m$ and $b_i\sigma_8$ for each redshift bin $i$, and demonstrate the impact of the weights on these constraints when holding other parameters fixed. We find the size of the systematic weights corrections to be smaller than those in Y3 by roughly half, likely because of improvements in photometry \cite{y6-gold}, masking \cite{y6-mask}, and the catalog cleaning (this work), as well as the approach to integrate the processes of masking, catalog selection, and weights corrections.

The methodologies developed here are directly applicable to upcoming surveys such as LSST, Euclid and SPHEREx; as these surveys push to greater depths and/or volumes, the control of systematics will be critical to fully exploit the statistical power they will bring.
The DES Y6 \maglimpp{} catalog and clustering measurements provide a core component of the final DES cosmological constraints and a template for systematic control in the next generation of large-scale structure surveys.

\begin{acknowledgements}

NW is supported by the Chamberlain fellowship at Lawrence Berkeley National Laboratory. 
AP acknowledges financial support from the European Union's Marie Skłodowska-Curie grant agreement 101068581, and from the \textit{César Nombela} Research Talent Attraction grant from the Community of Madrid (Ref. 2023-T1/TEC-29011).

\textbf{Author Contributions:}
\textbf{NW} developed the methodology, performed the analyses except where noted, and led the writing of the manuscript. \textbf{MRM} performed the analysis and methods development of the Y6 \isd{} weights, contributed to the associated \isd{} section of the manuscript, helped construct the SP maps and contributed to general methods development. \textbf{NW} and \textbf{MRM} co-led this project, which was performed in close coordination with the associated companion paper defining the Y6 footprint \cite{y6-mask}. \textbf{JEP} generated the mocks and contributed to methods development. \textbf{ISN} selected the baseline \maglim{} sample, helped construct SP maps and contributed to analyzing objects removed by the \maglimpp{} quality cuts. \textbf{AP} ran the $\wtheta$-only MCMC chains and provided guidance with \textbf{SA} as working group leads. \textbf{SA} contributed feedback on the manuscript. \textbf{SL} contributed to fitting mock skewnesses to data, \textbf{WR} ran the spatially-varying $n(z)$ tests, \textbf{MT} performed early tests with LSS in SP maps, \textbf{DH} contributed text to the introduction and feedback on the manuscript, \textbf{JP} helped compute catalog-level $\wtheta$ measurements, \textbf{JdV} helped produce various \dnf{} products, \textbf{JM} contributed to the early \maglim{} selection, \textbf{MC} provided guidance to the project, \textbf{CS} and \textbf{GB} provided the SOM code, and \textbf{EH} and \textbf{RC} ran exploratory analysis of the dependence of Y3 SP maps on external LSS tracers. \textbf{ISN} and \textbf{AR} provided detailed feedback on the manuscript as DES internal reviewers.

The remaining authors have made contributions to this paper that include, but are not limited to, the construction of DECam and other aspects of collecting the data; data processing and calibration; developing broadly used methods, codes, and simulations; running the pipelines and validation tests; and promoting the science analysis.

\textbf{Software:}
The analysis made use of the software tools {\sc SciPy}~\cite{Jones:2001}, {\sc Astropy}~\cite{astropy:2013,astropy:2018}, {\sc NumPy}~\cite{numpy:2020,Oliphant:2006},  {\sc Matplotlib}~\cite{matplotlib:2007}, {\sc Scikit-Learn}~\cite{scikit-learn}, {\sc CAMB}~\cite{Lewis:1999,Howlett:2012}, {\sc GetDist}~\cite{getdist:2019}, {\sc HealPix}~\cite{Gorski_2005}, {\sc HealPy}~\cite{healpy},  {\sc HealSparse}~\cite{healsparse},  {\sc SkyProj}~\cite{skyproj},  
and \cosmosis~\cite{Zuntz:2015}. Elements of the DES modeling pipeline additionally use {\sc Cosmolike}~\cite{Krause:2016jvl} including {\sc CosmoCov}~\cite{Fang_2020}, {\sc Halofit}~\cite{2012ApJ...761..152T, Bird:2011rb}, {\sc Fast-PT}~\cite{McEwen:2016fjn,Fang_2020}, and {\sc Nicaea}~\cite{Kilbinger:2008gk}.

{\bf Funding and Institutional Support:} 
Funding for the DES Projects has been provided by the U.S. Department of Energy, the U.S. National Science Foundation, the Ministry of Science and Education of Spain, 
the Science and Technology Facilities Council of the United Kingdom, the Higher Education Funding Council for England, the National Center for Supercomputing 
Applications at the University of Illinois at Urbana-Champaign, the Kavli Institute of Cosmological Physics at the University of Chicago, 
the Center for Cosmology and Astro-Particle Physics at the Ohio State University,
the Mitchell Institute for Fundamental Physics and Astronomy at Texas A\&M University, Financiadora de Estudos e Projetos, 
Funda{\c c}{\~a}o Carlos Chagas Filho de Amparo {\`a} Pesquisa do Estado do Rio de Janeiro, Conselho Nacional de Desenvolvimento Cient{\'i}fico e Tecnol{\'o}gico and 
the Minist{\'e}rio da Ci{\^e}ncia, Tecnologia e Inova{\c c}{\~a}o, the Deutsche Forschungsgemeinschaft and the Collaborating Institutions in the Dark Energy Survey. 

The Collaborating Institutions are Argonne National Laboratory, the University of California at Santa Cruz, the University of Cambridge, Centro de Investigaciones Energ{\'e}ticas, 
Medioambientales y Tecnol{\'o}gicas-Madrid, the University of Chicago, University College London, the DES-Brazil Consortium, the University of Edinburgh, 
the Eidgen{\"o}ssische Technische Hochschule (ETH) Z{\"u}rich, 
Fermi National Accelerator Laboratory, the University of Illinois at Urbana-Champaign, the Institut de Ci{\`e}ncies de l'Espai (IEEC/CSIC), 
the Institut de F{\'i}sica d'Altes Energies, Lawrence Berkeley National Laboratory, the Ludwig-Maximilians Universit{\"a}t M{\"u}nchen and the associated Excellence Cluster Universe, 
the University of Michigan, NSF NOIRLab, the University of Nottingham, The Ohio State University, the University of Pennsylvania, the University of Portsmouth, 
SLAC National Accelerator Laboratory, Stanford University, the University of Sussex, Texas A\&M University, and the OzDES Membership Consortium.

Based in part on observations at NSF Cerro Tololo Inter-American Observatory at NSF NOIRLab (NOIRLab Prop. ID 2012B-0001; PI: J. Frieman), which is managed by the Association of Universities for Research in Astronomy (AURA) under a cooperative agreement with the National Science Foundation.

The DES data management system is supported by the National Science Foundation under Grant Numbers AST-1138766 and AST-1536171.
Data access is enabled by Jetstream2 and OSN at Indiana University through allocation PHY240006: Dark Energy Survey from the Advanced Cyberinfrastructure Coordination Ecosystem: Services and Support (ACCESS) program, which is supported by U.S. National Science Foundation grants 2138259, 2138286, 2138307, 2137603, and 2138296.
The DES participants from Spanish institutions are partially supported by MICINN under grants PID2021-123012, PID2021-128989 PID2022-141079, SEV-2016-0588, CEX2020-001058-M and CEX2020-001007-S, some of which include ERDF funds from the European Union. IFAE is partially funded by the CERCA program of the Generalitat de Catalunya.

We  acknowledge support from the Brazilian Instituto Nacional de Ci\^encia
e Tecnologia (INCT) do e-Universo (CNPq grant 465376/2014-2).

Part of this research was carried out at the Jet Propulsion Laboratory, California Institute of Technology, under a contract with the National Aeronautics and Space Administration (80NM0018D0004).

This document was prepared by the DES Collaboration using the resources of the Fermi National Accelerator Laboratory (Fermilab), a U.S. Department of Energy, Office of Science, Office of High Energy Physics HEP User Facility. Fermilab is managed by Fermi Forward Discovery Group, LLC, acting under Contract No. 89243024CSC000002.

This research used resources of the National Energy Research Scientific Computing Center (NERSC), a Department of Energy User Facility using NERSC award HEP-ERCAP0031464.
We acknowledge the use of Spanish Supercomputing Network (RES) resources provided by the Barcelona Supercomputing Center (BSC) in MareNostrum 5 under the allocation 2025-2-0046.

\end{acknowledgements}


\section*{Data Availability}
The galaxy catalogs, weights, and clustering measurements that are the main products of this work will be made public
as part of the DES Y6 coordinated release following journal acceptance of the DES Y6 Cosmology Results papers (\url{https://www.darkenergysurvey.org/des-y6-cosmology-results-papers/}). The DES Y6 Gold catalog derived from the DES Data Release 2 (DR2) is publicly available at \url{https://des.ncsa.illinois.edu/releases}.


\bibliographystyle{apsrev4-2}
\bibliography{refs, y6kp, des, des_y1kp_short, y3kp}




\appendix
\section{LSS in Weights}
\label{sec:lss_in_weights_app}

It is possible that the template library used to identify and mitigate systematics in the selection function contains some dependence on actual LSS modes. As noted in Sec.~\ref{sec:template_rejection}, this can have a pernicious effect by nulling actual LSS fluctuations and biasing cosmological inference. 

This occurred while investigating potential systematics in the DES Y3 \threextwopt{} analysis. When the DES Y3 \redmagic{} sample displayed an abnormally large galaxy clustering signal and it was determined that this was unlikely to be cosmological in origin, there was a significant effort to identify if there were residual angular systematics in the galaxy clustering signal that were not adequately characterized and removed with the initial galaxy weights. This included significant expansion of the library of systematic templates to $\mathcal{O}(100)$ survey property maps, so as to loosen the prior on what spatial patterns the modeled selection function was allowed to take. 

\begin{figure}[htbp]
    \centering
    \includegraphics[width=0.45\linewidth]{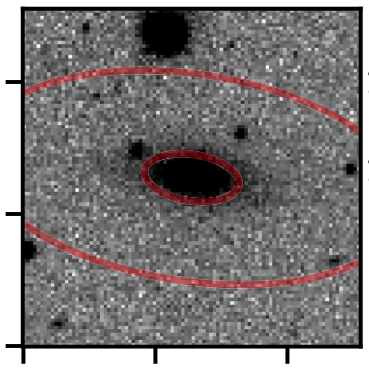}
    \hfill
    \includegraphics[width=0.45\linewidth]{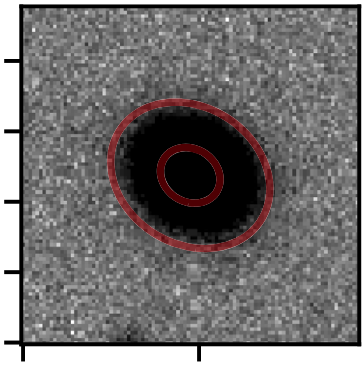}
    \caption{Illustration of a catastrophic failure mode detected in Y3 \texttt{Balrog} injections. The error on recovered fluxes from injected objects have a long tail, with a small fraction of objects having catastrophically misestimated sizes and fluxes. This tends to occur in strongly crowded fields as illustrated in the left subplot, with the $50\%$ and $99\%$ contours in red, which correlates with the estimated \texttt{SExtractor FLUX\_RADIUS}. The right subplot shows a correctly recovered profile. See \citet{Everett_2022} for details, reproduced from their Fig.~17.}
    \label{fig:fluxrad}
\end{figure}

The extended library produced significantly larger galaxy weights which were inconsistent with expectations for chance correlations with LSS, suggesting that true systematic contamination was being captured. However, subsequent investigations instead determined that this was due to the ability of the expanded template library to fit actual LSS modes, not just additional systematic modes, as determined by cross-correlating the weights (and later the templates) with external LSS tracers. This motivated the null tests in Secs.~\ref{sec:template_rejection} and \ref{sec:more_weights_tests} for the Y6 analysis. 

The extended library was able to fit true LSS modes because of the inclusion of both \texttt{FWHM\_WMEAN} and \texttt{FWHM\_FLUXRAD\_WMEAN} maps as templates. These are different summary statistics related to the PSF, with the former relating to the PSF model and the latter to average size of sources used to estimate the PSF \cite{Sevilla_Noarbe_2021}. They are highly correlated ($\rho\approx97\%$), but it wasn't \textit{a priori} clear which is more likely to trace a systematic in the galaxy fields, and hence both were included in the expanded library. However, injection tests with \texttt{Balrog} have identified a small tail of objects whose size (and flux) can be catastrophically over-estimated in heavily crowded fields \cite{Everett_2022}. 
The \texttt{FWHM\_FLUXRAD\_WMEAN} maps {include this effect moreso} than the \texttt{FWHM\_WMEAN} maps, such that \textbf{the difference of these maps acts as a tracer of the background object density.} 

\begin{figure}
    \centering
    \includegraphics[width=1\linewidth]{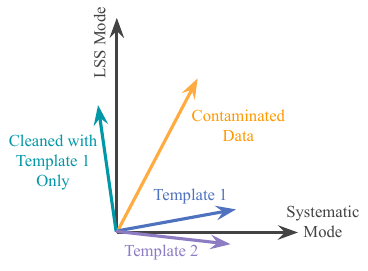}
    \caption{Schematic illustration of how including a second, correlated template with different sensitivity to true LSS can provide model freedom to fully null LSS modes. The x-axis represents a true systematic mode (a $N_{\rm pix}$-dimensional vector, or map, here shown in 1D ) and the y-axis an LSS mode. The observed contaminated data (orange) is a combination of each. Regressing the density against templates is analogous to removing any portion of the contaminated data that projects onto the basis formed by the templates. While either mode \textit{individually} does a reasonable job of removing the systematic contribution, if both Templates 1 and 2 are used \textit{together}, they fully deproject the LSS mode, despite both individually roughly tracing the systematic.  This illustrates the danger of including strongly correlated templates that might depend differently on actual LSS when removing angular systematics via deprojection or galaxy weights. }
    \label{fig:lss-templates_vectors}
\end{figure}

This is apparent when viewing the template library as a basis of modes to be nulled or deprojected; including both maps provides an additional basis vector that traces actual LSS, enabling it to be fit and removed.
Consider two templates, $t_0 = \alpha_0 \mathcal{F} + \epsilon_0 \delta_{LSS}$ and $t_1 = \alpha_1 \mathcal{F} - \epsilon_1 \deltasub{LSS}$, where $\mathcal{F}$ represents the true contamination and both templates contain small leakage from an LSS mode $\deltasub{LSS}$ ($\alpha_i F \gg \epsilon_i \deltasub{LSS}$). 

In linear regression or mode deprojection schemes wherein all modes that can be fit by any linear combination of the templates are nulled, cleaning with either of the templates \textit{individually} results in a reasonable estimate with minor residual error:
\begin{align}
\hat{\delta}_{\rm LSS} &\approx \deltasub{obs}- \hat{{\mathcal{F}}}\\
&= (\deltasub{LSS} + \mathcal{F}) -\frac{\langle{t_i}(\deltasub{LSS} + \mathcal{F})\rangle}{\langle t_i t_i \rangle} t_i\\
&\approx \left(1- \frac{\epsilon_i}{\alpha_i}\right)\deltasub{LSS} - \frac{\epsilon_i\langle\deltasub{LSS}\deltasub{LSS}\rangle}{\alpha_i \langle\mathcal{F}\mathcal{F}\rangle} \mathcal{F}.
\end{align}

Using \textit{both} templates, however, provides a degree of freedom to fit and fully remove the LSS mode, since $\deltasub{LSS} = (\alpha_1 t_0 - \alpha_0 t_1)/(\alpha_1 \epsilon_0 + \alpha_0 \epsilon_1)$, even when
the LSS leakage in the templates is very small ($\epsilon_0, \epsilon_1 \ll \alpha_0, \alpha_1$).  This overfitting manifests as significant detection of large and anti-correlated coefficients for $t_0$ and $t_1$, where the combination effectively amplifies the small $\delta_{\text{true}}$ contributions, and this was observed in the DES Y3 results when using both \texttt{FWHM} maps.

Fig.~\ref{fig:lss-templates_vectors} illustrates this schematically, assuming a toy model with a single LSS mode ($\deltasub{LSS}$) and a single systematic mode $\mathcal{F}$ (a mode can be thought of as just a spatial pattern or map, which can be represented as an $N_{\rm pix}$-dimensional vector). The observed map (orange) is a sum of true LSS and systematic modes, which lie along the $y$- and $x$-axes, respectively, such that deprojecting the $x$-component would recover the true LSS map. Deprojecting Template 1 (blue), results in a map (turquoise) that is close to the true LSS map, albeit with small amounts of LSS suppression and residual contamination. Deprojecting both Templates 1 and 2 fully nulls both the systematic and LSS modes, despite both templates being very reasonable approximations of the systematic.\footnote{Note this is not \textit{just} because the number of templates matches the number of dimensions of the data in this simplified model; it will hold in any scenario where the difference between two maps (or indeed any function accessible by the model space provided for mitigation) is significantly correlated with actual LSS.} 
Because of this, it would be far better to use either imperfect template in isolation, than to use them in combination. 

This explains why using the first 50 principle components of the PCA templates in Y3 resolved the issues of LSS in the weights \cite{DES:2021bat} -- the dimensions that correlated with LSS were in the very minor differences in the responses to background clustering of highly correlated summary statistics that were used as templates in the extended library, and hence concentrated in the highest principle components (corresponding to the smallest eigenvalues) describing only very small amounts of variation in the template space.

The possibility of introducing LSS into the templates presents a parallel challenge to the statistical mode suppression that comes from adding more templates or model freedom when modeling angular systematics in order to identify and mitigate unknown systematics, and one that cannot be easily characterized using mocks. 

This risk can be identified in several ways. One is to use the null checks described in Sec.~\ref{sec:template_rejection} to identify templates that might contain LSS (either individually or in combination), and remove them from the template library. After fitting for the weights, correlating them with external tracers can give indications of if they contain LSS as in Sec.~\ref{sec:weights_corr_external_tracers}. For more interpretable systematics models (e.g., which are linear in the templates) obtaining pairs of template coefficients that are large and strongly anti-correlated is a sign that they might be fitting LSS, especially when the templates are different summary statistics of the same property (e.g., median vs.\ mean for co-added exposures).

To mitigate this failure mode, adding a zero-centered prior (i.e., regularization) should help reduce the amount of LSS that is removed in these cases by penalizing fits with large template amplitudes (this already happens with \enet{}), but since the templates are fitting signal that is actually present in the data as opposed to mitigating chance correlation, it is unlikely to be particularly effective (nevertheless as shown in \cite{Weaverdyck:2020mff}, such a prior will very likely improve the accuracy of sky LSS maps even in the absence of LSS in the templates and is recommended over a vanilla OLS regression or mode deprojection). Adopting strongly correlated priors for templates that describe similar systematics may be a way to avoid imposing strong theoretical constraints on which templates may contaminate while mitigating this particular failure mode, as would using a PCA basis with the highest modes penalized or removed as done the DES Y3 analysis.

At minimum, we recommend avoiding the temptation to include multiple templates for the same systematic when there is the possibility of sensitivity to LSS. Including both the SFD dust map \cite{Schlegel_1998} and a DESI-specific dust map \cite{Zhou_2025_dust} would be one example of this (especially given that much of the error in the SFD dust map is thought to come from LSS \cite{Zhou_2025_dust}).


\section{Mocks}\label{sec:mocks}
We generate lognormal density mocks for each bin in the manner outlined in \texttt{FLASK} (\cite{2016MNRAS.459.3693X}), with angular power spectra generated by the modeling pipeline described in \cite{DES:2021rex}. We use a fiducial cosmology of $\Omega_m=0.3$, $h_0=0.69$, $\Omega_b=0.048$, $n_s=0.97$, $A_s=2.19\times10^{-9}$, $\Omega_{\nu}h^2=0.00083$, a fiducial linear galaxy bias of $b=[1.42,1.66,1.70,1.62,1.78,1.75]$, and no magnification bias. We largely follow the same procedures as in the Y3 \threextwopt{} analysis, and refer the reader to App. A of \cite{DES:2021bat} for details. We improve upon the Y3 approach by directly calibrating the density distribution skewness values ($k_0$) from the data rather than assuming a fixed theoretical value, which we describe below.

 We create pixelized maps of the overdensity for bin $i$ as $\delta_m = \delta^i_g/b_i$, where $b_i$ is the fiducial linear galaxy bias, and fit the distribution of values to
a shifted lognormal function, varying both the skewness ($k_0$) and Gaussian standard deviation ($\sigma_G$) of the field \cite{2014MNRAS.444.3473T}:
 
 \begin{multline}\label{eq:lognorm}
     P(\delta_m) = \frac{1}{\sqrt{2\pi}(\delta_m/k_0 + 1)\sigma_G} \\
     \times \exp{\left[ 
        -\frac{\left[ |k_0|\ln{(\delta_m/k_0 + 1)} + \sigma_G^2/(2|k_0|) \right]^2}
        {2\sigma_G^2} 
     \right]},
 \end{multline}

Discreteness in the overdensity makes for unstable results if fitted at high resolution (e.g., \nside{}=1024); however, as noted in \citet{2018PhRvD..98b3507G}, a smoothed and unsmoothed lognormal field can be well described using the same skewness parameter, and indeed we find that fixing $k_0$ to its best-fit obtained at $\nside{}=128$ enables us to flexibly and robustly fit the observed distribution at higher resolutions by varying only $\sigma_G$. Using these best-fit values of $k_0 = [0.76, 0.75, 0.79, 0.77, 0.71, 0.79]$, we create lognormal mocks at $\nside{}=1024$ with angular power corresponding to our fiducial cosmology, as computed via \texttt{CAMB} \cite{Lewis:1999}.

We create several batches of $\sim\mathcal{O}(1000)$ such mocks for different purposes, including estimating covariances of data and fit parameters in the absence of contamination when computing \isd{} weights (Sec.~\ref{sec:isd}), as well as the additive bias term to correct for mode nulling for different weights methods under different assumptions (Sec.~\ref{sec:bias_and_cov_corrections}).

\section{$\chi$-by-eye with $w(\theta)$}
\label{sec:wtheta_correlation}
\begin{figure*}[htbp]
    \centering
    \begin{minipage}{0.39\textwidth}
        \includegraphics[width=\textwidth]{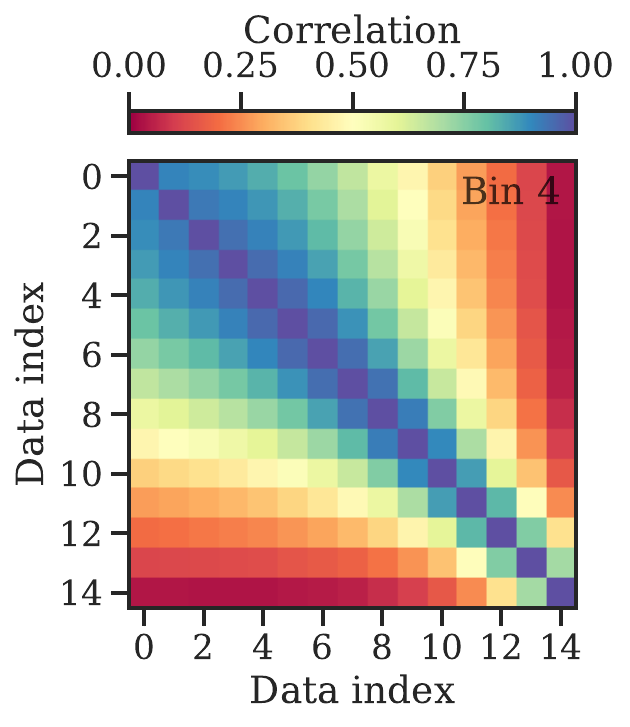}
    \end{minipage}%
    \hfill
    \begin{minipage}{0.59\textwidth}
        \includegraphics[width=\textwidth]{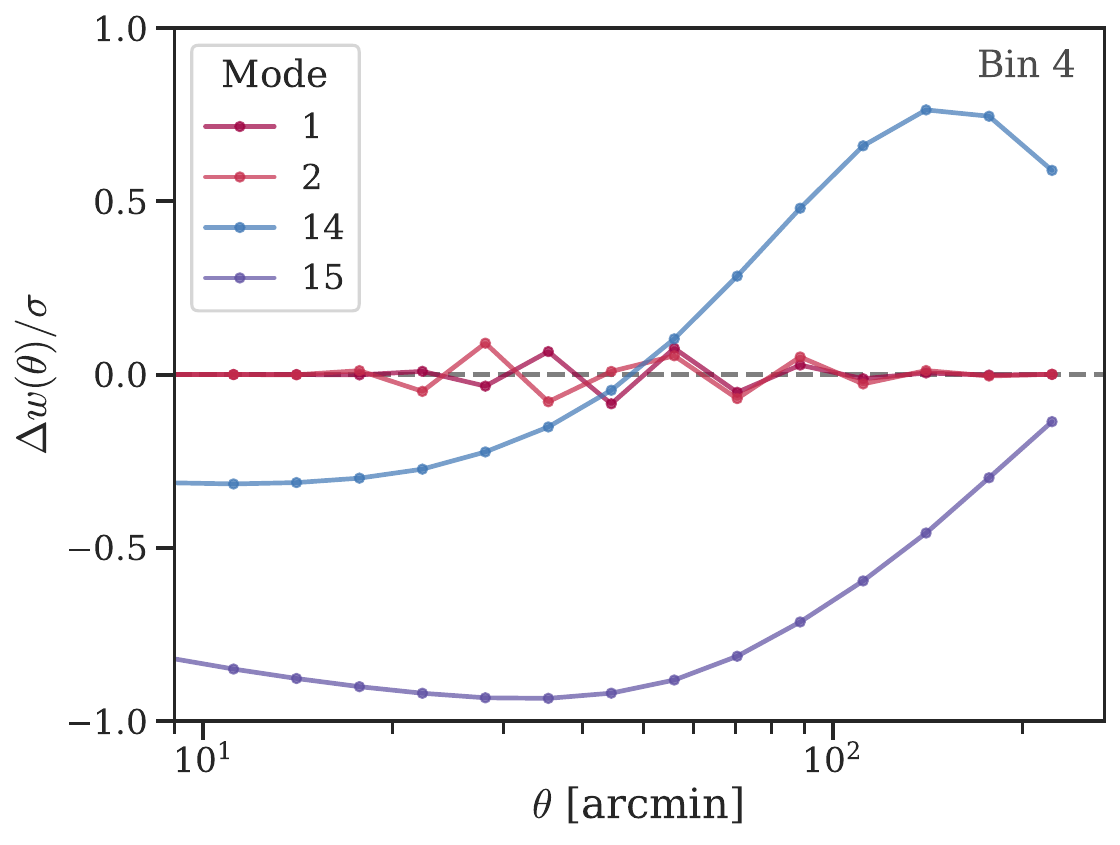}
    \end{minipage}
    \caption{\textit{Left}: Correlation matrix of $w(\theta)$ data points for Bin {4}. \textit{Right}: First and last two eigenmodes of the correlation matrix, scaled such that $\chi^2=1$ for each mode. Strongly correlated shifts of $w(\theta)/\sigma$ (blue, purple) are the least-well constrained data modes, whereas high-frequency wiggles are among the best constrained. This can lead to counter-intuitive results of significance when observing the residual plots, where low-amplitude scatter is typically assumed to be negligible. }\label{fig:wcorr_modes}
\end{figure*}

Because of the strong correlation between data points, it is quite challenging to infer by eye the significance of differences in plots of $w(\theta)$ and we in general caution against over-interpreting large shifts in $w(\theta)$, especially in comparison to shifts in $\gamma_t$ or $\xi_{\pm}$, which show much less correlation.
The left plot of Fig.~\ref{fig:wcorr_modes} displays the correlation matrix ($\rho_{ij} = \mathcal{C}_{ij}/\sqrt{\mathcal{C}_{ii}\mathcal{C}_{jj}}$) of $w(\theta)$ data points for redshift bin 4, showing high correlation across points. On the right, we show the first and last eigenmodes of the correlation matrix, scaled by their respective eigenvalues, such that each fluctuation corresponds to a $\chi^2=1$. The least constrained mode corresponds to a roughly uniform shift in $w(\theta)/\sigma$ (as typically normalized in residual plots, here $\sigma_i=\sqrt{\mathcal{C}_{ii}}$), whereas the best-constrained mode corresponds to high-frequency oscillations. Even a very small amplitude scatter of $\Delta w(\theta)/\sigma \ll 1$ can result in high significance, contrary to ones intuition with less correlated datasets. {This is typical across all of the \maglimpp{} redshift bins.}

\section{Discussion of Bin Two}\label{sec:bin2}
All redshift bins successfully passed the series of validation tests described in this work before unblinding, with no clear anomalies. However when performing the pre-unblinding tests with all parts of the data vector combined for the main \threextwopt{} cosmological analysis (c.f. Sections IV.H and Appendix B in \citet{y6-3x2pt}), we failed the posterior predictive distribution ($\Delta$PPD) goodness-of-fit criteria and identified anomalous behavior in the nuisance parameters associated with the $n(z)$ and magnification for lens bin 2. After subsequent investigation, the decision was made to exclude statistics that use lens bin 2 from the analysis, as this sufficiently improved the goodness-of-fit statistics to meet the unblinding criteria with relatively little impact on constraining power. 

The exact cause of the poor goodness-of-fit in \threextwopt{} is not fully understood, with both $\gamma_t$ and $\wtheta$ showing tension when conditioned on the other parts of the data vector. Interpretation of these results is complicated by the fact that simulated tests with internally-consistent datavectors found that $\Delta$PPD criterion used for unblinding has a high false-positive rate for $p(w|\gamma_t,\xi_{\pm})$, with formally consistent data vectors often returning low $p$-values. 

Bin 2 shows no obvious anomalies when examining the galaxy clustering measurements alone. Careful examination of Figures 10, 11, 12, 14, 16, and 17, along with Table II, reveals that bin 2 neither exhibits the most extreme behavior among all bins nor displays any significant deviations across these validation tests. Although the root cause of the anomalous behavior observed in the 3×2pt analysis remains unknown, our systematic investigation finds no evidence connecting this issue to imaging artifacts or foreground contamination.

\begin{figure*}
    \centering
    \includegraphics[width=1\linewidth]{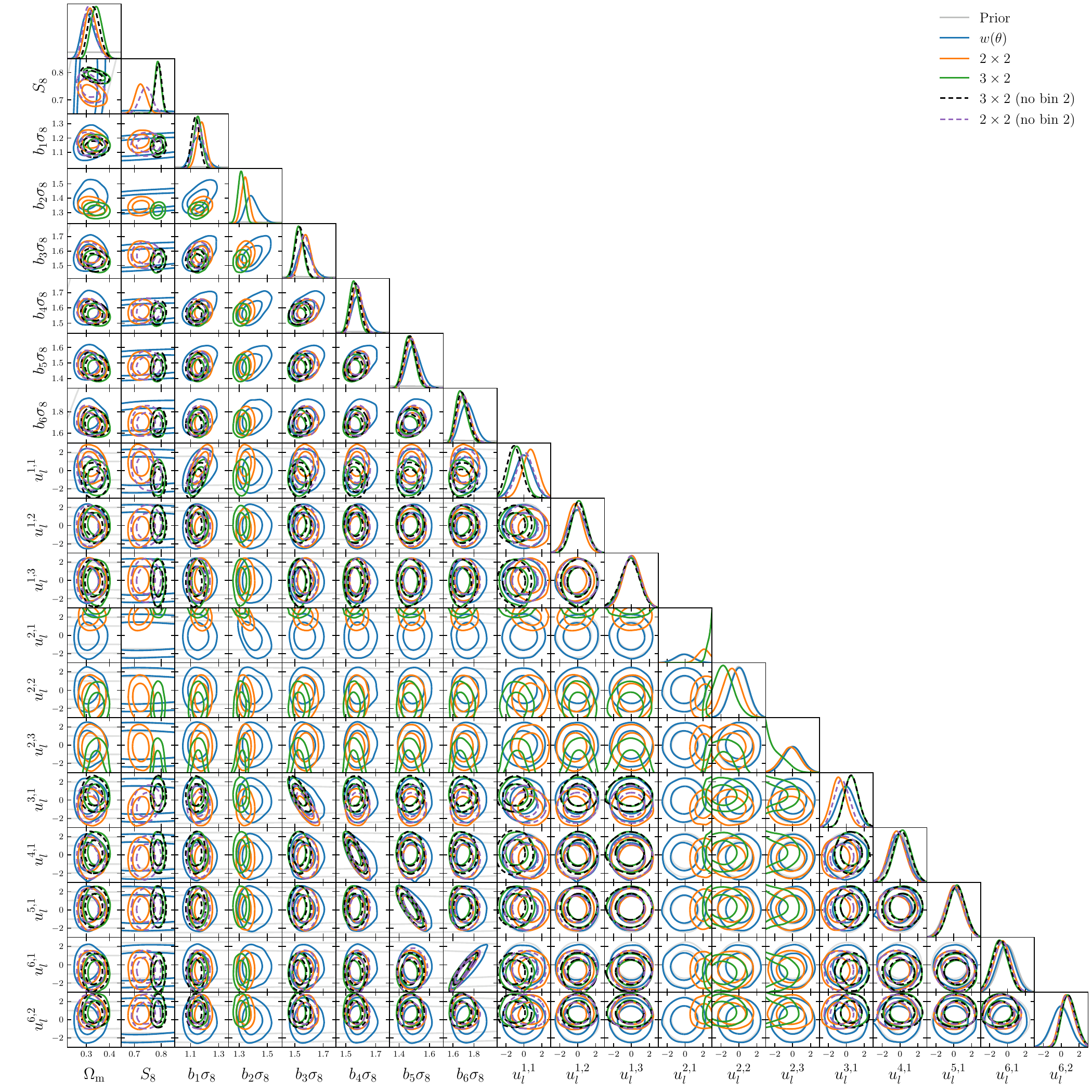}
    \caption{Joint posteriors on key cosmological and nuisance parameters with all bins for $\wtheta$ alone (blue) and in combination with other probes in the \twoxtwopt{} (orange) and \threextwopt{} (green) analyses. Priors are given in gray. For $w(\theta)$ only, the nuisance parameters $u_l^{i, j}$ are completely dominated by the \sompz{} prior; as more data is added the parameters for the second bin ($u_l^{2, j}$) show increasingly strong tension with the prior. This indicates some tension between the data and our model specification, either in the $n(z)$ or in other components that can be absorbed by the $n(z)$ nuisance parameters. Dashed black (purple) curves show the fiducial \threextwopt{} (\twoxtwopt{}) constraints from \citet{y6-3x2pt} after bin 2 has been removed, for comparison.
    }
    \label{fig:triangle}
\end{figure*}

Fig.~\ref{fig:triangle} shows posteriors on key cosmological parameters, bias amplitudes and redshift nuisance parameters ($u_l^{i,i}$, for bin $i$ and mode $j$) when using different combinations of DES data. We only show redshift nuisance parameters for the modes where we observe significant differences between the inferences, with bin 2 showing particularly strong shifts away from the priors derived from \sompz{} + WZ. This could reflect errors in the actual $n(z)$ characterization for this bin, or it could be that they're absorbing the effects of some other systematic. Since $\gamma_t$ and $\wtheta$ have different window functions for the same $n(z)$ and are thus sensitive to slightly different mean effective redshifts, redshift nuisance parameters could absorb predicted amplitude differences. Thus in addition to actual $n(z)$ systematics, larger than expected galaxy bias evolution \cite{pandey:2023tjn}, or anisotropic redshift variation \cite{lizancos_impact_2023} could plausibly contribute (though our limited tests of the latter showed no detection). {There could also be} a complex interplay of photometric systematics with the photo-$z$-dependent base \maglim{} selection function, which will be explored further in a future work.

\section{Maps}\label{sec:maps}
\begin{figure*}
    \centering
    \includegraphics[width=1\linewidth]{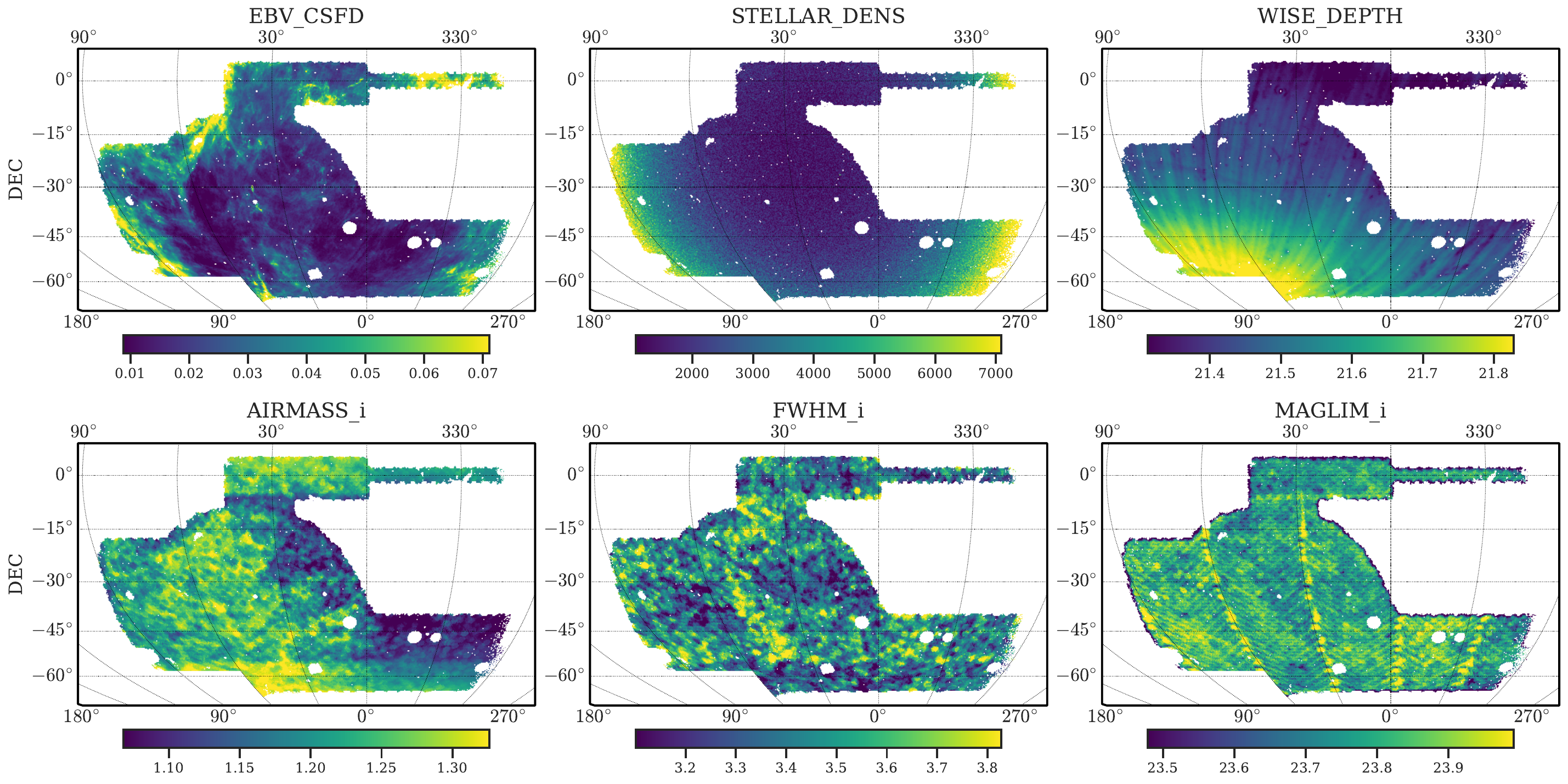}
    \caption{Example maps of survey property maps used as systematic templates. The top row are band independent maps, whereas the bottom row contains $i$-band versions of the band-dependent maps used for each of $griz$ in the fits (see \cite{y6-mask} for details of these maps).}
    \label{fig:spmaps}
\end{figure*}

\begin{figure*}
    \centering
    \includegraphics[width=1\linewidth]{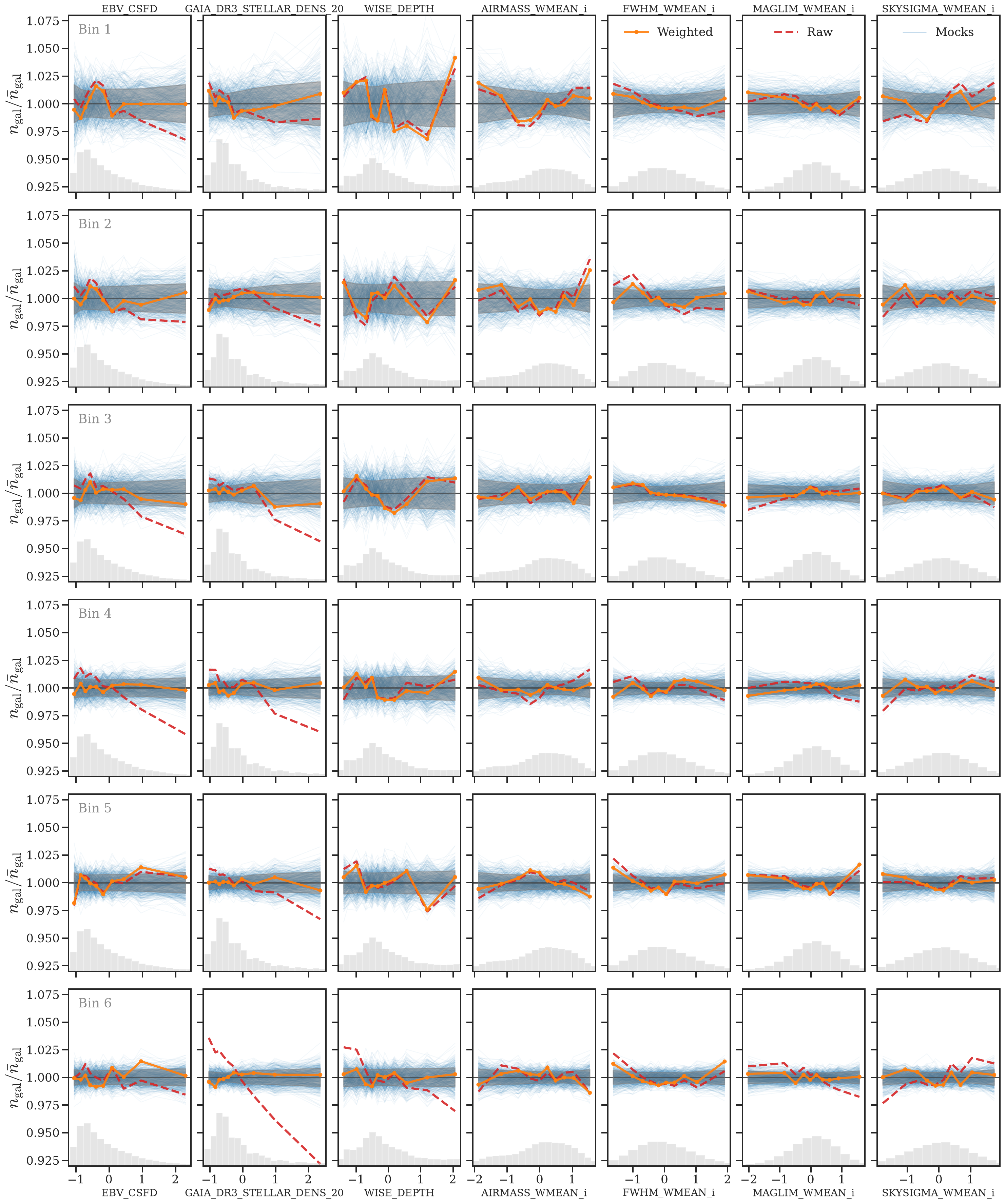}
    \caption{Same as Fig.~\ref{fig:ngal_sp_1d} but for all redshift bins (1-6 from top to bottom) for $i$-band survey property maps. Orange (red) curve shows the (un-)corrected density contrast in bins of each SP map before (red dashed) and after (orange solid) weighting. Gray shading indicates the $1\sigma$ range expected for the same from uncontaminated mocks (blue). After weighting, density fluctuations are controlled to $\lesssim2\%$ across bins of all SP maps and consistent with expectations from no residual contamination (c.f. Fig.~\ref{fig:chi2resid})}
    \label{fig:ngal_sp_1d_allbins}
\end{figure*}

Fig.~\ref{fig:spmaps} shows example survey property maps used as systematic templates for galaxy weights. The top row has band-independent quantities whereas the bottom row has $i$-band quantities, though we use templates for each of the $griz$ bands in the weights. {Note the \texttt{FWHM} maps are in units of \nside{} 16384 pixels; see \cite{y6-mask} for more details of the maps.} Fig.~\ref{fig:ngal_sp_1d_allbins} shows the trends of overdensity before and after weighting for all redshift bins, for $i$-band templates.

\begin{figure*}
    \centering
    \includegraphics[width=1\linewidth]{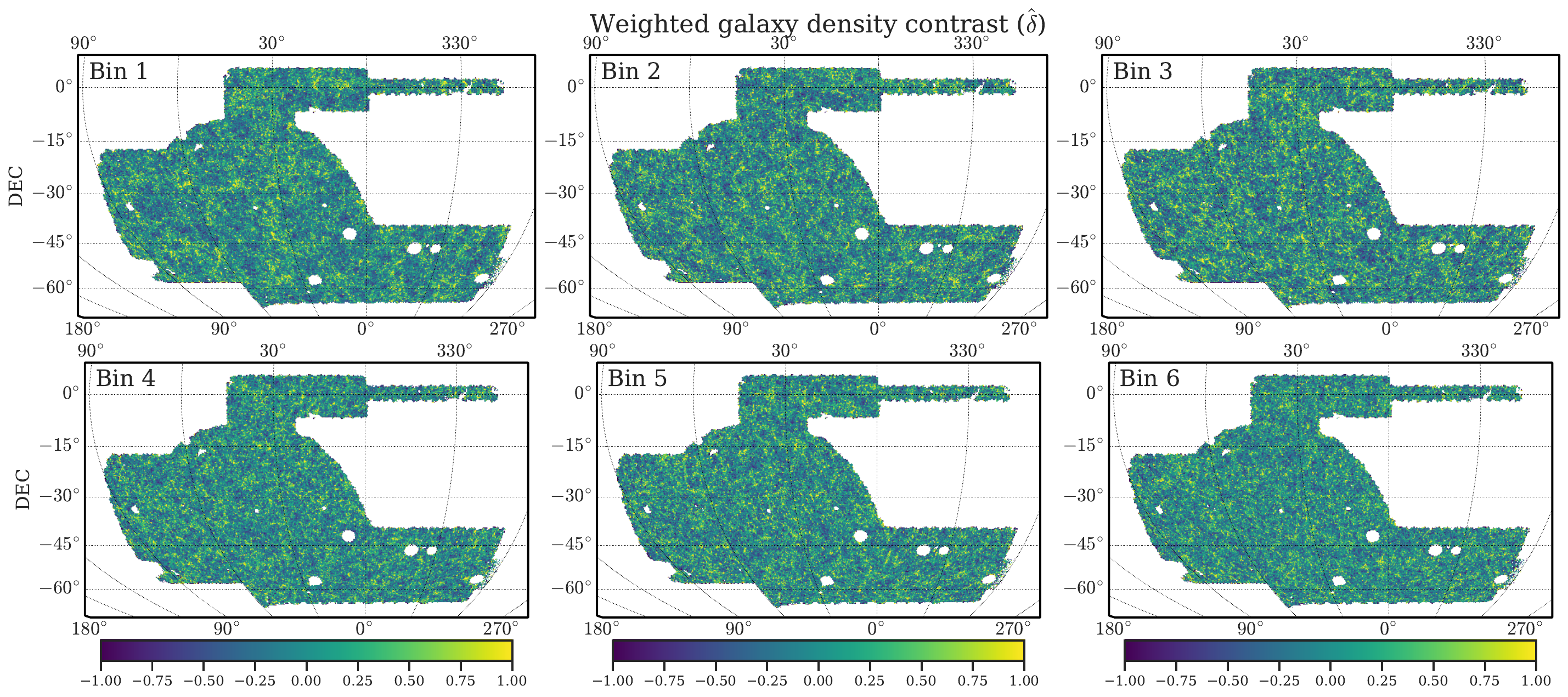}
    \caption{Estimated \maglimpp{} galaxy density contrast ($\hat{\delta}$) in each redshift bin at \texttt{NSIDE} 256, after applying the weights described in this work. No obvious systematic features are apparent.}
    \label{fig:overdensity_footprint}
\end{figure*}

As a visual check, Fig.~\ref{fig:overdensity_footprint} shows the overdensity field at \texttt{NSIDE} 256 in each redshift bin after applying the systematic galaxy weights from Sec.~\ref{sec:weights}. We see no obvious systematic features in the maps.

\section{Additional Weights Tests}
\label{sec:more_weights_tests}

\begin{figure*}
    \centering
    \includegraphics[width=\linewidth]{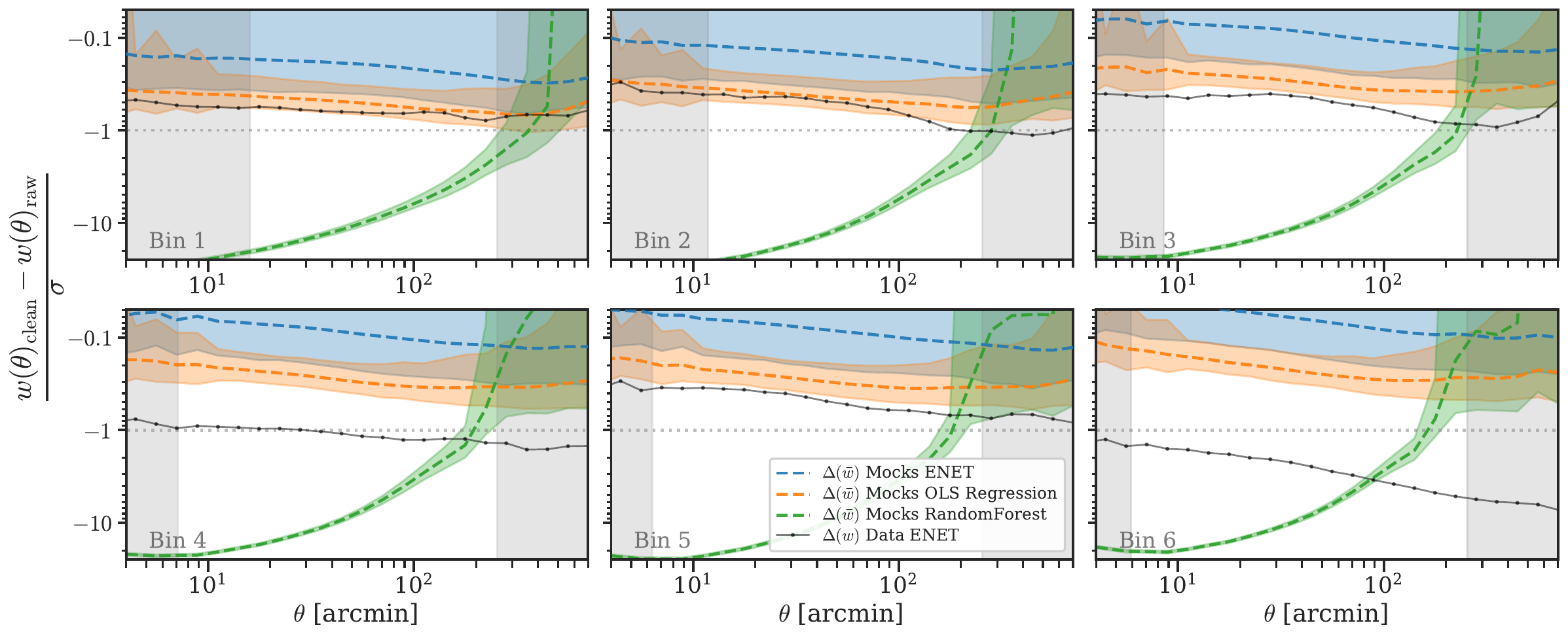}
    \caption{Same as Fig.~\ref{fig:overcorrection_mocks}, but also showing the average overcorrection from ordinary least squares (OLS) regression (orange) and a random forest regression (green), normalized by the covariance and using a log-scale. Dashed curves indicate the mean overcorrection to $\wtheta$ through application of the respective weights scheme, with shading indicating the central $68\%$ spread across realizations. The black line shows the observed difference between unweighted and \enet{}-weighted $w(\theta)$ on the data. \enet{} is the most robust to overfitting, whereas the random forest is extremely susceptible because of its unconstrained model space, despite having built-in cross-validation to optimize hyperparameters to minimize overfitting. Suppression from OLS regression is large enough that it would be difficult to identify if the observed contamination in the data were real or simply due to chance in the lowest three redshift bins. Dotted lines indicate $1\sigma$ error from the fiducial covariance and gray shaded regions indicate scale cuts.}
    \label{fig:overcorrection_mocks_all}
\end{figure*}

In developing this work, we have been guided by the philosophy that it's better to throw out a small amount of data and fit ``most" of it more correctly and parsimoniously, rather than to resort to unwieldy models with large amounts of model freedom so that it can fit more heterogeneous data, but at the expense of increased bias and variance.

We also focused using our model degrees freedom to minimize how informative our priors are on the spatial patterns of contamination (i.e., more spatial templates) and using reasonable perturbative assumptions for how a given survey property value impacts the observed density field. 

Here we test the impact of alternative weighting schemes. As noted in Sec.~\ref{sec:weights}, there is a tension between allowing enough model flexibility to fit and correct the selection function, but not so much as to significantly bias the inference through the removal of true LSS modes. Several works have compared the performance of machine learning methods such as a neural network or random forest algorithm against ``linear" weights (e.g., \cite{Chaussidon_2021, Rezaie:2019vlz}), but it's worth noting that the ``linear" weights in those works are such in the most restrictive sense -- they are derived from an OLS regression using the same set of base templates as the machine learning methods, and thus linear \textit{in the data} -- but in practice ``linear regression" only needs to be linear in the \textit{model parameters}, and can fit highly non-linear functions of the \textit{data} (e.g., fitting a polynomial). Furthermore, there is a very wide range of possible spatial templates that could plausibly trace systematic selection effects, but only a relatively small number of these are investigated (of course including too many templates increases the risk of the types of issues described in App.~\ref{sec:lss_in_weights_app}).

Here we allow considerably more freedom in the model than the linear and low-order polynomial models in Sec.~\ref{sec:bias_and_cov_corrections} by using a random forest algorithm to estimate the selection function from the same base template library. We use the \texttt{Scikit-Learn} \texttt{RandomForestRegressor} with \texttt{min\_samples\_leaf} $ =20$ \texttt{n\_estimators} $ = 200$, and use cross validation to optimize the hyperparameter \texttt{max\_depth} $\in[5, 10, 15]$ for predictive power as we did with \enet{}, keeping other parameters to their default. 
We also compare to an OLS regression with no regularization, which is equivalent to mode deprojection \cite{Elsner_2016, Alonso_2019},  after debiasing from overcorrection \cite{Weaverdyck:2020mff}.

Fig.~\ref{fig:overcorrection_mocks_all} demonstrates how the enormous amount of model freedom in the random forest results in significant bias on the inferred $\wtheta$ on uncontaminated mocks due to mode nulling, with far greater systematics coming from the weights process than the estimated impact from the raw contaminants in our data.  This is despite using cross-validation to optimize hyperparameters to avoid overfitting. While we only demonstrate a basic model here, where expanding the hyperparameter space could improve the results somewhat, our tests found it is unlikely to do so to an acceptable degree.
For comparison, we also show the bias imparted by the \enet{} weights (blue) and the OLS weights (orange), with dashed lines showing the mean and shaded curves the central $68\%$ spread. We normalize the bias to the uncertainty ($1\sigma$ indicated with a dotted line), and use a log scale to illustrate both the small impact from \enet{} and large impact from the random forest. Note that the mean and spread of the overcorrection from OLS is comparable to what we observe on the data with \enet{} in the first several bins, i.e., the error imparted by running mode deprojection would be comparable to the level of estimated systematics in these low redshift bins, making it hard to know if such weights would be actually improving or degrading the $\wtheta$ estimates.

\label{lastpage}
\end{document}